\documentclass{imsart}
\usepackage{url}
\usepackage{amsthm}
\usepackage{amssymb}
\usepackage{enumerate}
\usepackage{lmodern}

\usepackage{graphicx}
\usepackage{amsfonts}
\usepackage{amsmath}
\usepackage[utf8]{inputenc}
\usepackage{url}
\usepackage{color}
\usepackage{amsthm}
\usepackage{amssymb}
\usepackage{enumerate}
 \usepackage{paralist}
\usepackage{graphicx}

\usepackage{bbm}
\usepackage{mathtools}
\RequirePackage{amsthm,amsmath,amsfonts,amssymb}
\RequirePackage{natbib}
\RequirePackage[colorlinks,citecolor=blue,urlcolor=blue]{hyperref}
\RequirePackage{graphicx}

 {
     \theoremstyle{plain}
     
 }
 
\newtheorem {Proposition}{Proposition}[section]
\newtheorem {Lemma}[Proposition] {Lemma}
\newtheorem {Theorem}[Proposition]{Theorem}
\newtheorem {Corollary}[Proposition]{Corollary}
\newtheorem {Remark}[Proposition]{Remark}

\newtheorem {Definition}{Definition}[section]

\def\N{\mathbb{N}}

\def\R{\mathbb{R}}

\numberwithin{equation}{section}
\begin{document}
\begin{frontmatter}
\title{Nonparametric  Multiple-Output Center-Outward Quantile Regression\thanksref{t1}}

\runauthor{del Barrio, Gonz\'alez Sanz and Hallin}
\runtitle{Nonparametric Center-Outward Quantile Regression}
\thankstext{t1}{The research of Eustasio del Barrio is partially supported by FEDER,
Spanish Ministerio de Economía y Competitividad, grant MTM2017-86061-C2-1-P and Junta de
Castilla y León, grants VA005P17 and VA002G18. The research of Alberto González-Sanz is partially supported by the AI Interdisciplinary Institute ANITI, which is funded by
the French “Investing for the Future – PIA3” program under the Grant agreement ANR-19-PI3A-0004. }

\begin{aug}

\author[A]{\fnms{Eustasio} \snm{del Barrio}\ead[label=e1]{tasio@eio.uva.es}},
\author[B]{\fnms{Alberto} \snm{Gonz\'alez-Sanz}\ead[ label=e2]{alberto.gonzalezsanz@math.univ-toulouse.fr}}
\and
\author[C]{\fnms{Marc} \snm{Hallin}\ead[label=e3]{ mhallin@ulb.ac.be}}
\address[A]{IMUVA, Universidad de Valladolid, Spain.
\printead{e1}}

\address[B]{IMT, Universit\'e de Toulouse,
France. \printead{e2}}
\address[C]{ECARES and D\' epartement de Math\' ematique Universit\' e libre de Bruxelles, Brussels, Belgium.\\ 
\printead{e3}}
\end{aug}

\begin{abstract} Based on  the novel concept of multivariate center-outward quantiles introduced recently  in \cite{hallinChernozhukov2017} and  \cite{Hallin2020DistributionAQ}, 
we are considering the problem of nonparametric multiple-output quantile regression. Our approach  defines nested {\it conditional center-outward quantile regression contours}  and {\it regions}  with given conditional probability content irrespective of the underlying distribution; their  graphs 
  constitute nested {\it center-outward quantile regression  tubes}. 
 Empirical counterparts  of these concepts are constructed, yielding interpretable empirical regions and contours which are shown to   consistently reconstruct  their population versions  in the Pompeiu-Hausdorff topology.   
   Our method is entirely non-parametric and performs well in simulations including heteroskedasticity and  nonlinear trends;   its  power as a data-analytic tool is illustrated on some real datasets.
\end{abstract}


\end{frontmatter}


\maketitle

\begin{abstract}
...
\end{abstract}

\section{Introduction}
\subsection{Quantile regression,   single-  and multiple-output}
%

Forty-five years after its introduction by \cite{Koenker1978},   quantile regression---arguably the most powerful  tool in the statistical study of the dependence of a variable of interest $Y$ on cova\-riates~${\bf X}=(X_1,\ldots, X_m)$---has become part of statistical daily practice, with countless  applications in all domains of scientific research, from economics and social sciences to astronomy, biostatistics,  and medicine.  Unlike classical regression, which, somewhat narrowly, is focused on conditional means ${\rm E}[Y\vert {\bf X}]$, quantile regression indeed is dealing with the complete conditional distributions ${\rm P_{Y\vert {\bf X}={\bf x}}}$ of~$Y$ conditional on ${\bf X}={\bf x}$.  Building on that pioneering  contribution, a number of   quantile regression methods, parametric, semiparametric, and nonparametric,  have been developed for an extremely broad range of statistical topics, including time series, survival analysis, instrumental variables, measurement errors, and functional data---to quote only a few. Sometimes, a simple parametrized regression model allows for a parametric approach, yielding, for instance, linear quantile regression. In most situations, however,   parametric models are  too rigid and a more agnostic nonparametric approach is in order. We refer to \cite{koenker_2005} for an introductory text and to \cite{handbook} for a   comprehensive survey. 

 In single-output models (univariate variable of interest $Y$),  this nonparametric approach is well understood and well studied, and the  history of non-parametric estimation of conditional quantile functions  goes back, at least, to the seminal paper by  \cite{stone1977}. The results are much scarcer, however,  in the ubiquitous  multiple-output case ($d$-dimensional variable of interest $\bf Y$, with~$d>1$),  and the few existing ones are less satisfactory---the simple reason for this being the absence of a fully satisfactory concept of multivariate quantiles. 


A major difficulty with  quantiles in dimension $d>1$, indeed,  is   the fact that ${\mathbb R}^d$, contrary to $\mathbb R$,   is not canonically ordered. A number of attempts have been made to overcome that issue, the most remarkable of which is   the theory of statistical depth.  That theory has generated an abundant literature which we cannot summarize here---we refer to \citep{SerfZuo} or \citep{SerfStatNeerl,SerflingDepthI,SerflingDepthII} for general expositions and authoritative surveys. 

Several  depth concepts  coexist. The most popular of them is Tukey's halfspace depth \citep{Tukey75}, but   all depth concepts  (including the most recent ones: see, e.g.,  \cite{KonendPd22}) are sharing the same basic properties.  Tukey's halfspace depth characterizes, for each distribution $\rm P$ over ${\mathbb R}^d$ (for simplicity, assume  $\rm P$ to be Lebesgue-absolutely continuous) {\it depth regions} ${\mathbb D}_{\rm P}(\delta)$ (resp., {\it depth contours} ${\mathcal D}_{\rm P}(\delta)$) as collections of points with depth (relative to~$\rm P$) larger than or equal to $\delta$ (resp., equal to $\delta$), $\delta\in(0, 1/2]$. Depth regions are convex, closed, and nested as $\delta$ increases, and have been proposed as notions of quantile regions and contours---an interpretation that is supported by the L$_1$ nature of Tukey depth \citep{Hallin2010}. 

Among the merits of this interpretation is that it has imposed the idea that quantiles, in dimension $d\geq 2$, should rely on some center-outward   ordering with central region of depth~$\delta=1/2$ rather than a southwest-northeast extension of the classical univariate ``left-to-right'' linear ordering of the real line. Unfortunately, depth regions 
 fail to satisfy the quintessential property of quantile regions: the $\rm P$-probability content  ${\rm P}\big[{\mathbb D}_{\rm P}(\delta)\big]$ of the quantile region~${\mathbb D}_{\rm P}(\delta)$ indeed very much depends on $\rm P$. This is not a minor weakness: the univariate median~$Y_{1/2}^{\rm P}$ of an absolutely continuous distribution~$\rm P$, for instance, is characterized by the fact that~$\rm P\big[(-\infty,Y_{1/2}^{\rm P}]\big] =~\!1/2$ irrespective of ${\rm P}$---who  would call   {\it median} a quantity~${Y}_{\text{\rm med}}^{\rm P}$ such that ${\rm P}_1\big[(-\infty, {Y}_{\text{\rm med}}^{{\rm P}_1}]\big] = 0.4$ while~${\rm P}_2\big[(-\infty, {Y}_{\text{\rm med}}^{{\rm P}_2}]\big] = 0.6$~?\vspace{1mm} None of the depth concepts in the literature  is satisfying that essential property of quantiles, though, and depth regions, therefore, cannot be considered as {\it bona fide} quantile regions. 

A review of the various existing multiple-output quantile regression  models (linear and nonparametric, depth-based and others) can be found in   \cite{Halhandbook}. A nonparametric quantile regression model based on a directional form of   Tukey depth is developed in \cite{Halin2015} but suffers the same lack of control over the probability contents of the quantile regions involved as the depth-based quantile concept itself.  So does also the directional concept of M-quantiles  proposed by \cite{Merlo22}. 
\begin{figure}[t!]
    \centering
    \includegraphics[width=6.5 cm,height=6.5 cm]{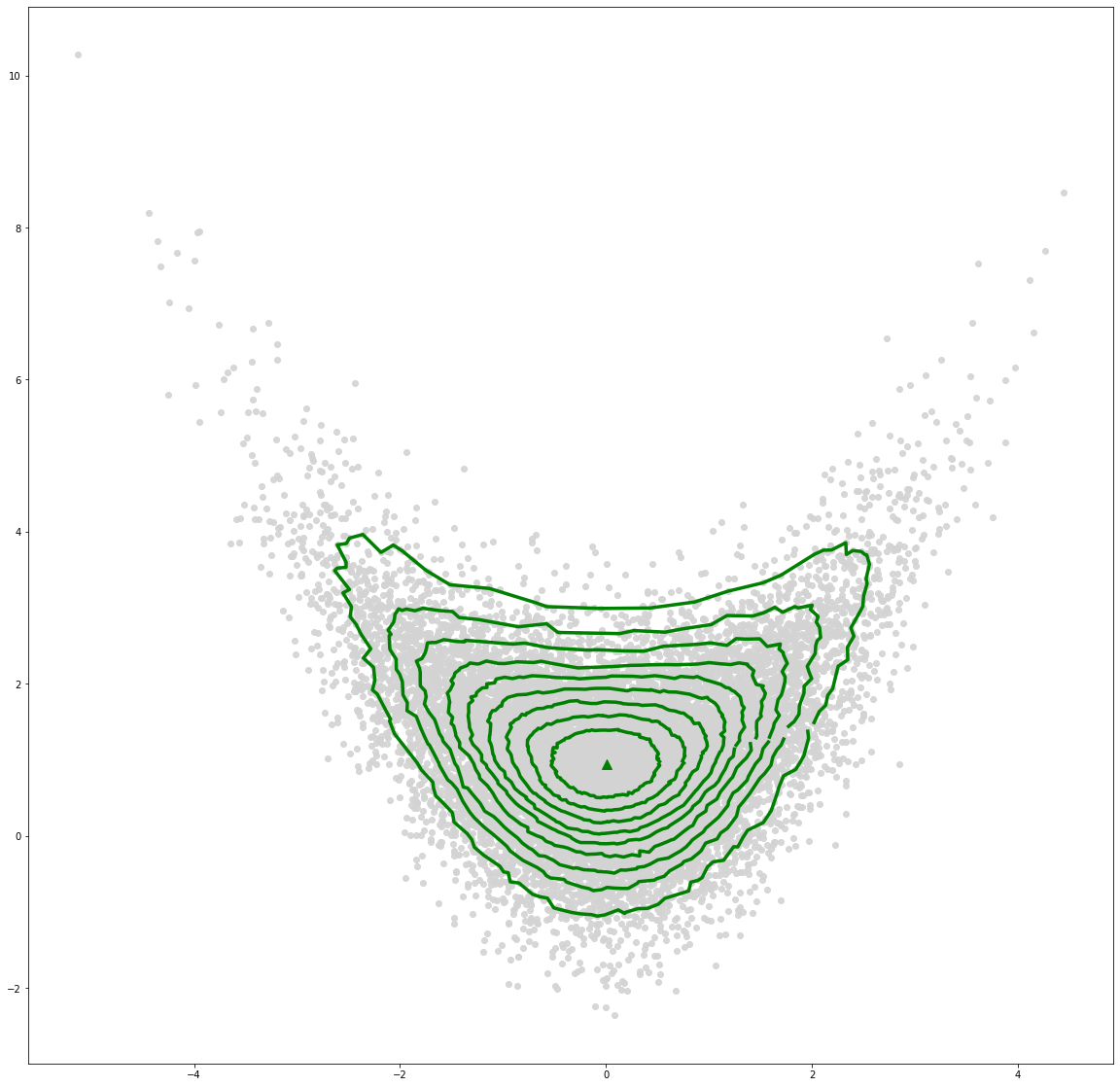}\ 
    \includegraphics[width=6.5 cm,height=6.5cm]{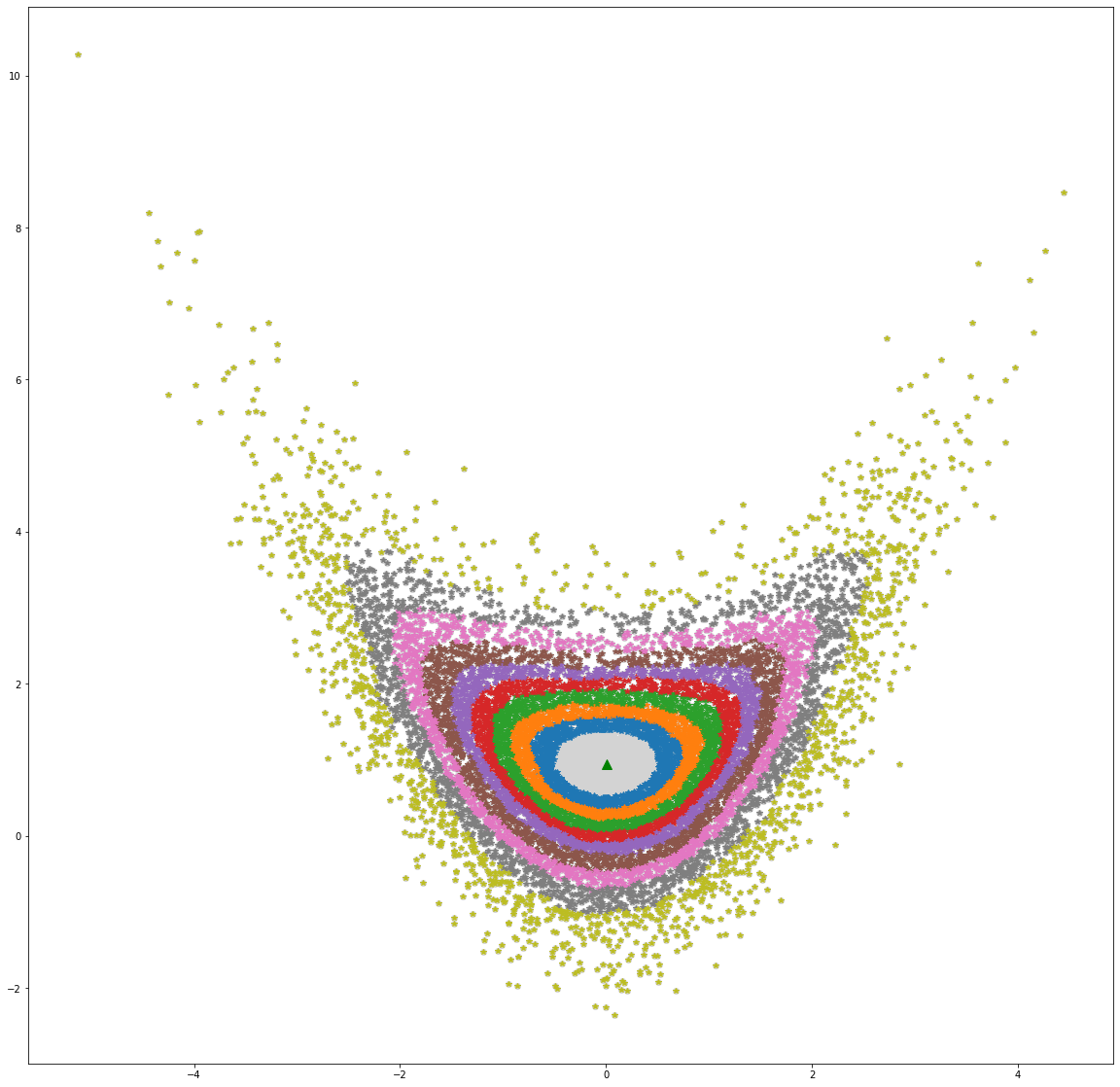}
    \caption{Quantile contours  $\mathcal{C}_{\pm}(\tau)$ (left panel) and regions $\mathbb{C}_{\pm}(\tau)$ (right panel) for the banana-shaped Gaussian mixture of Section~\ref{Sec:BananaReg} and the quantile orders   $\tau =0.1, \ 0.2, \dots, 0.9$. }
    \label{fig:corazon}
\end{figure}

Recently, based on measure transportation ideas, new concepts of quantiles in dimension~$d>1$ have been introduced in  \citep{hallinChernozhukov2017,Hallin17,Hallin2020DistributionAQ} under the names of {\it Monge-Kantorovich depth} and {\it center-outward quantile function}. Center-outward quantile functions define nested closed regions ${\mathbb C}_{\rm P}(\tau)$ and continuous contours ${\mathcal C}_{\rm P}(\tau)$ indexed by~$\tau\in ([0,1)$  such that, for any absolutely continuous $\rm P$, ${\rm P}\big[{\mathbb C}_{\rm P}(\tau)\big] = \tau$ irrespective of ${\rm P}$. Unlike the previous concepts, thus, and unlike the depth-based ones (as   proposed  in  \cite{Hallin2010} or \cite{Linglong}), these measure-transportation-based quantiles do satisfy the essential property that the $\rm P$-probability contents of the resulting quantile regions do not depend on $\rm P$. Moreover, the corresponding  quantile regions are not necessarily convex  and, as shown in Figure~\ref{fig:corazon}, they are able to capture the ``shape'' of the underlying distribution. We refer to  
\cite{HallinReview} for a survey of measure-transportation-based center-outward quantiles, the dual concepts of multivariate  ranks and signs, and their many applications in inference problems (\cite{GhosSen19,HallinVARMA, HallinIndAoS,DebSen21,HallinKendall,HallinMANOVA,HallinBernou}, among others). 

Motivated by this long list of successful applications,  we are proposing in this paper   a novel and meaningful solution,  based on the concept of center-outward quantiles,  to the problem of nonparametric multiple-output quantile regression. Namely, for a pair of multidimensional random variables $(\mathbf{X},\mathbf{Y})$ with values in $ \R^m\times\R^d$ ($\mathbf{Y}$ the variable of interest,~$\mathbf{X}$ the vector of covariates) and  joint distribution\footnote{For simplicity, in this introduction, we tacitly assume  all distributions to be Lebesgue-absolutely continous.} 
$\mathbb P$, we define (Section~\ref{section:Conditional})  the \emph{center-outward quantile map~$\mathbf{Q}_{\pm}$ of $\mathbf{Y}$ conditional  on $\mathbf{X}=\mathbf{x}$} as 
\begin{equation}\label{qmapdef}
{\bf u}\in
 {\mathbb S}_d
\mapsto \mathbf{Q}_{\pm}({\bf u} |\,
\mathbf{x})\in\R^d
\end{equation}
(${\mathbb S}_d$ the open unit ball in $\R^d$),  with the essential property that, letting  
\begin{equation}\label{condqregdef}
{\mathbb C}_\pm (\tau \vert  
\mathbf{x})\!\coloneqq\!  \mathbf{Q}_{\pm} (\tau\,\overline{\mathbb S}_d  \vert  
\mathbf{x}) \quad \tau\in (0,1), \ \  \mathbf{x} \in\mathbb{R}^m ,
\end{equation}
we have 
\begin{equation}\label{qmapptop}
{\mathbb P}
\left[\mathbf{Y}\in 
 {\mathbb C}_\pm (\tau \vert \, 
\mathbf{x})\big\vert\, \mathbf{X}=\mathbf{x} \right] = \tau\quad\text{for all $\mathbf{x}\in\R^m$,  $\tau\in (0,1)$, and $\mathbb P$},
\end{equation}
justifying the interpretation of~$\mathbf{x}\mapsto {\mathbb C}_\pm (\tau \vert \, 
\mathbf{x})$ as the value at $\mathbf{x}$ of a {\it regression quantile region of order $\tau$} of $\mathbf Y$ with respect to $\mathbf X$. For~$\tau =0$,  
\begin{equation}\label{qmed}{\mathbb C}_\pm (0\, \vert \, 
\mathbf{x})\coloneqq \bigcap_{\tau\in(0,1)}{\mathbb C}_\pm (\tau \vert \, 
\mathbf{x})
\end{equation} 
yields the value at $\mathbf X = \mathbf x$ of the {\it regression median} $\mathbf x\mapsto {\mathbb C}_\pm (0\, \vert \, 
\mathbf{x})$ of $\mathbf Y$ with respect to $\mathbf X$.
The same conditional quantile map   characterizes nested (no ``quantile crossing'' phenomenon)  {\it ``regression quantile tubes of order $\tau$''} (in $\R^{m+d}$)
\begin{equation}\label{qtube} {\mathbb T}_\pm (\tau)\coloneqq 
\left\{ \left(\mathbf{x}, \mathbf{Q}_{\pm} (\tau\,\overline{\mathbb S}_d \big\vert \,
\mathbf{x})\right)
\big\vert \mathbf{x}\in\R^m
\right\},\quad \tau\in (0,1)
\end{equation}
which are such that 
\begin{equation} \label{qtubetop}
{\mathbb P}\left[(\mathbf X , \mathbf Y)\in {\mathbb T}_\pm (\tau)
\right]= \tau\quad\text{ irrespective of ${\mathbb P}$,  $\tau\in (0,1)$} .
\end{equation} 
For $\tau =0$, define 
$$ {\mathbb T}_\pm (0)\coloneqq
\left\{ \left(\mathbf{x},  \mathbf{y})\right)
\big\vert \mathbf{x}\in\R^m,\,  \mathbf{y}\in {\mathbb C}_\pm (0\, \vert \, 
\mathbf{x})
\right\}=\bigcap_{\tau\in (0,1)}{\mathbb T}_\pm (\tau)$$ 
(the {\it graph} of~$\mathbf{x}\mapsto {\mathbb C}_\pm (\tau \vert \, 
\mathbf{x})$); with a slight abuse of language, also call $ {\mathbb T}_\pm (0)$ the {\it regression median} of $\mathbf Y$ with respect~to~$\mathbf X$.\

None of the earlier  attempts  to define multiple-output regression  quantiles---neither the depth-based definitions in \cite{Halin2015} or \citet{PAINDAVEINE2011193}, the directional concepts of marginal M-quantiles \citep{Breckling88} considered by  \cite{Merlo22}, nor   the measure-transportation-based approach   proposed by \cite{Carlier2016}---is characterizing quantile regions that satisfy  requirements~\eqref{qmapptop} and \eqref{qtubetop}.

\cite{Carlier2016} deserves special attention, though, as  the first attempt to break with directional and depth-based approaches to multiple-output quantile regression by  means of  innovative measure transportation ideas. Focusing on linear  quantile regression, their method is based on a  concept of multivariate quantile functions  defined over the unit cube~$[0,1]^d$ rather than the unit ball ${\mathbb S}_d$. While   yielding (under the assumption of a linear regression) an asymptotic reconstruction of the  distributions of $\mathbf{Y}$ conditional on $\mathbf{X}=\mathbf{x}$, however, their choice of    the unit cube  does not directly allow for the definition of quantile regions of given order similar to the center-outward regression quantile regions~${\mathbb C}_\pm (\tau \vert \, 
\mathbf{x})\subset~\!\R^d$ or the quantile regression tubes~${\mathbb T}_\pm (\tau)\subset\R^{m+d}$. 

   \begin{figure}[t!]
    \centering
    \includegraphics[width=12 cm,height=10cm, trim= 0mm 0mm 0mm 70mm, clip ]{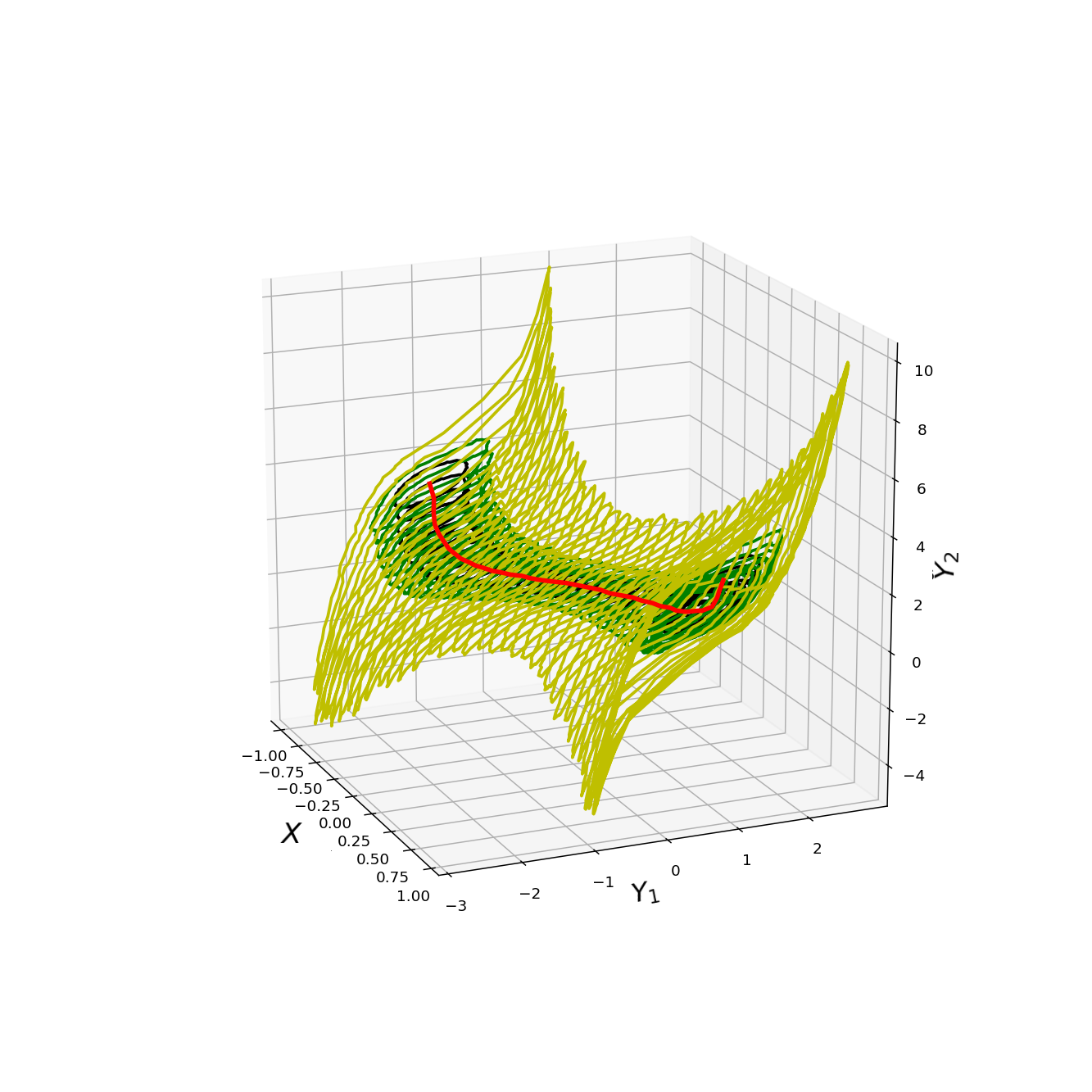}\vspace{-20mm}
    \caption{Quantile regression (two-dimensional variable of interest $\bf Y$; univariate regressor $X$), showing the  conditional center-outward medians (red) and  the conditional quantile contours $ (\mathbf{x}, {\mathbb C}_\pm (\tau \vert \,
\mathbf{x}))$of order $\tau=0.2$ (black),   $\tau =0.4$ (green), and $\tau =0.8$ (yellow). 
     }
    \label{fig:sinus}
\end{figure}\vspace{-00mm}

As for the depth-based quantile regions, moreover, they  are necessarily convex, even for distributions with highly non-convex shapes as in Figure~\ref{fig:corazon}. To circumvent this convexity problem, \cite{feldman2021calibrated} propose a clever  transformation of  the data  turning its distribution into a latent one  with convex level sets  by fitting a conditional variational 
    auto-encoder (\cite{NIPS2015}). The  probability contents of  the quantile regions  resulting from this machine-learning type of {\it ``convexity reparation,''} however, remain out of control (they still depend on ${\mathbb P}$).


Figure~\ref{fig:sinus} provides, for $m=1$ and $d=2$, a  visualization of  the regression median and quantile tubes of orders~$\tau=0.2$,  $0.4$, and $0.8$, along with some of the corresponding {\it conditional regression quantile contours}~${\mathcal C}_\pm(\tau\vert{\mathbf x})
\coloneqq 
\left\{ \left(\mathbf{x}, \mathbf{Q}_{\pm} (\tau\,{\mathcal S}_{d-1} \big\vert \,
\mathbf{x} )\right)
\right\}$ (${\mathcal S}_{d-1}$ the unit sphere in $\R^d$), 
 for the  model 
\begin{equation}\label{eyelidmod}
  \mathbf{Y}=  \left(\!\begin{array}{c}Y_1 \\ Y_2\end{array}\!\right)=\left(\!\begin{array}{c}\sin(\frac{2\pi}{3} X)
  + 0.575\, e_1\vspace{2mm} \\ \cos(\frac{2\pi}{3} X)+X^2+
   \frac{e_2^3}{2.3}+\frac{1}{4}e_1+2.65\, X^4\end{array}\!\right)\end{equation}
where $X\sim {\rm U}_{[-1,1]}$,  $\mathbf{e}= \sqrt{1+\frac{3}{2}\sin({\pi} X/2)^2}\mathbf{v}$,   $\mathbf{v}\sim \mathcal{N}(\mathbf{0}, \mathbf{Id})$, and $X$ and $ \mathbf{v}$ are mutually  independent.  Actually, since exact values cannot be obtained for these population concepts, a very large sample  of $n=400,040$ observations was generated from~\eqref{eyelidmod}, and the consistent estimation procedure described in Section~\ref{Section;3} was performed to obtain the picture. 
Note the non-convexity of the conditional contours for $\tau = 0.8$, the non-linearity of the regression median, and the marked heteroskedasticity of the regression.


\subsection{Outline of the paper}  The paper is organized as follows. Section~\ref{Section;2}  is dealing with the population concept of {\it conditional center-outward quantile map} and the resulting {\it   center-outward regression quantile contours} and {\it regions}, Section~\ref{Section;3} with their  estimation. In Section~\ref{empiricaldef}, we show how to construct   empirical quantile contours and regions, the consistency of which is established in Section~\ref{Section;3}. Numerical results are provided in Section~\ref{numsec}.   
Monte Carlo experiments in Section~\ref{toysec} show the ability of our method to pick heteroskedasticity, nonlinear trends, and the shape of  conditional distributions; comparisons also are made with the results of the depth-based method of \cite{Halin2015}. Real datasets  are analyzed in Section~\ref{realsec},  illustrating the power of our method as a data-analytic tool. Section~\ref{conclsec} concludes with  references to the recent literature on the numerical aspects of optimal transportation and   perspectives for future developments.

\section{Nonparametric center-outward quantile regression}\label{Section;2}

\subsection{Notation} 
For convenience, we are listing here the main notation to be used throughout the paper. 
Unless otherwise stated, we denote by $(\Omega,\mathcal{A}, \mathbb{P})$  the triple defining the underlying probability space.  Let   $\ell_d$ be the $d$-dimensional Lebesgue measure, $\mathcal{B}_d$  the Borel $\sigma$-field, and $\mathcal{P}(\R^d)$  the space of Borel probability measures on $\R^d$. The support of a probability $\rm P\in \mathcal{P}(\R^d)$ is denoted as $\operatorname{supp}(\rm P)$; its closure as $\overline{\operatorname{supp}}(\rm P)$.  
Throughout,  $(\mathbf{X},\mathbf{Y})$ denotes  an  $\R^{m+d}$-valued random vector with  probability distribution ${\mathbb P}={\rm P}_{\mathbf{X},\mathbf{Y}}\in \mathcal{P}(\R^{m+d})$, $m$-dimensional $\mathbf{X}$-marginal  ${\rm P}_{\mathbf{X}}$  and $d$-dimensional $\mathbf{Y}$-marginal  ${\rm P}_{\mathbf{Y}}$.   The  distribution of $\mathbf{Y}$  conditional  on $\mathbf{X}=\mathbf{x}$ is denoted as ${\rm P}_{\mathbf{Y}\, |\mathbf{X}=\mathbf{x}}$\footnote{The existence of the  regular conditional probability is a direct consequence of the disintegration theorem (see, e.g., Theorem 2.5.1 in~\cite{LehRom}).}. The open unit ball, the closed unit ball, and the unit hypersphere in $\R^d$ are denoted by $\mathbb{S}_{d}$, $\overline{\mathbb{S}}_{d}$, and $\mathcal{S}_{d-1}$,  respectively. We denote by $\mathrm{U}_d $ the spherical uniform over $\mathbb{S}_{d}$---that is, the product of a uniform distribution over the  unit hypersphere $\mathcal{S}_{d-1}$ (for the directions) and a uniform distribution over $[0,1]$ (for the distance to the center).

\subsection{Conditional center-outward quantiles, regions, and contours.}\label{section:Conditional}
Let us provide precise definitions for the concepts we briefly presented in the Introduction and  properly introduce 
 center-outward quantiles, regions, and contours. 

For any  $\rm P\in \mathcal{P}(\R^d)$, denote by $\mathbf{Q}_{\pm}=\nabla\varphi$ and call {\it center-outward quantile map} the (Lebesgue-a.e.) unique  gradient of a convex  function $\varphi: {\mathbb S}_d\to\R$ such that $\mathbf{Q}_{\pm}(\mathbf{U})\sim \rm P$, for any $\mathbf{U}\sim \mathrm{U}_d $---in the measure transportation convenient terminology, $\mathbf{Q}_{\pm}$ is {\it pushing~$\rm P$ forward to} $\mathrm{U}_d$, which we denote as  $\mathbf{Q}_{\pm}\# \rm P = \mathrm{U}_d $. This  only  defines $\mathbf{Q}_{\pm}$ at $\varphi$'s points of differentiability  (recall that convex functions are differentiable at almost every point in the interior of their domain: see Theorems~26.1 and~25.5 in  \cite{rockafellar1970}). At $\varphi$'s points of  non-differentiability $\mathbf{u}$,  let us  define $\mathbf{Q}_{\pm}(\mathbf{u})$ as the  subdifferential $\partial \varphi(\mathbf{u})$ of $\varphi$, namely, 
%
%
%
 \begin{equation*}\label{subdiff}
   \mathbf{Q}_{\pm}(\mathbf{u})= \partial \varphi(\mathbf{u}) \coloneqq  \left\{ \mathbf{y} \in \R^d |\ \text{for all } \mathbf{z} \in \R^d, \ \varphi(\mathbf{z}) - \varphi(\mathbf{u}) \geq \langle \mathbf{y}, \mathbf{z}-\mathbf{u} \rangle \right\},\quad{\mathbf{u}\in{\mathbb S}_d};
\end{equation*}
 then, $\mathbf{Q}_{\pm}$ is an everywhere-defined set-valued function. Slightly abusing the notation, we also write $\partial \varphi $ for the set of all points $(\mathbf{u},\mathbf{y})\in\R^{m+d}$ such that~$\mathbf{y}\in~\!\partial \varphi(\mathbf{u})$.
With this notation, we can introduce the concepts of conditional center-outward quantiles, contours, and regions.  
\begin{Definition}
Call \emph{ conditional center-outward quantile map of $\mathbf{Y}$ given~$\mathbf{X}=\mathbf{x}$} 
    the center-outward quantile map  $\mathbf{u}\mapsto\mathbf{Q}_{\pm}(\mathbf{u}\vert \mathbf{X}=\mathbf{x})$, $\mathbf{u}\in{\mathbb S}_d$  of ${\rm P}_{\mathbf{Y}\, |\mathbf{X}=\mathbf{x}}$, $\mathbf{x}\in\R^m$. The corresponding \emph{conditional center-outward quantile regions and   contours of order}  $\tau\in (0,1)$ are  the sets
\begin{equation*}
  \mathbb{C}_{\pm}(\tau\, \big| \mathbf{x})\coloneqq \mathbf{Q}_{\pm}\left(\tau\,\overline{\mathbb{S}}_{d}|\mathbf{X}=\mathbf{x}\right)\ \ \text{and}\ \ \mathcal{C}_{\pm}(\tau\, \big| \mathbf{x})\coloneqq \mathbf{Q}_{\pm}\left(\tau\mathcal{S}_{d-1}\, |\mathbf{X}=\mathbf{x}\right),
\end{equation*}
respectively. The conditional center-outward quantile maps also characterize (see Defini\-tions~\eqref{qmed},   \eqref{qtube}, and \eqref{qtubetop}) conditional medians ${\mathbb C}_\pm (0\, \vert \, 
\mathbf{x})$ and regression quantile tubes~${\mathbb T}_\pm (\tau)$.
\end{Definition}
When no confusion is possible, we also write $\mathbf{Q}_{\pm}(\mathbf{u}\vert \mathbf{x})$ for  $\mathbf{Q}_{\pm}(\mathbf{u}\vert \mathbf{X}=\mathbf{x})$.
The   terminology   center-outward  {\em regression} quantile region, contour, and median is used for the map-\linebreak pings~$\mathbf{x}\mapsto  \mathbb{C}_{\pm}(\tau\, \big| \mathbf{x})$, $\mathbf{x}\mapsto  \mathcal{C}_{\pm}(\tau\, \big| \mathbf{x})$, and $\mathbf{x}\mapsto{\mathbb C}_\pm (0\, \vert \, 
\mathbf{x})$, $\mathbf{x}\in\R^m$.

Recall, however, that, in the absence of any assumptions on the conditional probabi\-lities~${\rm P}_{\mathbf{Y}\, |\mathbf{X}=\mathbf{x}}$,   the mappings $\mathbf{u}\mapsto \mathbf{Q}_{\pm}(\mathbf{u}\, |\mathbf{X}=\mathbf{x})$ typically are  set-valued, see \cite{RockWets98}. Whenever continuous, single-valued functions (typically, on the punctured unit ball ${\mathbb S}_d\setminus\{{\boldsymbol 0}\}$) are  needed,  we will make the following assumption. 
\\  \\ 
 \noindent {\bf Assumption (R)}\emph{ For ${\rm P}_{\mathbf{X}}$-a.e.\ $\mathbf{x}\in \R^m$, the conditional distribution ${\rm P}_{{\mathbf{Y}}\, |{\mathbf{X}}=\mathbf{x}}$ admits, with respect to the Lebesgue measure, a density ${\rm p}_{{\mathbf{Y}}\, |{\mathbf{X}}=\mathbf{x}}$  
 with convex support 
 $\operatorname{supp}(\rm P_{{\mathbf{Y}}\, |{\mathbf{X}}=\mathbf{x}})$; moreover,    for every $R>0$,  there exist constants $0<\lambda_R^{\mathbf{x}}\leq  \Lambda_R^{\mathbf{x}}<\infty$ such that  
\begin{equation}\label{upperlowwer}
 \lambda_R^{\mathbf{x}}  \leq {\rm p}_{{\mathbf{Y}}\, |{\mathbf{X}}=\mathbf{x}}({\bf y}) \leq 
 \Lambda_R^{\mathbf{x}} \quad\text{ for all ${\bf y}\in 
\operatorname{supp}(\rm P_{{\mathbf{Y}}\, |{\mathbf{X}}=\mathbf{x}}) \cap R\,\mathbb{S}_{d}$}.
\end{equation}}

In the classical single-output case ($d=1$),  consistent estimation of conditional quantiles similarly requires the continuity of  the conditional quantile maps (see  \cite{stone1977}). In   dimension~$d>1$, the continuity of center-outward quantile maps follows from  assumptions similar to Assumption~(R)---see \cite{FigalliCenter} and  \cite{BarrioSanzHal}.

\section{Empirical center-outward quantile regression}\label{Section;3}
We now proceed with the construction of empirical versions of the conditional center-outward quantile concepts defined in Section~\ref{section:Conditional} and their consistency properties. 

\subsection{Empirical conditional center-outward  quantiles}\label{empiricaldef}
Let $(\mathbf{X},\mathbf{Y})^{(n)}\!\coloneqq\!\big((\mathbf{X}_1,\mathbf{Y}_1), \dots,(\mathbf{X}_n,\mathbf{Y}_n)\big)$  be a sample
of $n$  i.i.d.\ copies of  $(\mathbf{X},\mathbf{Y})\sim
=~\!{\rm P}_{\mathbf{X}\mathbf{Y}}$.  
In this section, we develop an estimator of the conditional center-outward quantile\linebreak maps~$\mathbf{u}\mapsto\mathbf{Q}_{\pm}(\mathbf{u}\vert \mathbf{X}=\mathbf{x})$, $\mathbf{x}\in\R^m$. 
Our estimator is obtained in two   steps: in Step 1, we construct an empirical  distribution of $\mathbf Y$ conditional on ${\mathbf X}={\mathbf x}$  and, in Step 2, we compute the corresponding  empirical center-outward quantile map.\medskip

{\it Step 1.} For each value of ${\bf x}\in\R^m$, our estimation of the conditional distribution of~$\mathbf Y$ conditional on ${\mathbf X}={\mathbf x}$ involves a sequence of \emph{weight functions} $w^{(n)}:\R^{m(n+1)} \rightarrow \R^n$ measurable with respect to $\mathbf{x}$ and the sample~$ \mathbf{X}^{(n)}\coloneqq (\mathbf{X}_1,\dots, \mathbf{X}_n)$ of~$\mathbf{X}$ observations, of the form 
\begin{align}\label{wdef}
   ( \mathbf{x},\,\mathbf{X}^{(n)})\mapsto w^{(n)}\left(  \mathbf{x},\,\mathbf{X}^{(n)}\right)
\coloneqq \left(w_1(\mathbf{x};\mathbf{X}^{(n)}),\dots, w_n(\mathbf{x};\mathbf{X}^{(n)})\right)
\end{align} 
where $w_j^{(n)}: \R^{m(n+1)} \rightarrow \R$, $j=1,\ldots, n$ are such that
\begin{equation} \label{weightCond}
    w^{(n)}_j(\mathbf{x};\mathbf{X}^{(n)})\geq 0\quad  \text{and} \quad \sum_{j=1}^nw^{(n)}_j(\mathbf{x};\mathbf{X}^{(n)})=1\quad  \text{a.s. for all $n$.}
\end{equation}
We refer to a function $w^{(n)}$ of the form \eqref{wdef} satisfying \eqref{weightCond} as a \emph{probability weight function} and  define the \emph{empirical conditional distribution} of $\mathbf{Y}$ given $ \mathbf{X}=\mathbf{x}$ as
\begin{equation}\label{condemp}
     {\rm P}_{w(\mathbf{x})}^{(n)}\coloneqq \sum_{j=1}^n w^{(n)}_j(\mathbf{x};\mathbf{X}^{(n)})\delta_{\mathbf{Y}_j},
\end{equation}
where $\delta_{\mathbf{Y}_j}$ is the Dirac function computed at $\mathbf{Y}_j$.    Following  \cite{stone1977},  we say that the sequence $w^{(n)}$ is a \emph{consistent} weight function if,    whenever $(\mathbf{X},Y),(\mathbf{X}_1, {Y}_1), \dots, (\mathbf{X}_n,{Y}_n)$  are i.i.d., where  $Y$ is real-valued and such that  ${\rm E}|Y|^r<\infty$ for  $r>1$,
\begin{equation}
    \label{consisten_stone}
     {\rm E}\left|\sum_{j=1}^n w^{(n)}_j(\mathbf{X};\mathbf{X}^{(n)}){{Y}_j}-{\rm E}(Y|\mathbf{X})\right|^r\longrightarrow 0\quad\text{as $n\to\infty$}. 
\end{equation}
 
{\it Step 2.} To estimate the conditional quantiles, consider a \emph{regular grid} ${\mathfrak G}^{(N)}$ of $\mathbb{U}_d$ consisting of $N$ gridpoints denoted as ${\scriptstyle{\mathfrak G}}_1^{(N)}, \ldots,{\scriptstyle{\mathfrak G}}_N^{(N)}$. The number $N$ here is arbitrarily chosen as factorizing into a product of integers of the form $N=N_RN_S + N_0$ with $N_0=0$ or $1$. That  regular grid is created as the intersection between
\begin{compactenum}
	\item[--] the rays generated by an $N_S$-tuple  $\mathbf{u}_1, \dots,\mathbf{u}_{N_S}\in{\mathcal S}_{d-1}$ of unit vectors such\linebreak that~${N_S}^{-1}\sum_{j=1}^{N_S} \delta_{\mathbf{u}_j}$ converges weakly to the uniform over $\mathcal{S}_{d-1}$ as $N_S\to\infty$, and
	\item[--]  the $N_R$ hyperspheres with center $\mathbf{0}$ and radii ${j}/{(N_R+1)}$,   $j=1, \dots, N_R$,
\end{compactenum}
along with the origin  if $N_0=1$.  Based on this   grid, we define  the sequence of discrete uniform measures 
  \begin{equation*}
    { \rm U}_d^{(N)}\coloneqq\frac{1}{N} \sum_{j=1}^N\delta_{{\scriptstyle{\mathfrak G}}_j^{(N)}}\in \mathcal{P}(\R^d),
 \quad N\in\mathbb N \end{equation*}
over ${\mathfrak G}^{(N)}$  and  
  require that $N\rightarrow\infty$ with both $N_R\rightarrow\infty$ and $N_S\rightarrow+\infty$. By construc\-tion,~$ { \rm U}_d^{(N)}$ converges weakly to ${\mathrm U}_d$ as $N\to\infty$. Note that imposing $N_0=0$ or $1$ is not a problem, since $N$, unlike $n$, is chosen by the practitioner. 
   Having $N_0=0$ or~$1$ yields the fundamental advantage that all points of ${\mathfrak G}^{(N)}$ have multiplicity one and that Corollary 3.1 in \cite{Hallin2020DistributionAQ}, to be used below, applies. 

Our estimation of the conditional center-outward quantile maps relies on the  optimal transport pushing $ { \rm U}_d^{(N)}$  forward to $ {\rm P}_{w(\mathbf{x})}^{(n)}$---more precisely, adopting (since typically $N\neq~\!n$) the Kantorovich formulation of the optimal transport problem, on the solution of the linear program (solvable using efficient numerical methods such as 
 the auction
or Hungarian algorithms---see \cite{MAL-073} and references therein)\footnote{We are dropping the superscripts $^{(N)}$ and~$^{(n)}$ when no confusion is possible.}   
\begin{align}
    \begin{split}\label{kanto_reg}
    	&\min_{\pi \coloneqq\{\pi_{i,j}\}} \sum_{i=1}^N\sum_{j=1}^n   \frac{1}{2}\, \left\vert \mathbf{Y}_j- {\scriptstyle{\mathfrak G}}_i\right\vert ^2 \pi_{i,j} \ , \\
    	\text{s.t.} &\ \ \sum_{j=1}^n  \pi_{i,j}= \frac{1}{N}, \quad  i\in \{1, 2, \dots, N \},\\
    	&\ \sum_{i=1}^N \pi_{i,j}=  w^{(n)}_j(\mathbf{x};\mathbf{X}^{(n)}), \quad  j\in \{1, 2, \dots, n \}, \\
    	&\ \pi_{i,j}\geq 0,\quad  i\in \{1, 2, \dots, N \}, \  j\in \{1, 2, \dots, n \}.
\end{split}
\end{align}
Here,  any $Nn$-tuple $\pi \coloneqq\{\pi_{i,j}\vert i=1,  \dots, N \  j=1, \dots, n\}$ satisfying the constraints in~\eqref{kanto_reg} represents a transport plan from $ { \rm U}_d^{(N)}$   to $ {\rm P}_{w(\mathbf{x})}^{(n)}$---that is, a discrete distribution over~$\R^d\times \R^d$ with marginals $ { \rm U}_d^{(N)}$  and $ {\rm P}_{w(\mathbf{x})}^{(n)}$. Let  ${\pi}^*(\mathbf{x})=\{{\pi}_{i,j}^*(\mathbf{x})\vert i=1,  \dots, N \  j=1, \dots, n\}$ be a solution  of~\eqref{kanto_reg} (an optimal {\it transport plan}).    Theorem~2.12(i) in \cite{Villani2003}  implies that its support 
$
\operatorname{supp}({\pi}^*(\mathbf{x}))\coloneqq \{( {\scriptstyle{\mathfrak G}}_i, \mathbf{Y}_j)\vert {\pi}_{i,j}^*(\mathbf{x}) >0\}
$ 
 is   {\it cyclically monotone},\footnote{Recall from \cite{rockafellar1970}
that a set $S \subset \R^d \times \R^d$ is {\it cyclically monotone} if any finite   subset~$\{ (x_{k_1},y_{k_1}),\ldots, (x_{k_\nu},y_{k_\nu})\}\subset S$, $\nu\in{\mathbb N}$ satisfies 
 $  \sum_{\ell=1}^{\nu-1} \langle y_{k_\ell},x_{k_\ell +1}-x_{k_\ell} \rangle +\langle y_{k_\nu},x_{k_1}-x_{k_\nu} \rangle \leq 0$, 
where~$\langle\cdot,\cdot \rangle$ stands for the scalar product in $\R^d$.} hence is contained in the graph of the \emph{subdifferential} of a convex function. Therefore, the idea is to construct a smooth interpolation of~${\pi}^*(\mathbf{x})$ that maintains this property. 

Note that,  for any gridpoint 
 ${\scriptstyle{\mathfrak G}}_i$, $i\in \{1, \dots, N\}$, the constraints in \eqref{kanto_reg} imply that there exists at least one $j\in \{1, \dots, n\}$ such that~$({\scriptstyle{\mathfrak G}}_i,\mathbf{Y}_{j})\!\in~\!\!\operatorname{supp}({\pi}^*(\mathbf{x}))$. Since more than one such $j$ may exist,   we choose the one which  \textit{``gets the highest mass''}  from~${\scriptstyle{\mathfrak G}}_i$, and in case of ties we choose the \textit{``smallest''} one: namely, let 
\begin{align}
    \label{smalle2r}
    \mathbf{T}^*({\scriptstyle{\mathfrak G}}_i\vert\ \mathbf{x})\coloneqq \arg\inf\left\lbrace\Vert\mathbf{y}\, \Vert :\ \mathbf{y}\in \operatorname{conv}\!\left(\{\mathbf{Y}_J:\ J\in \arg\max_j {\pi}_{i,j}^*(\mathbf{x}) \}\right) \right\rbrace,
\end{align}
where  $\operatorname{conv}(A)$ denotes the convex  hull of a set $A$. Since $\operatorname{conv}\!\left(\{\mathbf{Y}_J:\ J\in \arg\max_j {\pi}_{i,j}^*(\mathbf{x}) \}\right)$ is closed and convex in $\R^d$,  \eqref{smalle2r} defines a unique $\mathbf{T}^*({\scriptstyle{\mathfrak G}}_i\vert\ \mathbf{x})$. Due to the cyclical monotoni\-city of~$\operatorname{supp}({\pi}^*(\mathbf{x}))$, there exists a convex function ${\varphi}^*(\cdot \ \big| \mathbf{x}): \R^d\to\R$ with subdifferential~$\partial{\varphi}^*(\cdot \ \big| \mathbf{x})$ such that,  for all $1\leq  i\leq N$, 
$$\emptyset\neq \{\mathbf{Y}_j:\ ({\scriptstyle{\mathfrak G}}_i,\mathbf{Y}_{j})\in \operatorname{supp}({\pi}^*(\mathbf{x})) \}
\subset \partial{\varphi}^*({\scriptstyle{\mathfrak G}}_i\ \big| \mathbf{x}).$$ 
Since   
sub-differentials are convex sets,  this entails
$$
    \operatorname{conv}\{\mathbf{Y}_j:\ ({\scriptstyle{\mathfrak G}}_i,\mathbf{Y}_{j})\in \operatorname{supp}({\pi}^*(\mathbf{x}))  \}\subset  \partial{\varphi}^*({\scriptstyle{\mathfrak G}}_i\ \big| \mathbf{x}).
$$
Consequently,   
$
    \{({\scriptstyle{\mathfrak G}}_i,\mathbf{T}^*({\scriptstyle{\mathfrak G}}_i\vert\ \mathbf{x})):  \ i=1, \dots, {N}  \} 
$
is cyclically monotone and satisfies
the\linebreak  assumptions of Corollary 3.1 in \cite{Hallin2020DistributionAQ}. This implies the  existence, \linebreak for all~$\mathbf{x}$,  of a continuous cyclically monotone map  $\mathbf{u}\mapsto \mathbf{Q}^{(n)}_{w,\pm}(\mathbf{u}\ \big| \mathbf{x})$, say, interpolating\linebreak  the~$N$-tuple  $\left({\scriptstyle{\mathfrak G}}_1,\mathbf{T}^*({\scriptstyle{\mathfrak G}}_1\vert\ \mathbf{x})\right),\ldots,\left({\scriptstyle{\mathfrak G}}_N,\mathbf{T}^*({\scriptstyle{\mathfrak G}}_N\vert\ \mathbf{x})\right)$, i.e.,  
 such that $\mathbf{Q}^{(n)}_{w,\pm}({\scriptstyle{\mathfrak G}}_i \vert\ \mathbf{x})= \mathbf{T}^*({\scriptstyle{\mathfrak G}}_i\vert\ \mathbf{x})$\linebreak  for~$i=1, \dots,{N}$. 

In particular, we proceed as in \cite{Hallin2020DistributionAQ} by choosing the smooth cyclically monotone interpolation with largest Lipschitz constant.  This continuous map ${\mathbf u}\mapsto\mathbf{Q}^{(n)}_{w,\pm}({\mathbf u}  \big| \mathbf{x})$ from $\mathbb{S}_d$ to $\R^d$  will be called
 the \emph{empirical  conditional center-outward quantile function} of $\mathbf{Y}$ given $\mathbf{X}=\mathbf{x}$. It defines the {\it empirical center-outward regression quantile regions}  and {\it contours}
%
\begin{equation}\label{empiricalregionsdef}
  \mathbb{C}_{w,\pm}^{(n)}(\tau\, \big| \mathbf{x})\coloneqq \mathbf{Q}^{(n)}_{w,\pm}\left(\tau\,\overline{\mathbb{S}}_{d} \ \big| \mathbf{x}\right)\ \ \text{and}\ \  \mathcal{C}_{w,\pm}^{(n)}(\tau\, \big| \mathbf{x})\coloneqq 
\mathbf{Q}^{(n)}_{w,\pm}\left( \tau\mathcal{S}_{d-1} \ \big| \mathbf{x}\right),\quad \tau\in (0,1)
\end{equation}
which we are proposing as estimators of   $\mathbb{C}_{\pm}(\tau\, \big| \mathbf{x})$ and $\mathcal{C}_{\pm}(\tau\, \big| \mathbf{x})$, respectively. The intersec-\linebreak tion~$\bigcap_{\tau\in (0,1)}  \mathbb{C}_{w,\pm}^{(n)}(\tau\, \big| \mathbf{x})\vspace{1mm}$ yields the \emph{empirical conditional  center-outward  regression median region}. The definition of 
 {\it empirical regression quantile tubes}
 \begin{equation*} {\mathbb T}_{w,\pm}^{(n)} (\tau)\coloneqq 
\left\{ \left(\mathbf{x}, \mathbf{Q}_{\pm}^{(n)} (\tau\,\overline{\mathbb S}_d \big\vert \,
\mathbf{x})\right)
\big\vert \mathbf{x}\in\R^m
\right\},\quad \tau\in (0,1)
\end{equation*}
naturally follows.

\begin{Remark}\label{Remark}{\rm 
Note that the results of this section and the next one still hold for any continuous  map  with  cyclically monotone graph satisfying 
$$ ( \mathbf{u}_i ,\mathbf{Q}^{(n)}_{w,\pm}(\mathbf{u}_i \ \big| \mathbf{x}))\in \operatorname{conv}\left(\{\mathbf{Y}_j:\ (\mathbf{u}_i,\mathbf{Y}_{j})\in \operatorname{supp}({\pi}^*(\mathbf{x})) \}\right) \quad \text{for all }i=1,\dots,N.$$
The reason for choosing the ``smallest'' $\mathbf y$ in \eqref{smalle2r} is to have a ``universal criterion.''

}
\end{Remark}

\subsection{Consistency of  empirical conditional center-outward quantiles, regression quantile regions,  and regression  quantile   contours}\label{subsection;2}

The objective of this section is to justify the definitions of Section~\ref{empiricaldef} by showing the consistency of   the empirical quantile regions and contours   defined in \eqref{empiricalregionsdef}. The asymptotic behavior of  these regions and contours, quite naturally,  depends on the regularity of the conditional distributions involved. In fact, as discussed before, when Assumption~(R) does not hold, the population conditional quantile maps are not necessarily defined for every~$\mathbf{u}\in~\!{\mathbb S}_d$, but only for a set of {${\rm U}_d$}-probability one. Consistency results can be obtained despite  this  a.s.\ definition, provided that     population  quantile maps are extended  into  set-valued maps. The following theorem shows\footnote{Note that, although  $N$ does not appear in the notation, $\mathbf{Q}^{(n)}_{w,\pm}$ depends on both $N$ and $n$.}, under mild assumptions, that any possible limit of~$\mathbf{Q}^{(n)}_{w,\pm}(\ \cdot\ | \mathbf{x})$  asymptotically belongs to the set  $\mathbf{Q}_{\pm}(\ \cdot\ |\mathbf{X}= \mathbf{x})$.

\begin{Theorem}\label{MainTheorem}
Let $(\mathbf{X},\mathbf{Y}),(\mathbf{X}_1,\mathbf{Y}_1), \dots,(\mathbf{X}_n,\mathbf{Y}_n)$ be pairs of i.i.d. random vectors with values in $\R^m\times \R^d$ and let $w^{(n)}$ be a consistent sequence of weight functions. Then,  for every~$\mathbf{u}\in \mathbb{S}_{d}$ and  $\epsilon>0$, 
\begin{equation*}
  \mathbb{P}\left(\text{ $ \mathbf{Q}^{(n)}_{w,\pm}(\mathbf{u}\, | \mathbf{X})\not\in {\mathbf{Q}}_{\pm}(\mathbf{u}\, |\mathbf{X})+\epsilon\mathbb{S}_{d}$ } \right)\longrightarrow 0\quad\text{as $n$ and $N\to\infty$},
\end{equation*}
and, for every $\tau\in (0,1)$, 
\begin{equation*}
  \mathbb{P}\left(\text{ $ \mathbb{C}_{\pm}^{(n)}(\tau \, \big| \mathbf{X})\not\subset \mathbb{C}_{\pm}(\tau \, \big| \mathbf{X})+\epsilon\mathbb{S}_{d}$ } \right)\rightarrow 0 \ \text{and}\ \   \mathbb{P}\left(\text{ $ \mathcal{C}_{\pm}^{(n)}(\tau \, \big| \mathbf{X})\not\subset \mathcal{C}_{\pm}^{(n)}(\tau \, \big| \mathbf{X})+\epsilon\mathbb{S}_{d}$ } \right)\rightarrow 0\end{equation*}
as $n$ and $N\to\infty$.
\end{Theorem}
Neater convergence results---avoiding   the notion of set-valued maps--- are obtained if it can be assumed that Assumption~(R)   holds, which implies that, for any $\mathbf{X}=\mathbf{x}$ and any~$\mathbf{u}\in \mathbb{S}_{d}\setminus\{ \mathbf{0}\}$, the set $ \mathbf{Q}_{\pm}(\mathbf{u}\, |\mathbf{X}=\mathbf{x})$ is  a singleton. Then,  the map $\mathbf{u}\mapsto \mathbf{Q}_{\pm}(\mathbf{u}\, |\mathbf{X}=\mathbf{x})$ can be seen as continuous   on $\mathbb{S}_{d}\setminus\{ \mathbf{0}\}$, see Theorem 25.5 in \cite{rockafellar1970}, hence single-valued on $\mathbb{S}_{d}\setminus\{ \mathbf{0}\}$  since the gradient of a convex function is single-valued at a point  if and only if it is continuous at this point. We then can state the following theorem (see Appendix~A.1 for the proof), the second part of which  describes the convergence of the contours in terms of the Pompeiu-Hausdorff distance~$d_{\infty}$. Recall   that he Pompeiu-Hausdorff distance between  two sets $A$ and  $B$ in $ \R^d$ is defined~as 
$$ d_{\infty}(A,B)\coloneqq \inf \{ \nu\geq 0: \ A\subset B+\nu\mathbb{S}_{d} \ \text{and} \ B\subset A+\nu\mathbb{S}_{d}  \}$$
 (see \cite{RockWets98}). 
 \begin{Theorem}\label{MainTheorem_discrete}
Let $(\mathbf{X},\mathbf{Y}),(\mathbf{X}_1,\mathbf{Y}_1), \dots,(\mathbf{X}_n,\mathbf{Y}_n)$ be pairs of i.i.d. random vectors with values in $\R^m\times \R^d$ and let $w^{(n)}$ be a consistent sequence of weight functions. Suppose moreover that Assumption~(R) holds. Then, for every compact $K\subset \mathbb{S}_{d}\setminus\{\mathbf{0} \}$, as $n$ and $N\to\infty$,
\begin{equation*}
  \sup_{\mathbf{u}\in K}\, | \mathbf{Q}^{(n)}_{w,\pm}(\mathbf{u}\, |\mathbf{X})- {\mathbf{Q}}_{\pm}(\mathbf{u}\, |\mathbf{X})|\stackrel{{\mathbb P}\;\;\;\;}{\longrightarrow 0} 
\end{equation*}
and, for every $\tau \in (0,1)$ and $\epsilon >0$, 
\begin{equation*}
\mathbb{P}\left(d_{\infty}\left(\mathcal{C}_{\pm}^{(n)}(\tau \, \big| \mathbf{X}),\mathcal{C}_{\pm}(\tau \, \big| \mathbf{X}) \right)>\epsilon\right)\rightarrow 0.
\end{equation*}
 
\end{Theorem}
Under the assumptions of Theorem~\ref{MainTheorem_discrete}, consistency in Pompeiu-Hausdorff distance  of the quantile contours holds in  case   the median is a single point\footnote{This is always  the case for $d=2$ and $d=3$: see \cite{FigalliCenter}.}---the continuity of quantile maps  then extends to the whole open unit ball. This, however, is not necessarily the case for $d>3$ (see \cite{FigalliCenter}), and  Pompeiu-Hausdorff   consistency may fail  due to the fact that our empirical version is  continuous over~$\mathbb{S}_d$ while $\mathbf{Q}_{\pm}( \mathbf{0} | \mathbf{x})$ could be a set rather than a single point: convergence then  holds along subsequences of  $\mathbf{Q}^{(n)}_{w,\pm}(\ \cdot\ | \mathbf{x})$ towards an element of~$\mathbf{Q}_{\pm}( \mathbf{0} | \mathbf{x})$. This, 
 obviously, has an impact on   convergence   in terms of the Pompeiu-Hausdorff distance---although 
  it does not affect the  control over the asymptotic probability contents of quantile regions. 
 More precisely, the following corollary holds (see Appendix A.1 for the proof). 

\begin{Corollary}\label{Remarkcontrol}
Under the conditions of Theorem \ref{MainTheorem_discrete},  as $n$ and $N\to\infty$,
\begin{equation}
\label{convergenceProbaControl}
 \mathbb{P}\left(\mathbf{Y}\in  \mathbb{C}_{\pm}^{(n)}(\tau\, \big|  \mathbf{X}) \big|\mathbf{X} \right) \stackrel{{\mathbb P} }{\longrightarrow}  \tau\quad \text{for all $\tau\in (0,1)$.}
\end{equation}
\end{Corollary}

\begin{Remark}
Under the weaker conditions of Theorem~\ref{MainTheorem}, in view of  \eqref{consequence_theorem_weak}, still some asymptotic control of the probability content of the empirical regions can be derived. More precisely, letting $N=N(n)$ be such that $N(n)\to \infty$ as~$n\to \infty$,  for all $\tau\in (0,1)$ and  every subsequence $n_k$, there exists a further subsequence $n_{k_j}$ such that 
$$ \limsup_m\mathbb{P}\left(\mathbb{C}_{\pm}^{(n_{k_j})}(\tau\, \big|  \mathbf{x}) \big|\mathbf{X}=\mathbf{x}\right)\leq \tau, \ \ \text{$\mathbf{x}$-a.e.\  in $\R^m$}.$$
\end{Remark}

The above results, as well as the proposed regularization, are valid for any consistent sequence of weight functions. This includes---along with adequate additional assumptions---most of the classic choices of weight functions. Here are three examples. 
\begin{enumerate}[(i)]
 \item\label{optionGauss} The kernel weight function, usually defined (see chapter 5 in \cite{gyodfi}) as 
    \[
        w^{(n)}_i(\mathbf{x};\mathbf{X}^{(n)})\coloneqq 
        {K\left(\frac{\mathbf{X_i}-\mathbf{x}}{h_n}\right)}/{\sum_{j=1}^n K\left(\frac{\mathbf{X_j}-\mathbf{x}}{h_n}\right)},\quad i=1,\ldots,n
    \]
    where $h_n$ is the bandwidth, and $K:\R^m\rightarrow \R$  the kernel.  Sufficient conditions for~$w^{(n)}_i$ to form a consistent sequence of weight functions  are \smallskip
    \begin{compactenum}
    \item[(a)] $h_n\rightarrow 0$, 
       \item[(b)]$c_1\min\Big(\mathbbm{1}_{\big[|\mathbf{x}\, |\leq r\big]},H(| \mathbf{x}\, |)\Big)\leq K( \mathbf{x})\leq c_2H(| \mathbf{x}\, |)$, where $c_1,c_2, r$ are positive constants, and $H:[0,\infty)\rightarrow\R$ is bounded, decreasing, and such that $H(t)t^m\to 0$  as~$t\rightarrow\infty$, and 
           \item[(c)]$\lim_{n\rightarrow\infty} {n^{\alpha}h_n^m}/{\log(n)}=\infty$ for any $\alpha\in (0,1)$;\smallskip
    \end{compactenum}
     when the kernel  $K$ is compactly supported, the assumptions are are much simpler, and we only need  (a) and $\lim_{n\rightarrow\infty}h_n^m n=\infty$, see  Theorem 5.1 in \cite{gyodfi}. The particular case  $K(\mathbf{x})=e^{-|\mathbf{x}|^2}$ is known as the Gaussian kernel.  \vspace{2mm}
    \item\label{optionclasic} The (classical) $k-$nearest neighbors weight function:   the $k-$nearest neighborhood of~$\mathbf{x}\in \R^m$  is  obtained (Chapter 6 in  \cite{gyodfi})  by ordering $\{ \mathbf{X}_1, \dots, \mathbf{X}_n\}$ according to increasing values of $|\mathbf{X}_j-\mathbf{x}\, |$. Denoting by $\{ \mathbf{X}_{(0,\mathbf{x})}, \dots, \mathbf{X}_{(n,\mathbf{x})}\}$ the reordered   sequence, the set of $k-$nearest neighbors of $\mathbf{x}$ is  
$
{\mathcal N}_k^{(n)}(\mathbf{x})\coloneqq \{\mathbf{X}_{(j,\mathbf{x})}: j\leq k\}
$
     and the $k$-nearest neighbors weight function is defined as
\[
    w^{(n)}_i(\mathbf{x};\mathbf{X}^{(n)})\coloneqq  \frac{1}{k}\mathbbm{1}_{ \mathbf{X}_{i} \in {\mathcal N}^{(n)}_k(\mathbf{x})},\quad i=1,\ldots,n.
\]
Sufficient conditions for this $k$-nearest neighbors~$w^{(n)}_i$ to form a consistent sequence of weight functions  are  (see \cite{stone1977})  
\begin{equation}
    \label{knearest}
    k\rightarrow +\infty  \ \text{ and  }\  {k}/{n}\rightarrow 0.
\end{equation}
If a $k$-nearest neighbors weight function $w^{(n)}$ is satisfying \eqref{knearest} and Assumption~(R) holds,   we thus have the convergence---in probability---of the conditional quantile map described in Theorem~\ref{MainTheorem_discrete}; without this assumption we still obtain the slightly weaker result of Theorem~\ref{MainTheorem}.
\end{enumerate}
The $k-$nearest neighbors in \eqref{optionclasic}  were   understood in the classical sense of the Euclidean distance in $\R^m$, which does not take into account the distribution ${\mathrm P}_{\mathbf{X}}$ of $\mathbf{X}$. An alternative $k-$nearest neighbors weight function can be derived from a notion of nearness based on the ordering induced by empirical center-outward distribution functions. This  new weight function is obtained as follows. 
\begin{enumerate}[(i)]
\setcounter{enumi}{2}
\item\label{optionNew} An alternative $k$-nearest neighbors  weight function. 
Fixing   ${\mathbf x}\in \R^m$, first  compute,  as in \cite{Hallin2020DistributionAQ}, the empirical  center-outward distribution function associated with 
\[
   \frac{1}{n+1}\sum_{j=1}^n \delta_{\mathbf{X}_j} +\frac{1}{n+1} \delta_{\mathbf{x}}\in \mathcal{P}(\R^m).\]
That distribution function  is the solution $T^*_{\mathbf x}$ of the minimization problem
\[ \min_{T\in \Gamma_{n+1}}\sum_{k=0}^n | \mathbf{X}_k-T(\mathbf{X}_k)|^2\]
	where $\mathbf{X}_0=\mathbf{x}$, $\Gamma_{n+1}$ is the set of all bijections $T$ between $\{\mathbf{x}, \mathbf{X}_1, \dots,\mathbf{X}_{n}\}$ and a  \emph{regular grid} ${\mathfrak G}^{(n+1)}$ of~$\mathbb{S}_m$, of the form described in Section~\ref{empiricaldef}, consisting of $(n+1)$ gridpoints denoted as~${\scriptstyle{\mathfrak G}}_0, {\scriptstyle{\mathfrak G}}_1, \ldots,{\scriptstyle{\mathfrak G}}_n$, obtained via a factorization of  $n$ into a product of non-negative integers of the form   $n+1=n_Rn_S + n_0$ with $n_0<\min(n_R,n_S)$. That  regular grid is created as the intersection between\vspace{2mm}
\begin{compactenum}
	\item[--] the rays generated by an $n_S$-tuple  $\mathbf{u}_1, \dots,\mathbf{u}_{n_S}\in{\mathcal S}_{m-1}$ of unit vectors such\linebreak that~${n_S}^{-1}\sum_{j=1}^{n_S} \delta_{\mathbf{u}_j}$ converges weakly,  as $n_S\to\infty$,  to the uniform over $\mathcal{S}_{m-1}$  and
	\item[--]  the $n_R$ hyperspheres with center $\mathbf{0}$ and radii ${j}/{(n_R+1)}$,   $j=1, \dots, n_R$,\vspace{2mm}
\end{compactenum}
along with $n_0$ copies of the origin whenever $n_0 >0$.  Based on this   grid, we define  the sequence of discrete uniform measures  
  \begin{equation*}
    { \rm U}_d^{(n+1)}\coloneqq\frac{1}{n+1} \sum_{j=1}^{n+1}\delta_{{\scriptstyle{\mathfrak G}}_j}\in \mathcal{P}(\R^m)
 \quad N\in\mathbb N \end{equation*}
 This map is defined only at the $(n+1)$ points $\mathbf{x},\mathbf{X}_1, \dots, \mathbf{X}_n$, but, as in the previous section, it can be countinuously extended (see also \cite{Hallin2020DistributionAQ}) to the whole space $\R^m$---call   $\mathbf{F}_{{\mathbf x};\pm}^{(n)}:\R^m\rightarrow\mathbb{S}_m$ this extension---with the properties that $\mathbf{F}_{{\mathbf x};\pm}^{(n)}$ coincides, on~$\{\mathbf{x},\mathbf{X}_1, \dots, \mathbf{X}_n\}$, with $T^*_{\mathbf x}$,  is the gradient of a differentiable convex function with domain $\R^m$, and  satisfies $\mathbf{F}_{{\mathbf x};\pm}^{(n)}(\R^m)\subset \overline{\mathbb{S}}_m$. We then  define the \emph{set of $k-$nearest center-outward neighbors} of $\mathbf{x}$ as 
 \begin{align}
 \begin{split}
     {\cal K}^{(n)}_k(\mathbf{x})&\coloneqq \{\mathbf{X}_j: \mathbf{F}_{\pm}^{(n)} (\mathbf{X}_j)\in {\cal N}_k(\mathbf{F}_{\mathbf{x};\pm}^{(n)}(\mathbf{x}))\},
 \end{split}
\end{align}
where, for each $\mathbf{a}\in \mathbf{B}_m$ and $k\in \N$, ${\cal N}_k(\mathbf{a})$ denotes the set of $k$-nearest neighbors (in the sense of Euclidean distance)  of~$\mathbf{a}$. 
Based on this, define  the {\it center-outward nearest neighbor weight function} 
\begin{equation}
    \label{center_weigh}
    w^{(n)}_j(\mathbf{x};\mathbf{X}^{(n)})\coloneqq \frac{1}{k}\mathbbm{1}_{ \mathbf{X}_{j} \in {\cal K}^{(n)}_k(\mathbf{x})}\ ,\ \ j=1, \dots, k
\end{equation}
and proceed as in Section~\ref{empiricaldef} with the estimation \eqref{kanto_reg} of the conditional quantile functions.   
\end{enumerate}

The next result shows that, for a suitable choice of $k = k(n)$,    center-outward nearest neighbors  weight functions form a  consistent sequence of weights (see the appendix for a proof).

\begin{Lemma}\label{Lemma_k_nearest}
If $k=k(n)$ is such that $k(n)\rightarrow \infty$ and ${k(n)}/{n}\rightarrow 0$ as $n\to\infty$, the sequence of weight functions defined in \eqref{center_weigh} is consistent in the sense of \eqref{consisten_stone}.
\end{Lemma}
This means, in particular, that Theorem~\ref{MainTheorem} applies when the   weight function~\eqref{center_weigh} is used under the assimptions of Lemma~\ref{Lemma_k_nearest}, and that the resulting estimators   are consistent.

Theorems~\ref{MainTheorem} and~\ref{MainTheorem_discrete} provide {\it weak} (in probability) {\it consistency} results under minimal assumptions.  
For   sequences of weights satisfying, as $n\to\infty$, 
\begin{equation}
    \label{consisten_strong}
    \sum_{j=1}^n w^{(n)}_j(\mathbf{X};\mathbf{X}^{(n)}){{Y}_j} \longrightarrow {\rm E}[Y|\mathbf{X}]\ \ \text{a.s}. 
\end{equation}
({\it strongly consistent} sequences), the conclusion in Theorem \ref{MainTheorem} can be upgraded to {\it strong} (almost sure) {\it consistency}. For the particular case of  $k$-nearest neighbors, \eqref{consisten_strong} and strong consistency hold if \eqref{knearest} is replaced with 
\begin{equation}
    \label{knearest2}
     {k}/{\log(n)}\rightarrow \infty  \quad \text{ and  }\quad {k}/{n}\rightarrow 0,
\end{equation}
see \cite{Devroye} and \cite{Devroye1982NecessaryAS}.

\section{Numerical results}\label{numsec}
This section is devoted to a numerical assessment of the performance of the estimation procedures described in Section~\ref{Section;3}. We  first analyze (Section~\ref{toysec}) some artificial  datasets---inclu\-ding  the motivating example of \cite{Halin2015}---then turn (Section~\ref{realsec}) to real-data cases. These examples showcase three important  features of our estimators: their ability to capture heteroskedasticity, to deal (non-parametrically) with highly nonlinear  regression, and to adapt to non-convex   noise.

\subsection{Simulated examples}\label{toysec}     
\subsubsection{Parabolic trend and periodic heteroskedasticity; spherical conditional densities.}\label{sphcasesec}
\begin{figure}[hb!]
    \centering
        \includegraphics[width=6.5 cm,height=7cm,trim={5cm 5cm 6cm 4cm},clip]{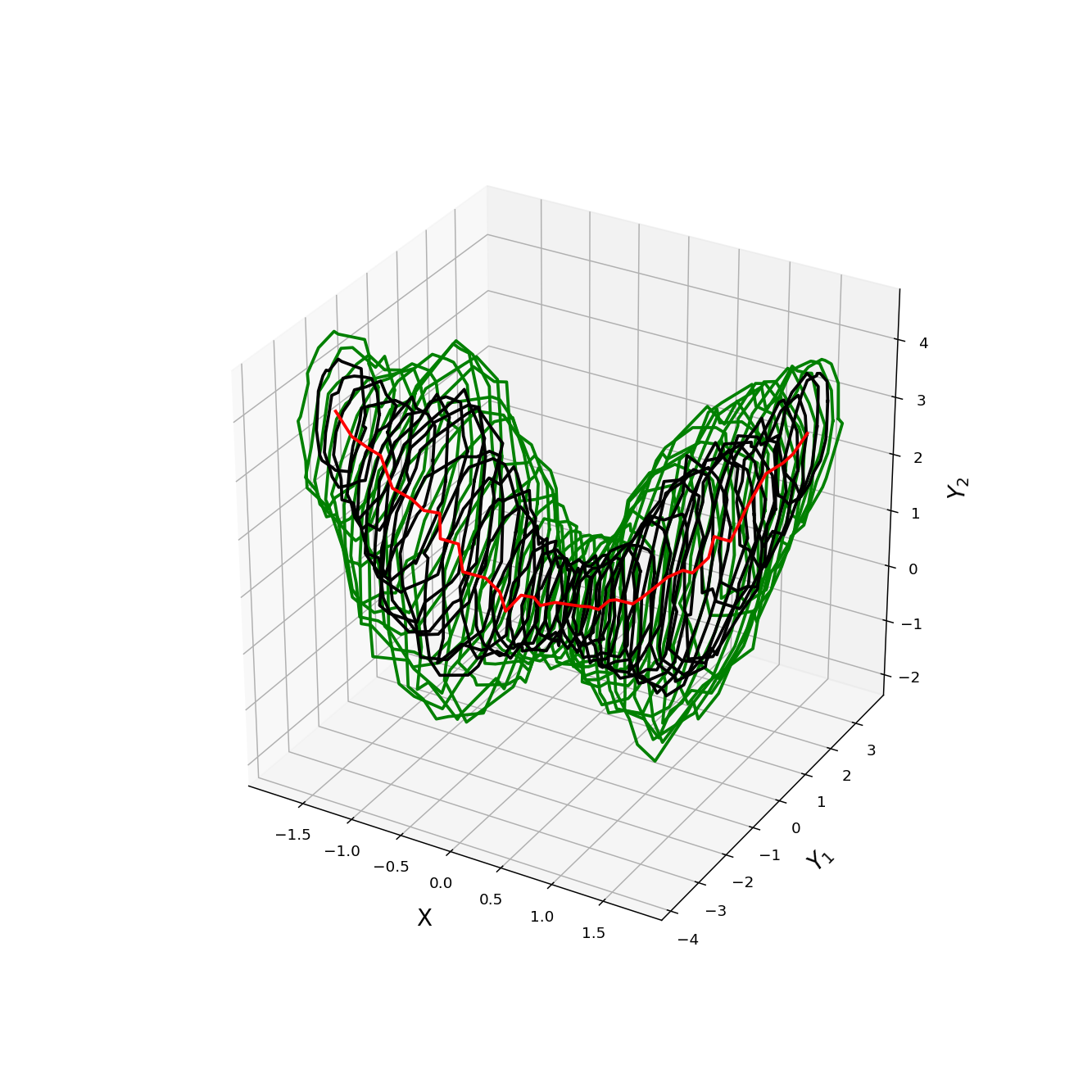}
    \includegraphics[width=6.5 cm,height=7cm,trim={5cm 5cm 6cm 4cm},clip]{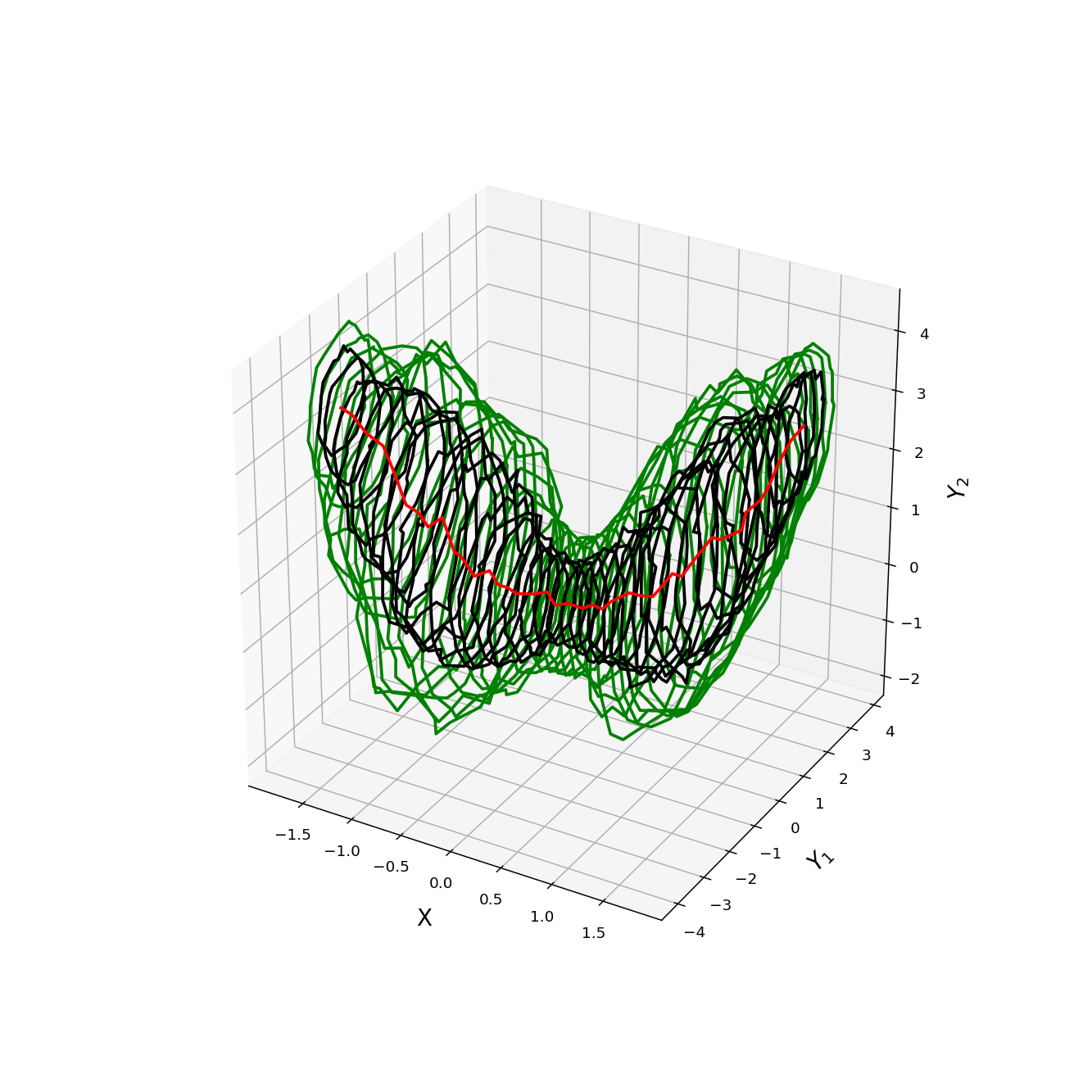}
    \includegraphics[width=6.5 cm,height=7cm,trim={5cm 5cm 6cm 4cm},clip]{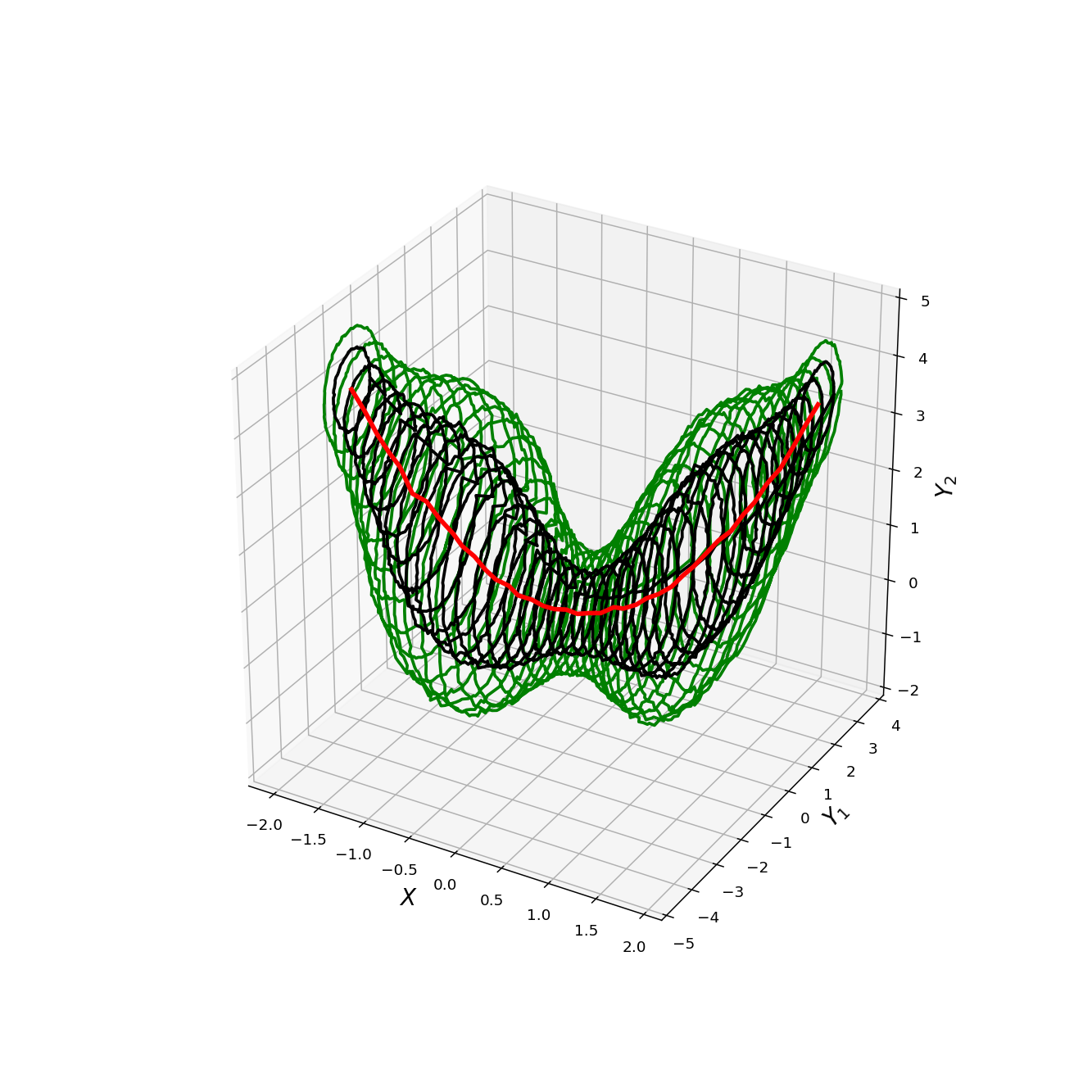}
    \includegraphics[width=6.5 cm,height=7cm,trim={5cm 5cm 6cm 4cm},clip]{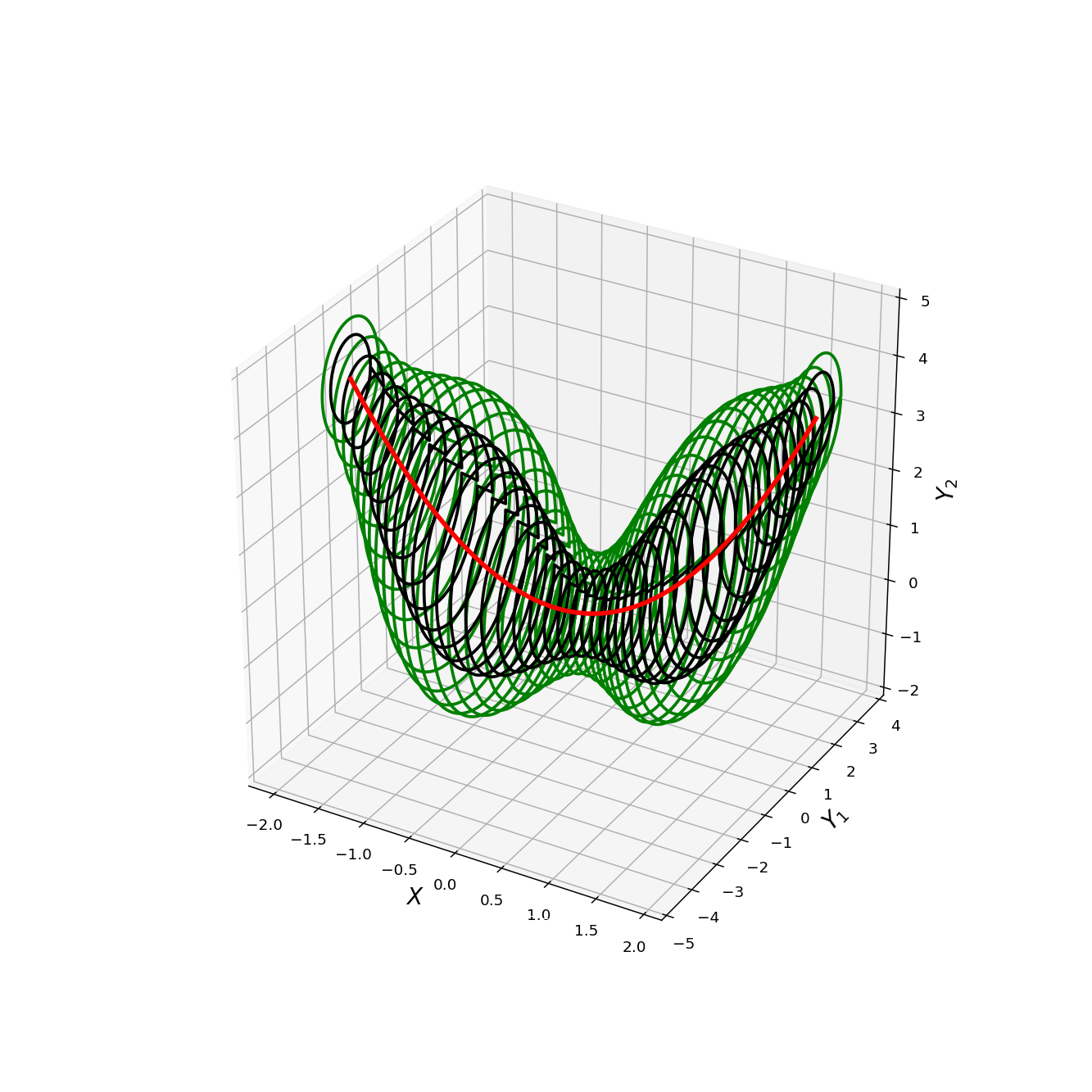}
   \caption{Estimated (sample sizes   $3, 601$ in the upper left panel,   $10, 000$ in the upper right panel, $128,020$ in the lower left panel) and population (lower right panel) quantile contours of order $\tau=0.2$ (black) and $0.4$ (green)  for   Model~\eqref{banana};  the (estimated)  conditional center-outward medians are shown in red. Estimations are based on the $k$-nearest neighbors weights \eqref{optionclasic}  
  with $N=k=401, 1,000$, and $6,401$ and $n=3,601, 10,000$, and $128,020$, respectively. }
    \label{fig:toyComparehallin}
\end{figure}
We start with analyzing the motivating example given in \cite{Halin2015}. The model (with $m=1$, $d=2$)  is 
\begin{equation}
    \label{banana}
  \mathbf{Y}=  \left(\!\begin{array}{c}Y_1 \\ Y_2\end{array}\!\right)
    =
    \left(\!\begin{array}{c}X \\ X^2\end{array}\!\right)+\left(1+\frac{3}{2}
    \sin\left(\frac{\pi}{2} X\right)^2\right)\mathbf{e}, \ 
     \ X\sim U_{[-2,2]}\ \text{and} \ \mathbf{e}\sim \mathcal{N}(\mathbf{0}, \mathbf{Id}),\vspace{0mm}
\end{equation}
with $X$ and $\mathbf{e}$ mutually independent. In this case, the population   conditional (on $X=x$) quantile contours are circles with radii depending on $x$ and can be computed exactly;    trend is parabolic and  heteroskedasticity  periodic.

      \begin{figure}[t!]
    \centering
 
     \includegraphics[width=6.5 cm,height=7cm,trim={5cm 5cm 6cm 4cm},clip]{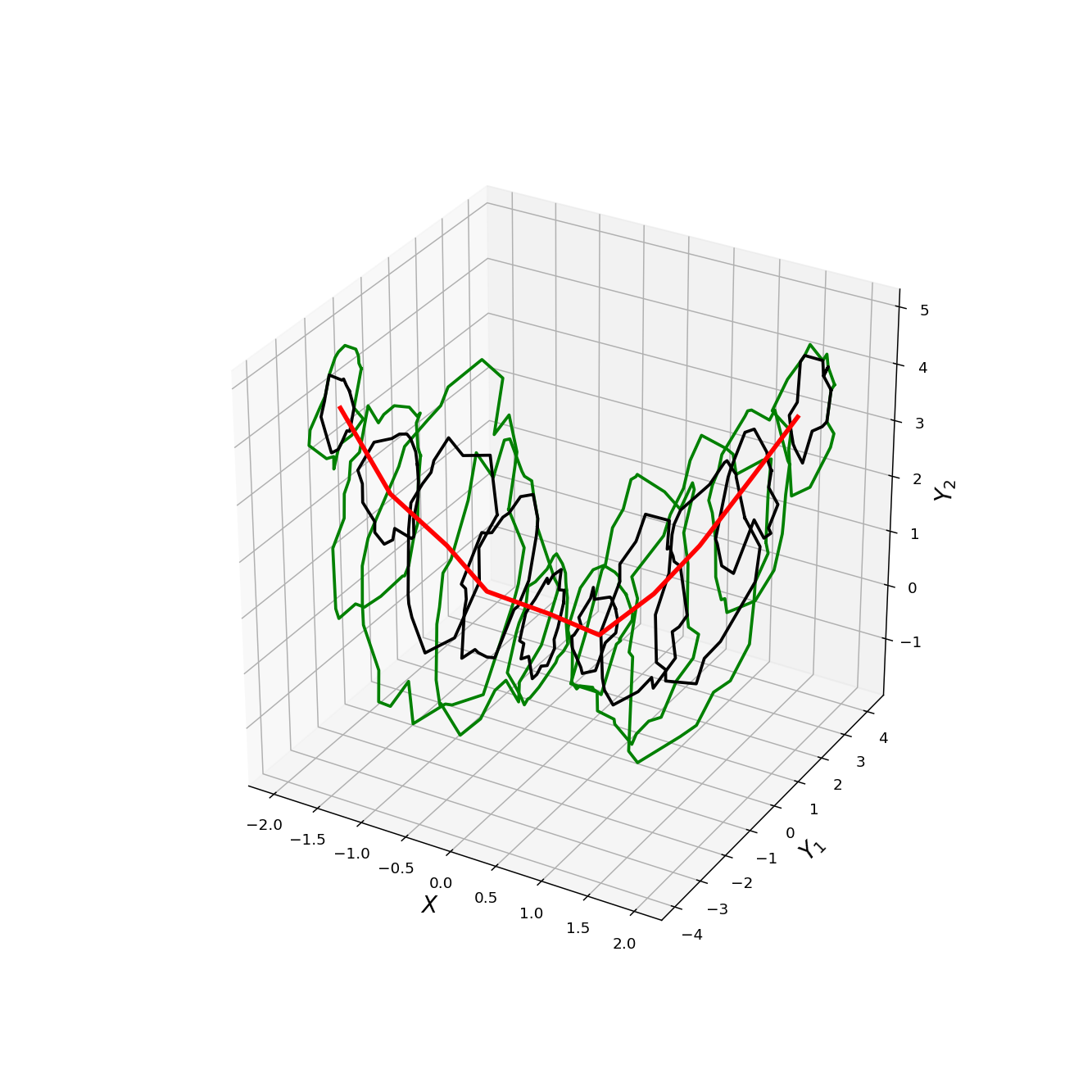}
     \includegraphics[width=6.5 cm,height=7cm,trim={5cm 5cm 6cm 4cm},clip]{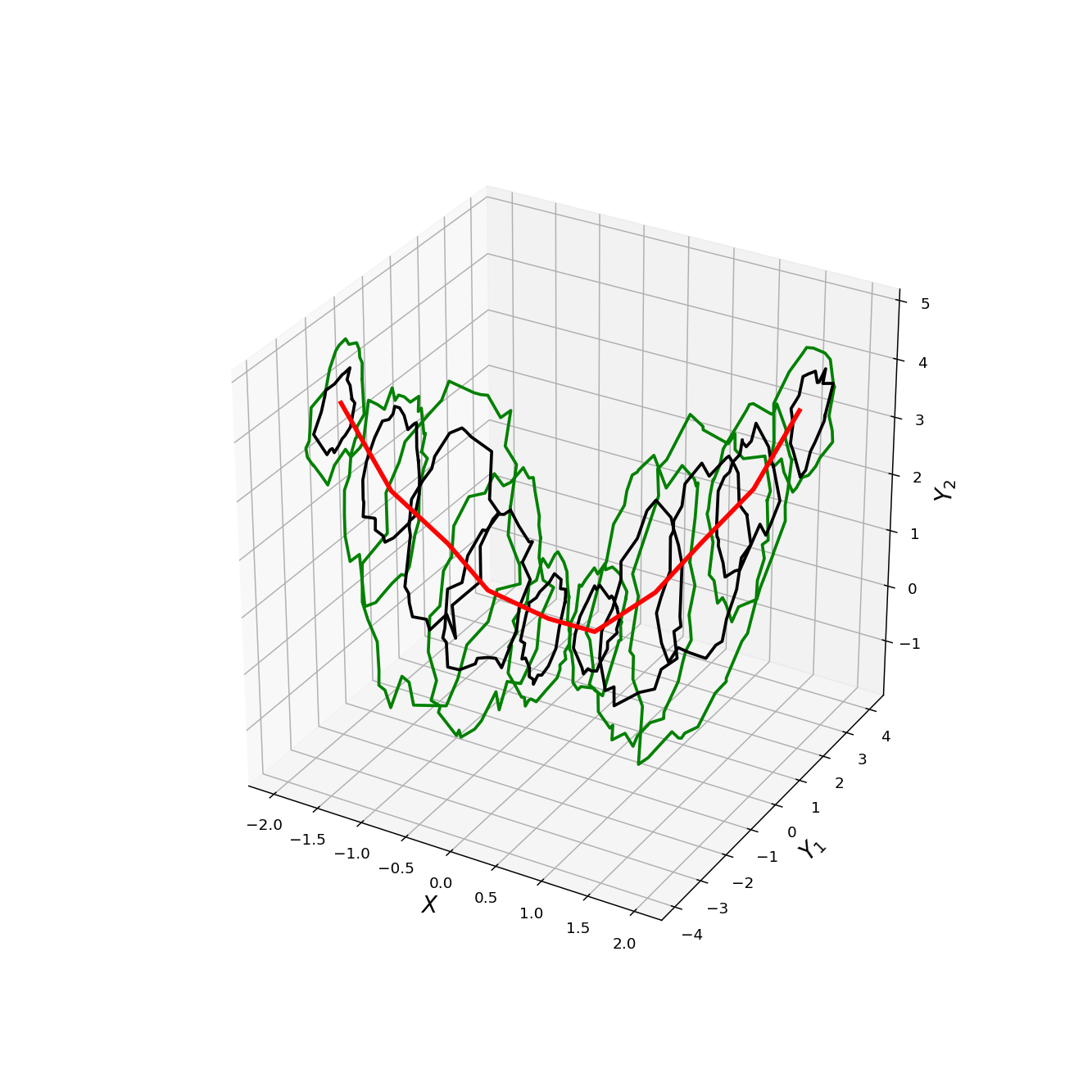}
        \includegraphics[width=6.5 cm,height=7cm,trim={5cm 5cm 6cm 4cm},clip]{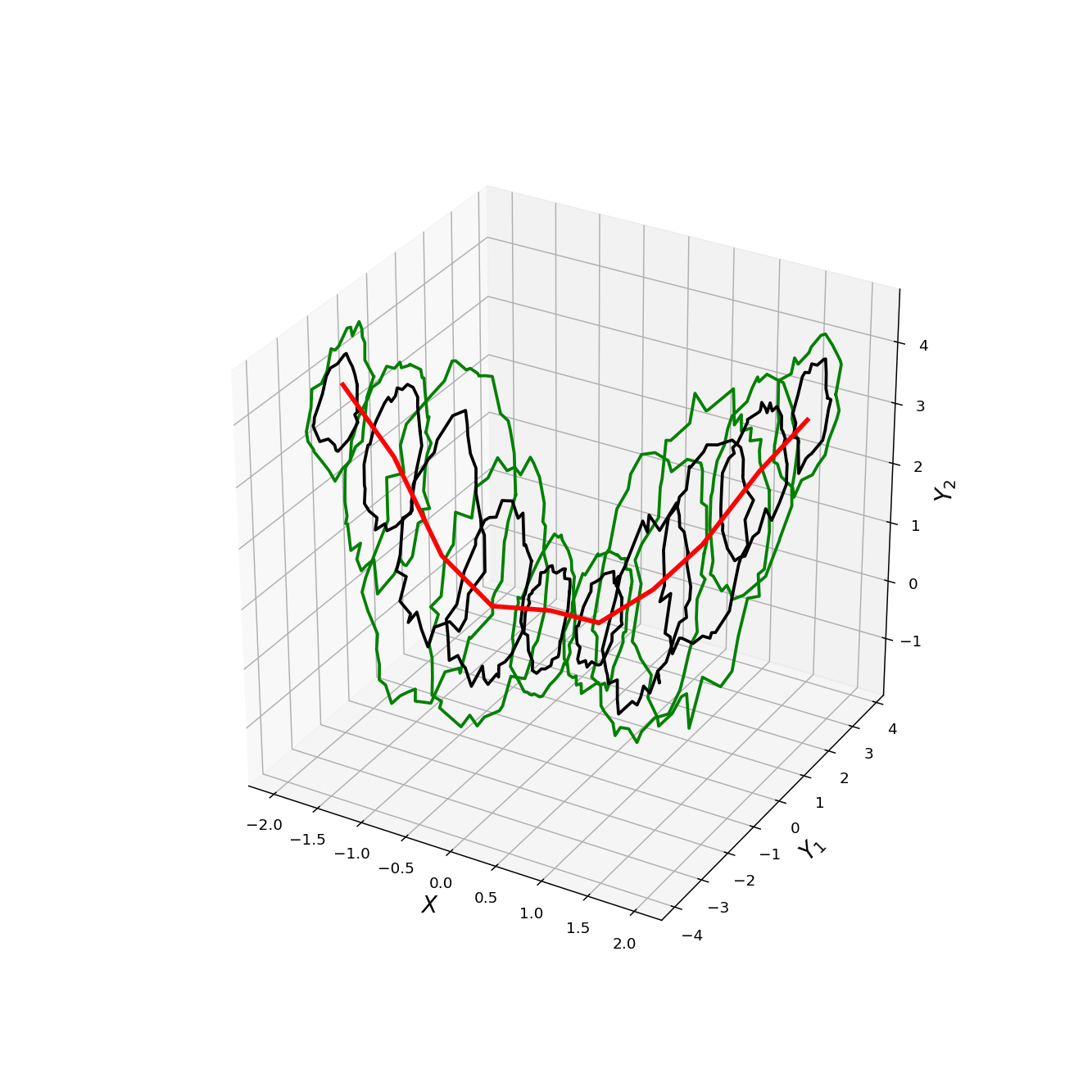}
            \includegraphics[width=6.5 cm,height=7cm,trim={5cm 5cm 6cm 4cm},clip]{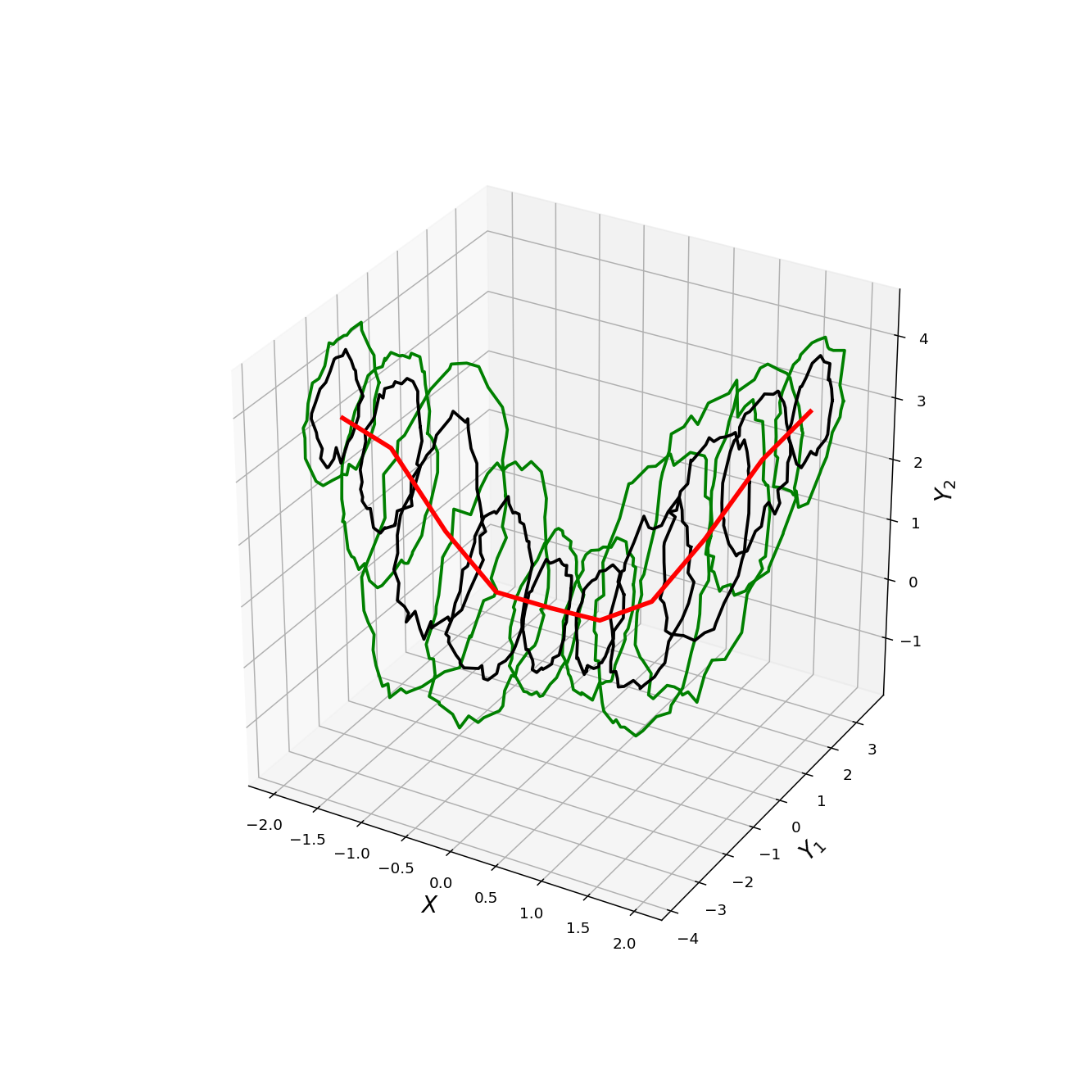}
            
    \caption{  Estimated quantile contours (Model~\eqref{banana})  based on Gaussian kernel weight functions for different choices of the bandwith $h$. The sample size is $n=3,601$;  bandwidth values are  $h=0.05$ (upper left panel),~$h=0.1$ (upper right panel), $h=0.2$ (lower left panel) and   $h=0.3$ (lower right panel). The estimated contour orders are $\tau=0.2$ (black) and $0.4$ (green);  the estimated conditional center-outward medians are shown in  red.\vspace{-5mm}}
    \label{fig:toyCompareGauss}
\end{figure}

{Figure~\ref{fig:toyComparehallin} illustrates the convergence of our estimated contours to the population counterparts. Compared to Figure~1 in  \cite{Halin2015},   our method produces less smooth contours, at least for smaller sample sizes. On the other hand,  our method is able to capture non-convex contour shapes---something the method in \cite{Halin2015} cannot, see Section \ref{Sec:BananaReg}. We  also underline that, from a computational point of view, our method is able to handle rather large datasets (in contrast, the R packaged \texttt{modQR} cannot handle sample sizes over $10, 000$, as explained in the documentation).

 Model~\eqref{banana}, as pointed out in \cite{Halin2015}, allows for testing the capacity of a method to estimate the trend while catching potential  heteroskedasticity.  A comparison with Figure~1 in  \cite{Halin2015} shows that both methods estimate the parabolic trend quite well, but that our method  performs much better  at  capturing heteroskedasticity. Estimations are based on the $k$-nearest neighbors weights \eqref{optionclasic}   of Section~\ref{Section;3}, 
  with $N=k=401, 1,000$, and $6,401$ and $n=3,601, 10,000$, and $128,020$, respectively.  Note that, in this univariate    covariate case, the weight choices  \eqref{optionclasic} and \eqref{optionNew} coincide.} 

\begin{figure}[t!]
    \centering
 
     \includegraphics[width=6.5 cm,height=7cm,trim={5cm 5cm 6cm 4cm},clip]{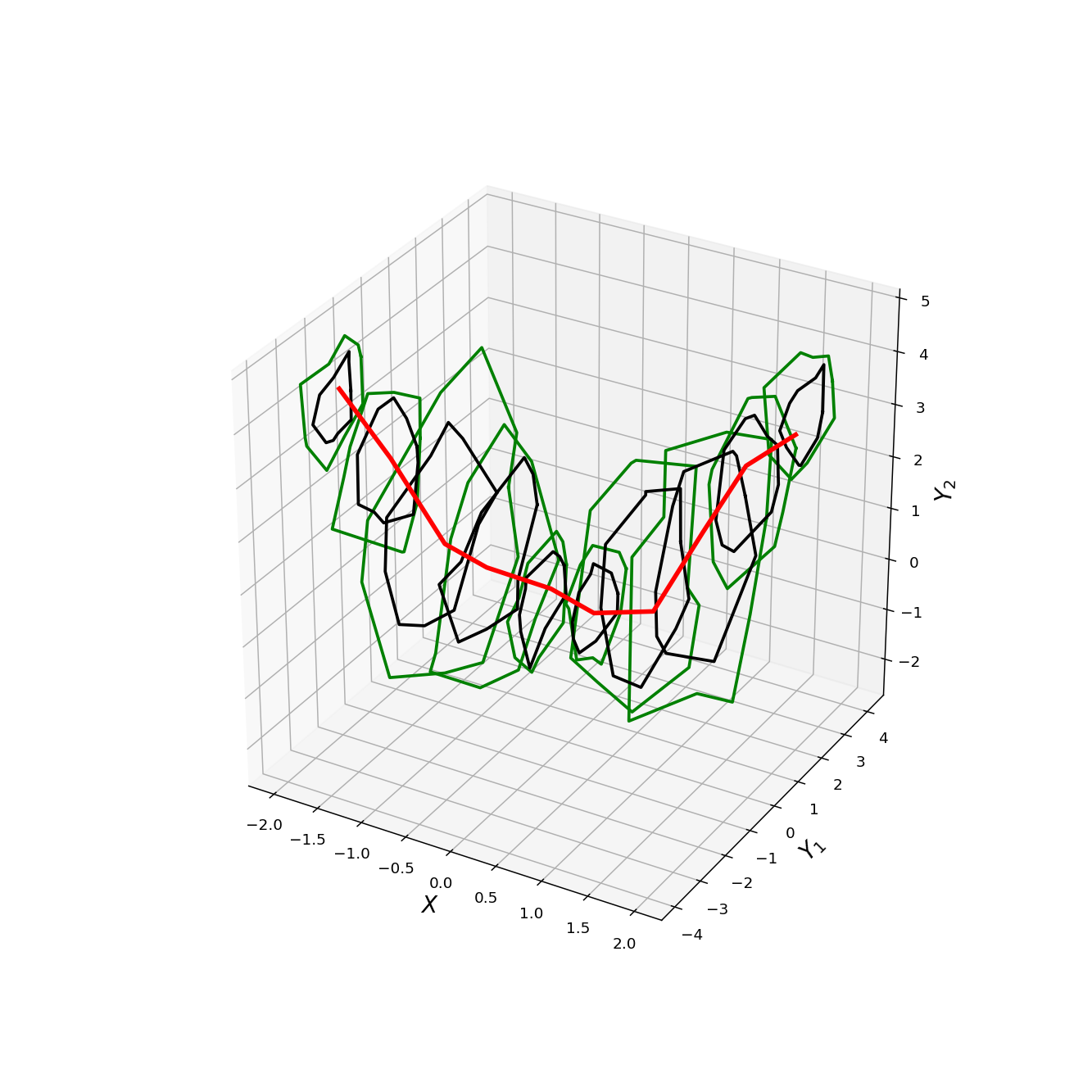}
     \includegraphics[width=6.5 cm,height=7cm,trim={5cm 5cm 6cm 4cm},clip]{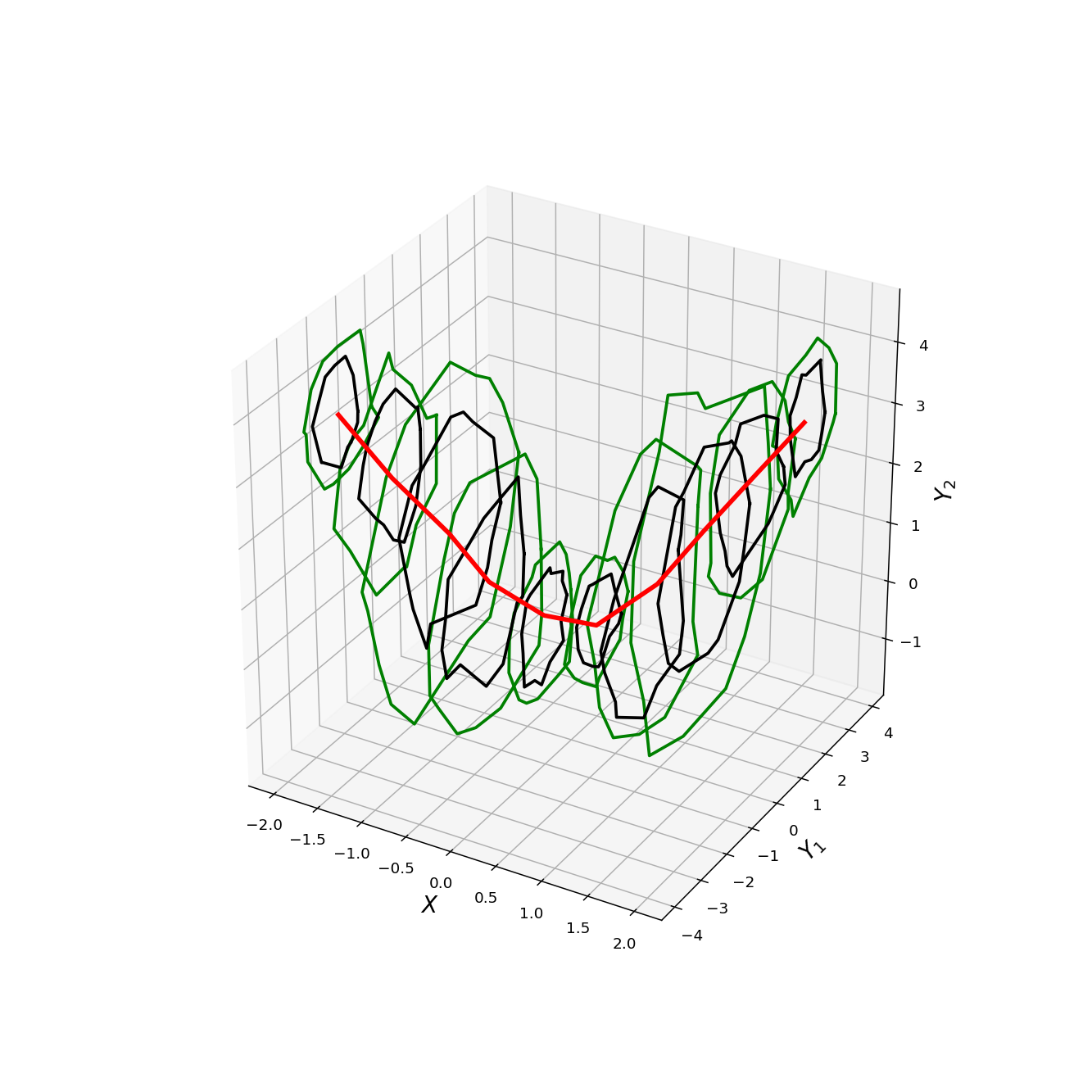}
        \includegraphics[width=6.5 cm,height=7cm,trim={5cm 5cm 6cm 4cm},clip]{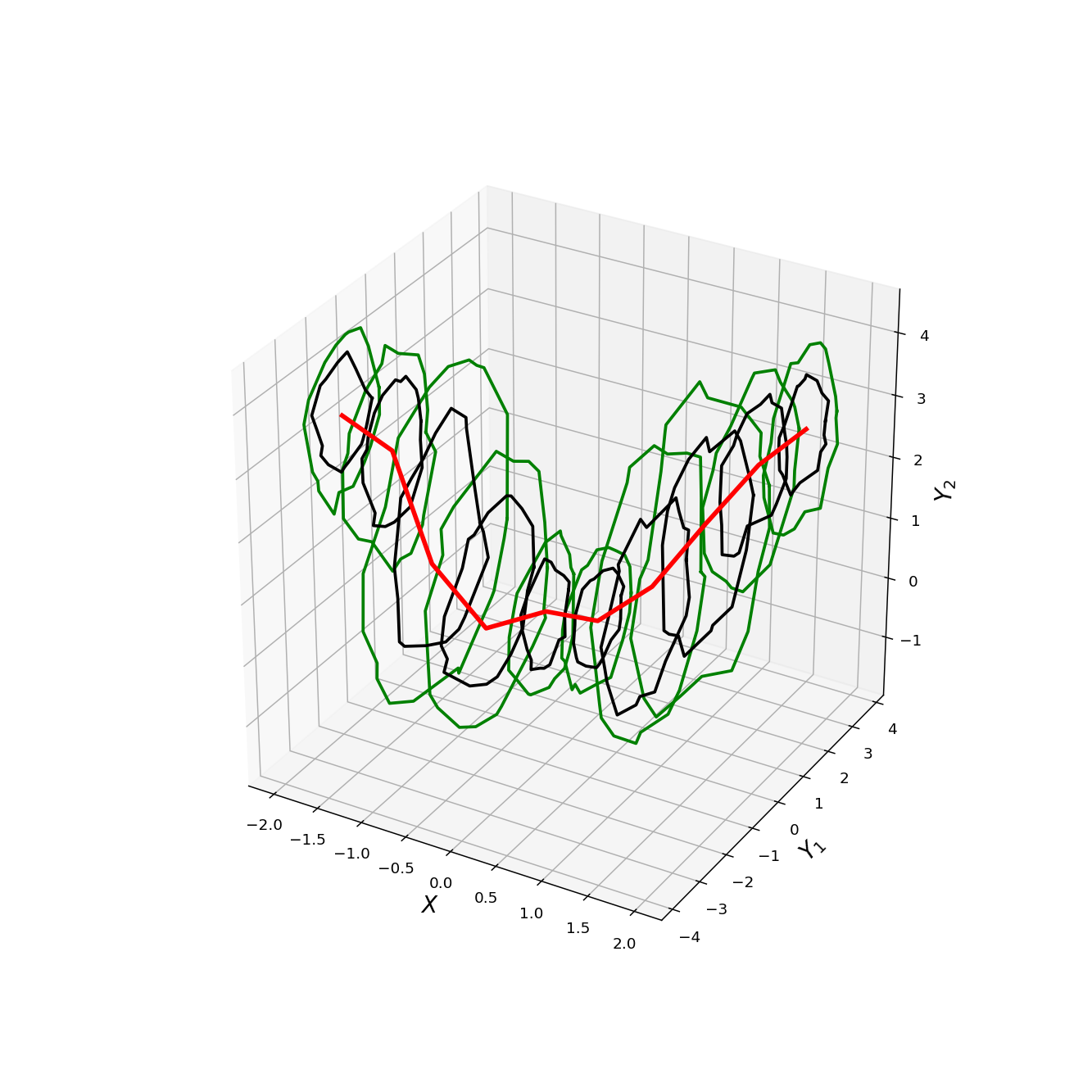}
            \includegraphics[width=6.5 cm,height=7cm,trim={5cm 5cm 6cm 4cm},clip]{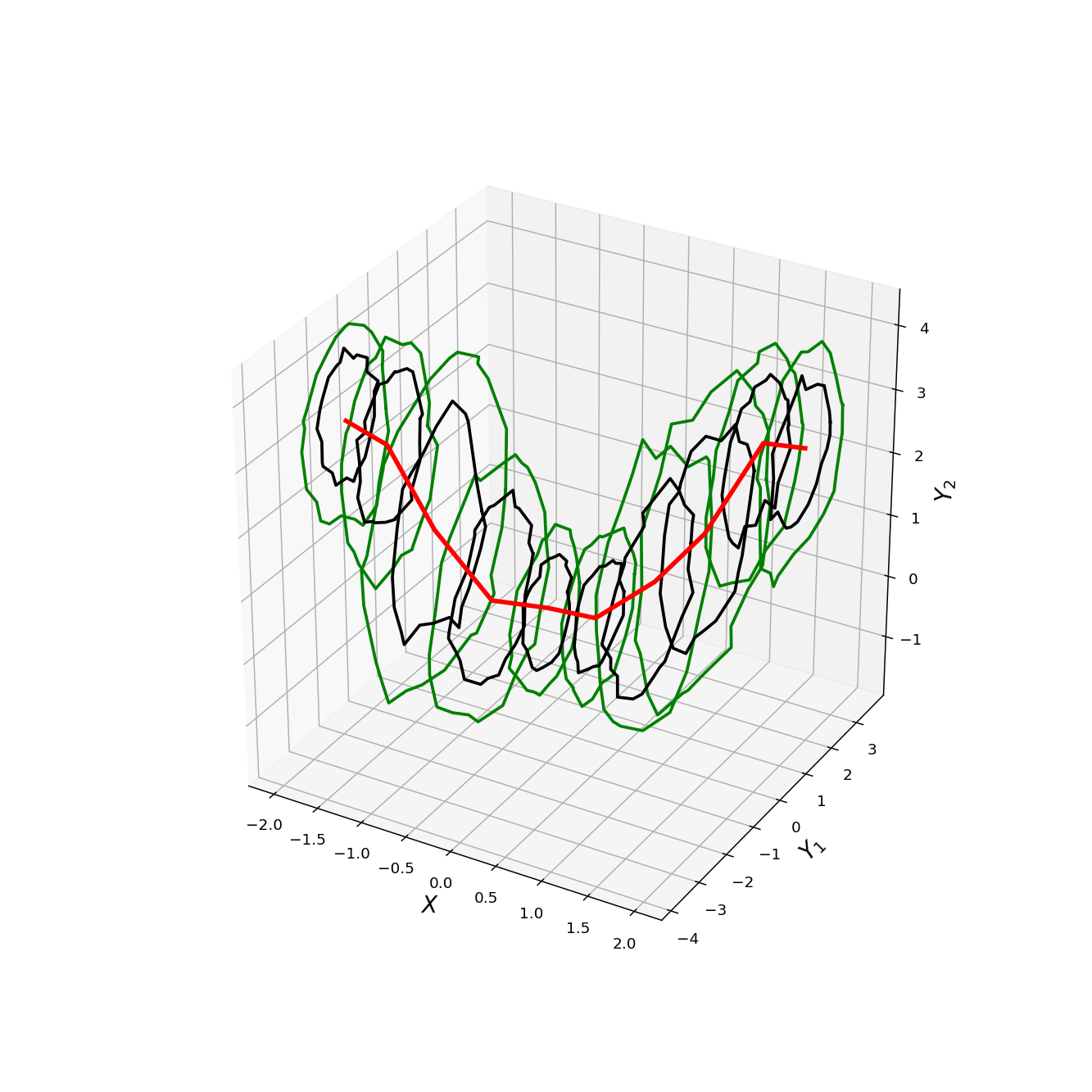}
            
    \caption{Estimated quantile contours (Model~\eqref{banana})  based on $k$-nearest neighbors weight functions   for different choices of $k$. The sample size is $n=3601$;  $k=101$ (upper left panel), $k=256$ (upper right panel), $k=401$ (lower left panel) and   $k=625$ (lower left panel). The estimated contour orders are $\tau=0.2$ (black) and $0.4$ (green);  the estimated conditional center-outward medians are shown in  red.\vspace{-5mm}}
    \label{fig:toyCompareKNN}
\end{figure}

The performance of the Gaussian kernel \eqref{optionGauss} (with various bandwidth choices) and classical $k$-nearest neighbors \eqref{optionclasic} (with various choices of $k$) weight functions are investigated in  Figures~\ref{fig:toyCompareGauss} and~\ref{fig:toyCompareKNN}, still  for  Model~\ref{banana}, with sample size $n=3,601$.  More precisely,  estimation in Figure~\ref{fig:toyCompareGauss} is based on the weight function 
\begin{equation}\label{GaussianW}
w^{(n)}_i(\mathbf{x};\mathbf{X}^{(n)})\coloneqq 
       {e^{-\left(\frac{|\mathbf{X_i}-\mathbf{x}|}{h_n}\right)^2}}/{\sum_{j=1}^n e^{-\left(\frac{|\mathbf{X_j}-\mathbf{x}|}{h_n}\right)^2}},\quad i=1,\ldots,n,
\end{equation}
for $h_n=0.05,\,0.1,\, 0.2$, and $0.3$;   the discretization of the spherical uniform is  based on a grid of size~$N=n$. 
Estimation in  Figure~\ref{fig:toyCompareKNN} is based on  $k$-nearest neighbors weights with~$k=101,\, 256$, and~$625.$ Here we chose $N=k$, which yields a one-to-one solution in~\eqref{kanto_reg}. In both figures the conditional contours (of order $\tau=0.2$ (black) and   $\tau=0.4$ (green)) and  the estimated conditional center-outward medians (red) are shown for $$x\in \{-2, -1.6, -1.1, -0.7, -0.2,  0.2,
  0.7,  1.1, 1.6, 2\}. $$ 

 For this sample size, the Gaussian kernel weights---due to the fact that they better exploit  the  information available on the $x$'s---yield better  results than the $k$-nearest neighbors ones.   But the Gaussian kernel has a drawback for large datasets; the optimization problem \eqref{kanto_reg} requires the whole dataset   and cannot be efficiently computed. For instance, the Gaussian kernel counterpart of Figure~\ref{fig:toyComparehallin} (where $n=128,020$) cannot be computed on a standard desktop computer:  large-sample datasets should be handled either with  nearest neighbors or   compactly supported kernel weights. On the other hand, the bandwith $h_n$ and the neighborhood size $k$ apparently have little impact on he result.  
 \begin{figure}[b!]
    \centering
    \includegraphics[width=10 cm,height=9 cm,trim={10cm 5cm 6cm 8cm},clip]{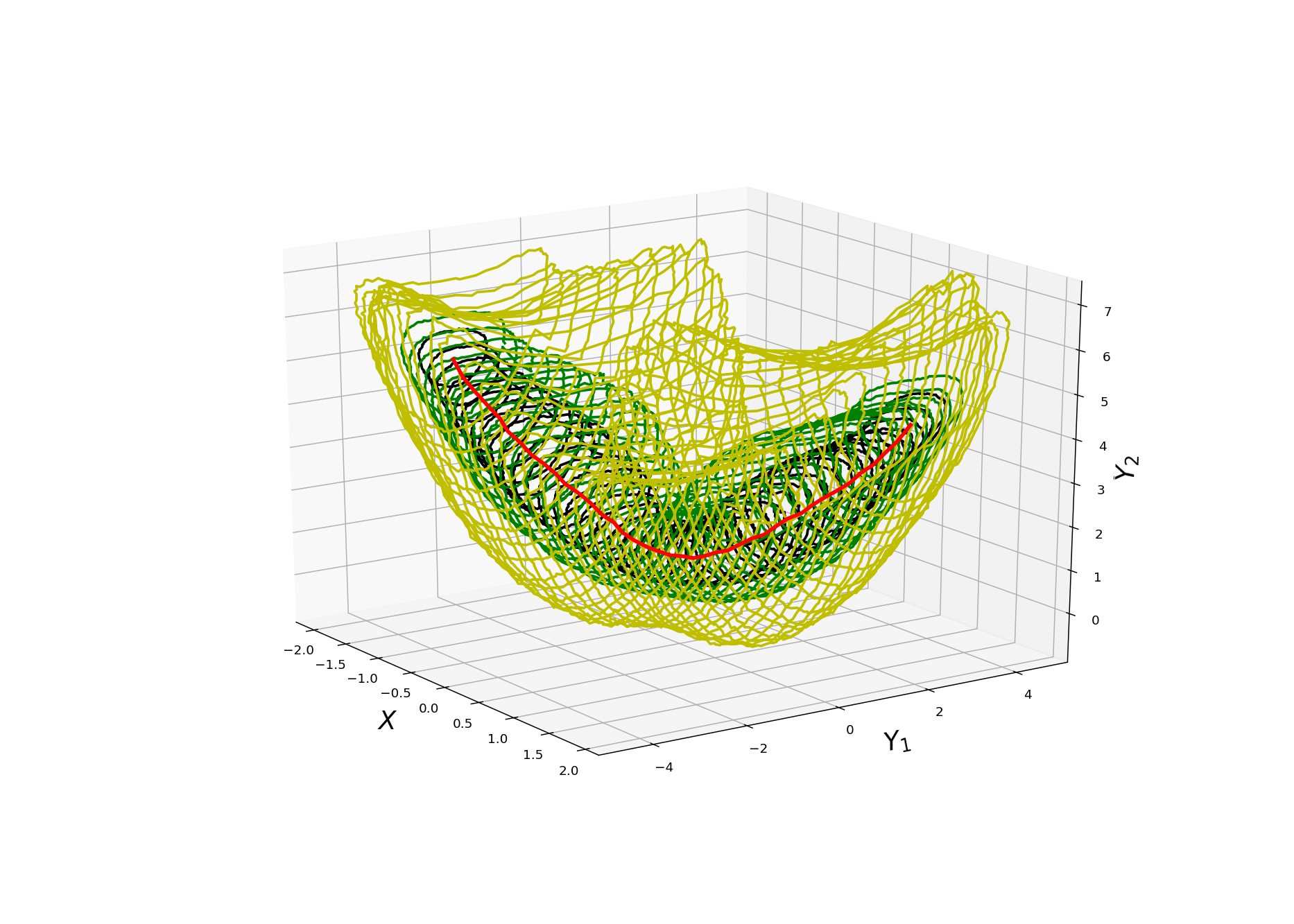}\vspace{-4mm}

    \caption{A numerical approximation   (estimation based on a simulated sample of size $n=576,040$)   of the quantile contours of order $\tau=0.2$ (black),  $0.4$  (green), and~$0.8$ (yellow) for   Model~\eqref{mixtmod};  the conditional center-outward medians are shown in red. \vspace{-4mm}
}
    \label{fig:banana}
\end{figure}


\subsubsection{Parabolic trend and periodic heteroskedasticity; banana-shaped conditional densities.}\label{Sec:BananaReg}

\begin{figure}[b!]
    \centering
 
     \includegraphics[width=6.5 cm,height=7cm,trim={5cm 5cm 6cm 4cm},clip]{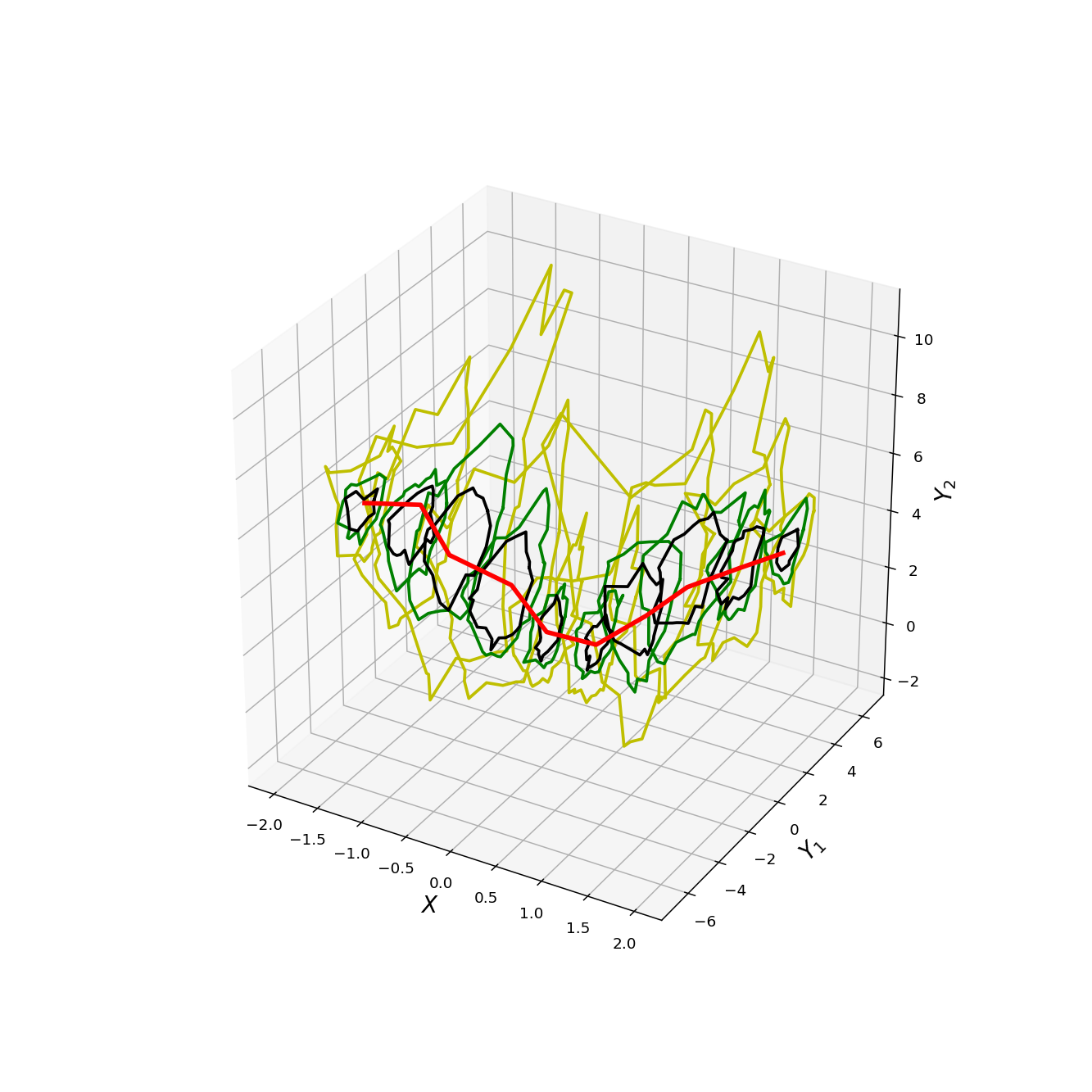}
     \includegraphics[width=6.5 cm,height=7cm,trim={5cm 5cm 6cm 4cm},clip]{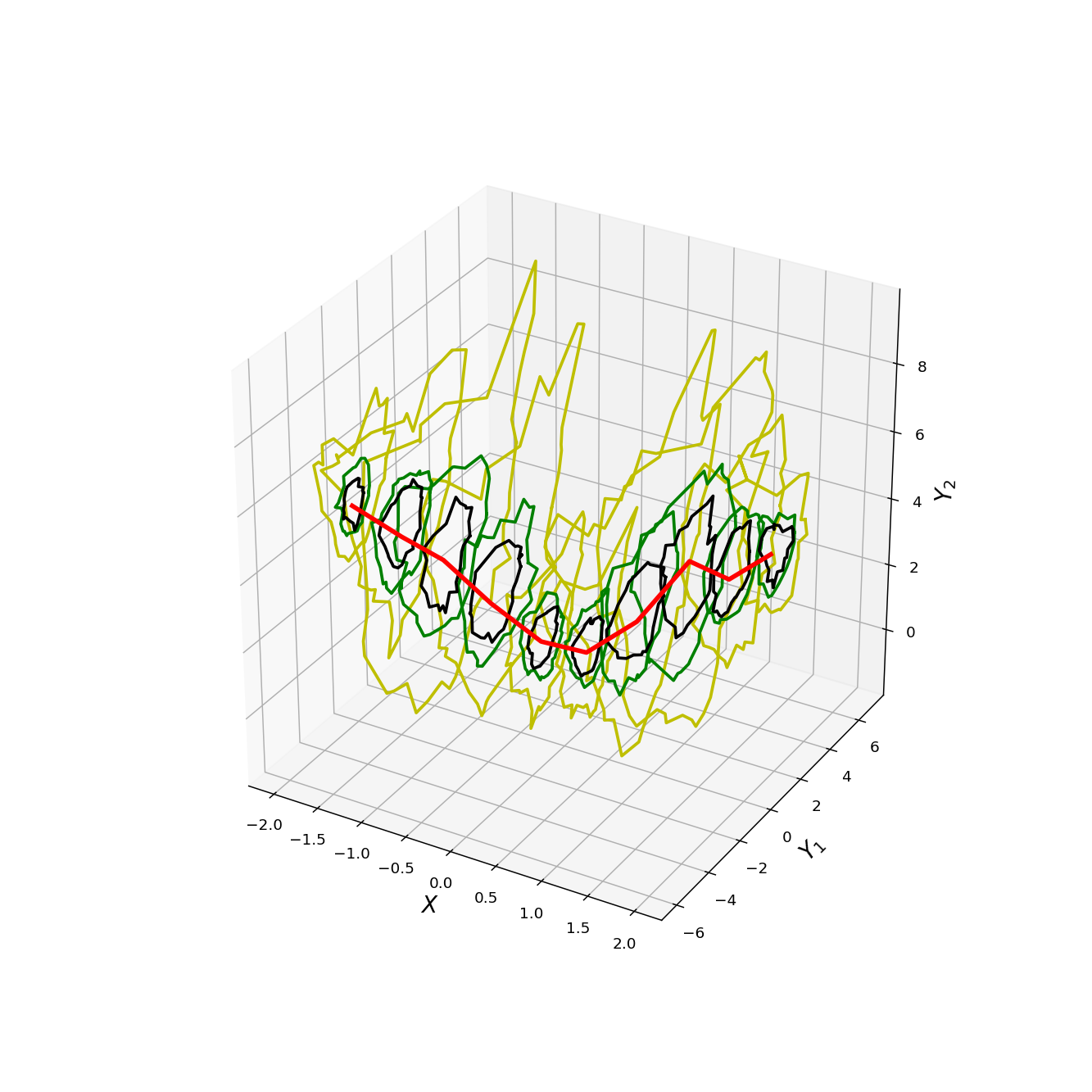}
        \includegraphics[width=6.5 cm,height=7cm,trim={5cm 5cm 6cm 4cm},clip]{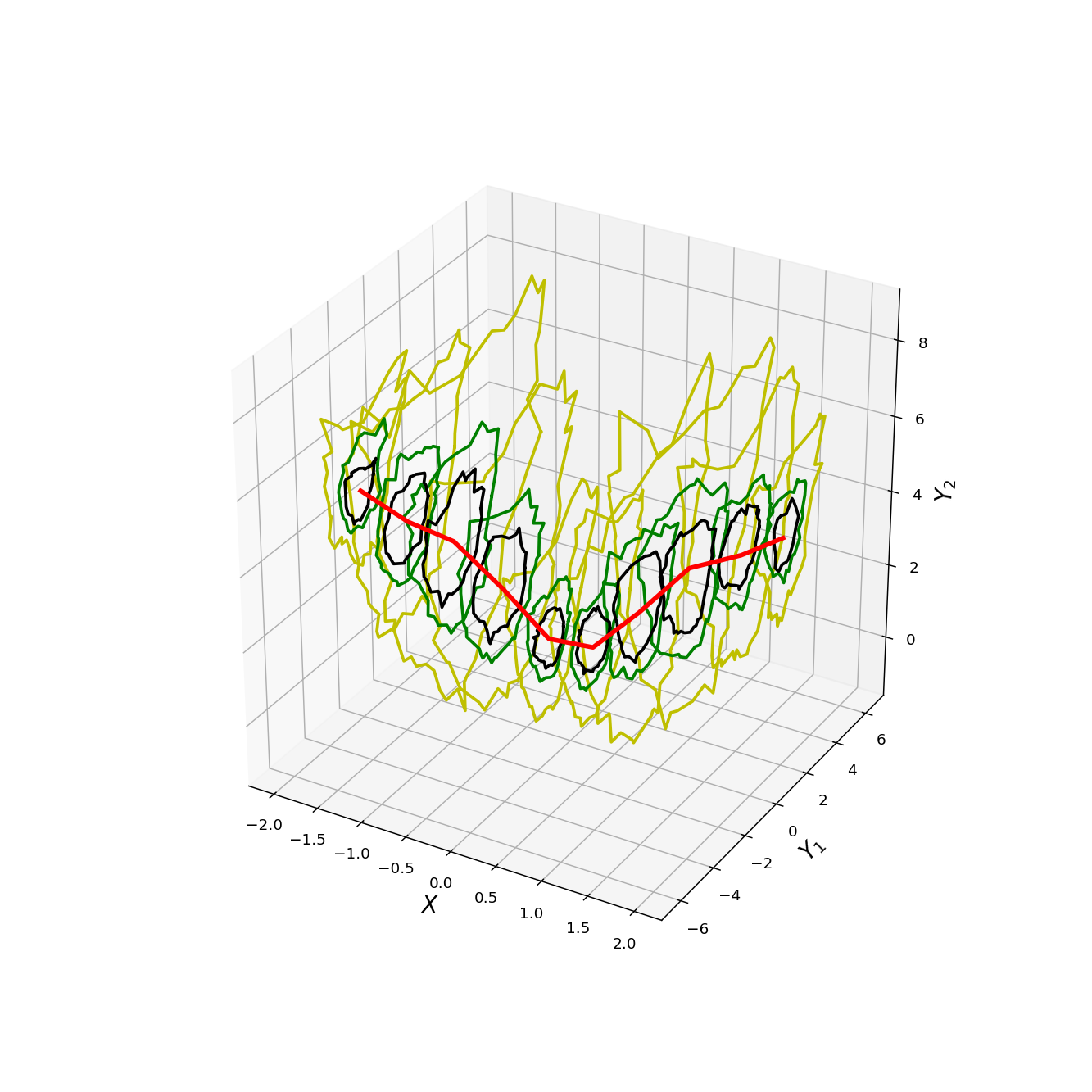}
            \includegraphics[width=6.5 cm,height=7cm,trim={5cm 5cm 6cm 4cm},clip]{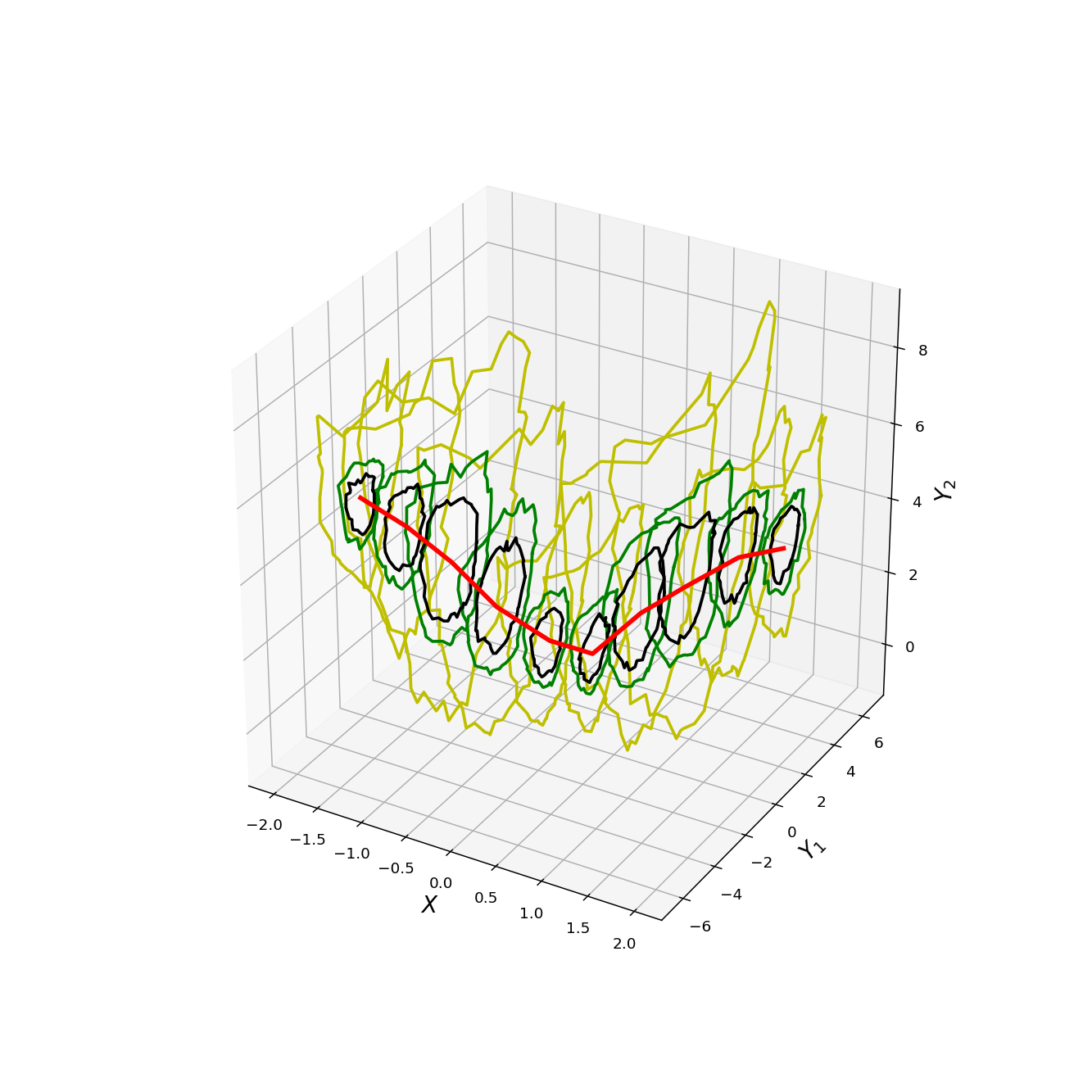}
            
    \caption{  Performance, in Model~\eqref{mixtmod}, of   Gaussian kernel weight functions-based estimation  for various choices of the bandwidth. The sample size is $n=3,601$, and the bandwidths are  $h=0.05$ (upper left panel),~$h=0.1$ (upper right panel), $h=0.2$ (lower left panel), and   $h=0.3$ (lower left panel). The empirical contour levels  are $\tau=0.2$ (black) , $0.4$ (green) and $0.8$ (yellow);  the estimated conditional center-outward medians are shown in red.}
    \label{fig:toyBananCompareGauss}
\end{figure}
\begin{figure}[t!]
    \centering
 
     \includegraphics[width=6.5 cm,height=7cm,trim={5cm 5cm 6cm 4cm},clip]{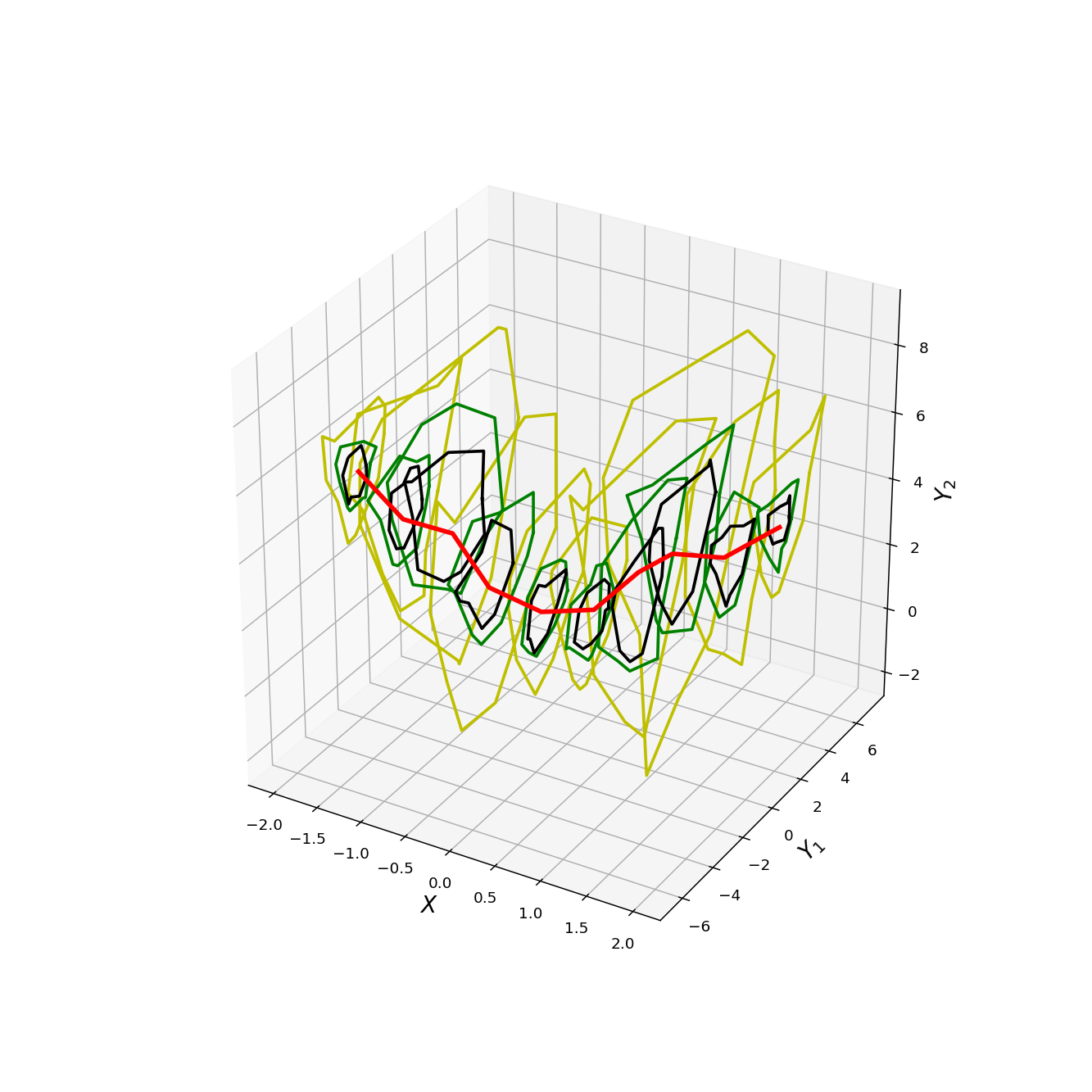}
     \includegraphics[width=6.5 cm,height=7cm,trim={5cm 5cm 6cm 4cm},clip]{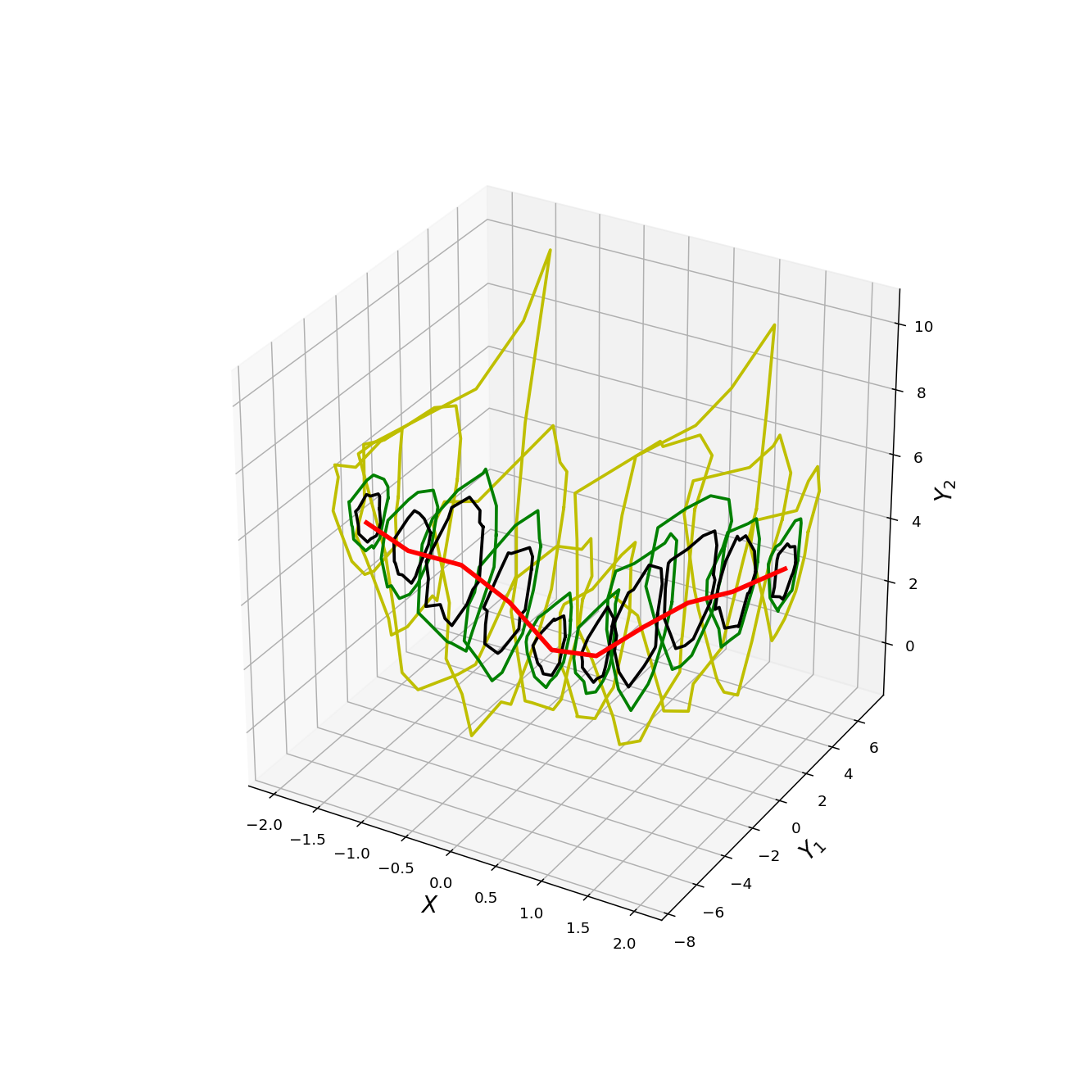}
        \includegraphics[width=6.5 cm,height=7cm,trim={5cm 5cm 6cm 4cm},clip]{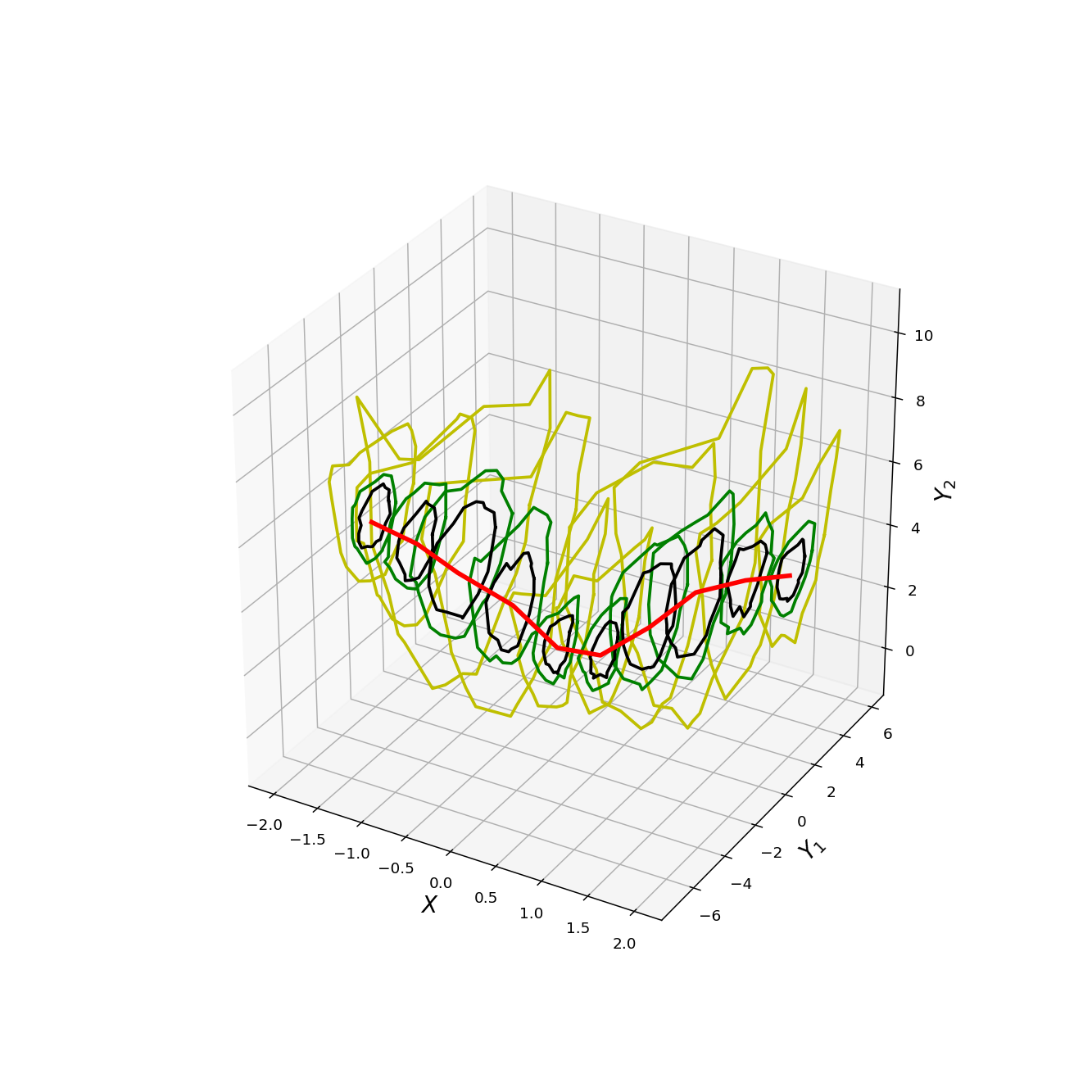}
            \includegraphics[width=6.5 cm,height=7cm,trim={5cm 5cm 6cm 4cm},clip]{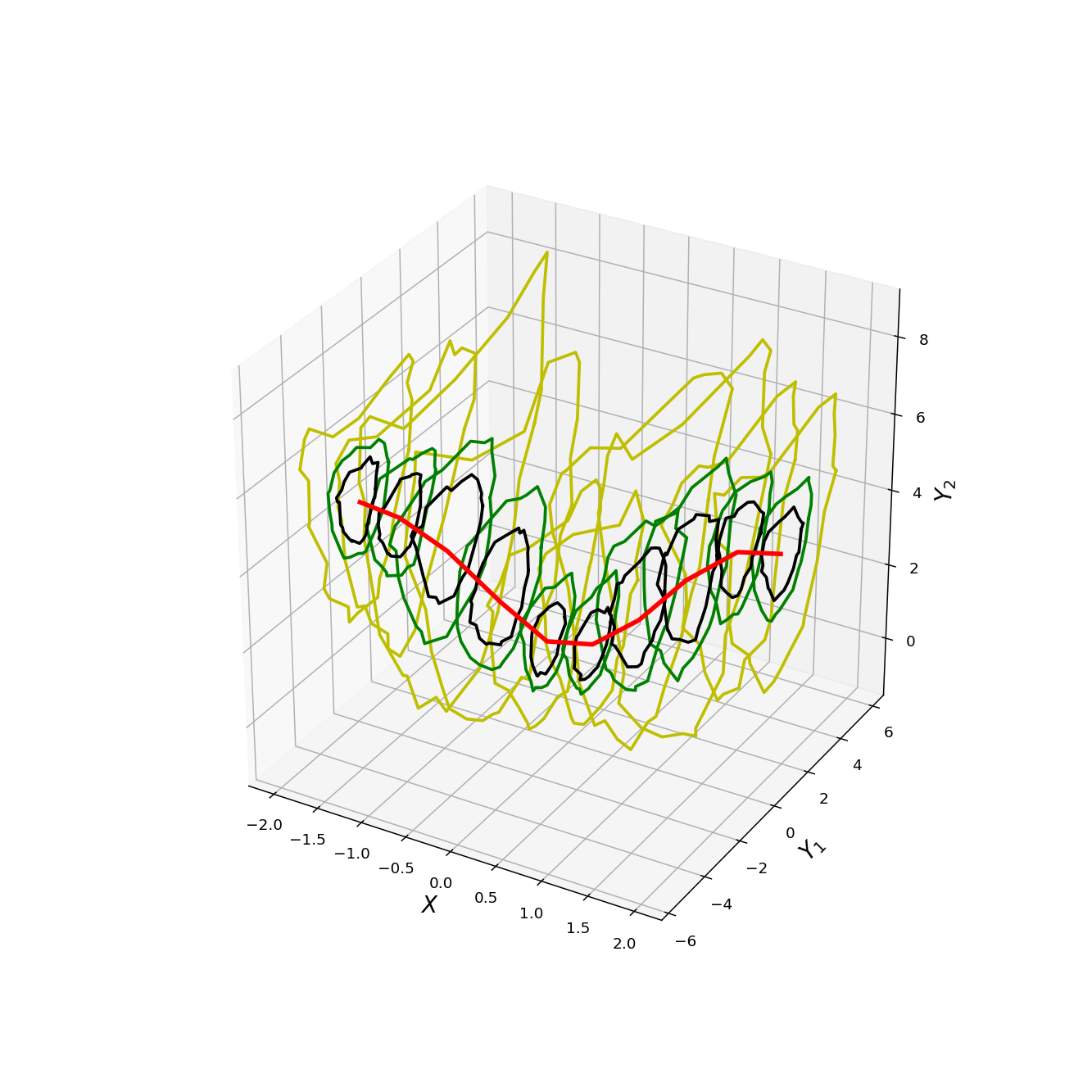}
            
    \caption{  Performance, in Model~\eqref{mixtmod}, of the $k$-nearest neighbors weight functions-based estimation for various choices of $k$. The sample size is $n=3601$;  $k=101$ (upper left panel), $k=256$ (upper right panel),~$k=401$ (lower left panel), and   $k=625$ (lower left panel). The empirical contour levels are $\tau=0.2$ (black), $0.4$ (green),  and $0.8$ (yellow);  the estimated conditional center-outward medians are shown in red.}
    \label{fig:toyBananCompareKNN}
\end{figure}
We  now consider a model in which the trend and heteroskedasticity  are the same   as in Model~\eqref{banana}, but the quantile contours  are  non-convex (conditional densities are  banana-shaped):
\begin{equation}\label{mixtmod}
   \mathbf{Y}=  \left(\!\begin{array}{c}Y_1 \\ Y_2\end{array}\!\right)
    =\left(\!\begin{array}{c}X \\ X^2\end{array}\!\right)
    +\left(\!\begin{array}{c}
  \left(1+\frac{3}{2}\sin\left(\frac{\pi}{2} X\right)^2\right){1.15}\, e_1\\ \left(1+\frac{3}{2}\sin\left(\frac{\pi}{2} X\right)^2\right)\left(\dfrac{e_2}{1.15}+0.5(e_1^2+1.21)\right)\end{array}\!\right),\vspace{-1mm}
    \end{equation} 
    with $X\sim U_{[-2,2]}$ and   $\mathbf{e}=(e_1,e_2)\sim \mathcal{N}(\mathbf{0}, \mathbf{Id})$,  $X$  and  $\mathbf{e}$ mutually independent.  The  dataset shown in Figure~\ref{fig:corazon} was generated from that model, with  $X=0$ and justifies the termonology ``banana-shaped.''

Population conditional quantile contours here cannot be computed analytically.  Figure~\ref{fig:banana} shows their estimations (same method as in Section~\ref{sphcasesec}, with   a $k$-nearest neighbors weight function,   $k=14,401$) for orders $\tau= 0.2$ (black), $0.4 $ (green), and~$0.8$ (yellow), along with the conditional medians (red). The sample size $n=576,040$  is very large, so that, in view of   consistency results, 
 these estimations can be considered as   close  numerical approximations of their theoretical counterparts. This example illustrates further the ability of our method to handle vary large sample sizes.
The  \cite{Halin2015} approach produces  contours that are convex by construction---hence cannot capture the ``banana shape'' of the conditional densities.
The same comparative analysis of weight functions  is performed   as for Model~\eqref{banana}. Figures~\ref{fig:toyBananCompareGauss}  and~\ref{fig:toyBananCompareKNN} show the results for  Gaussian kernel weights (various bandwiths) and $k$-nearest neighbors weights (various values of $k$), respectively. 
The  sample size is $n=3601$. In all the examples, for the ease of computation, $N=k$. Note that, for the $k$-nearest neighbor weights, this choice creates a one-to-one (between the sample and the grid) transport map.

\begin{figure}[b!]
    \centering
    \includegraphics[width=6.5 cm,height=7cm, trim={6cm 2cm 3cm 5cm},clip]{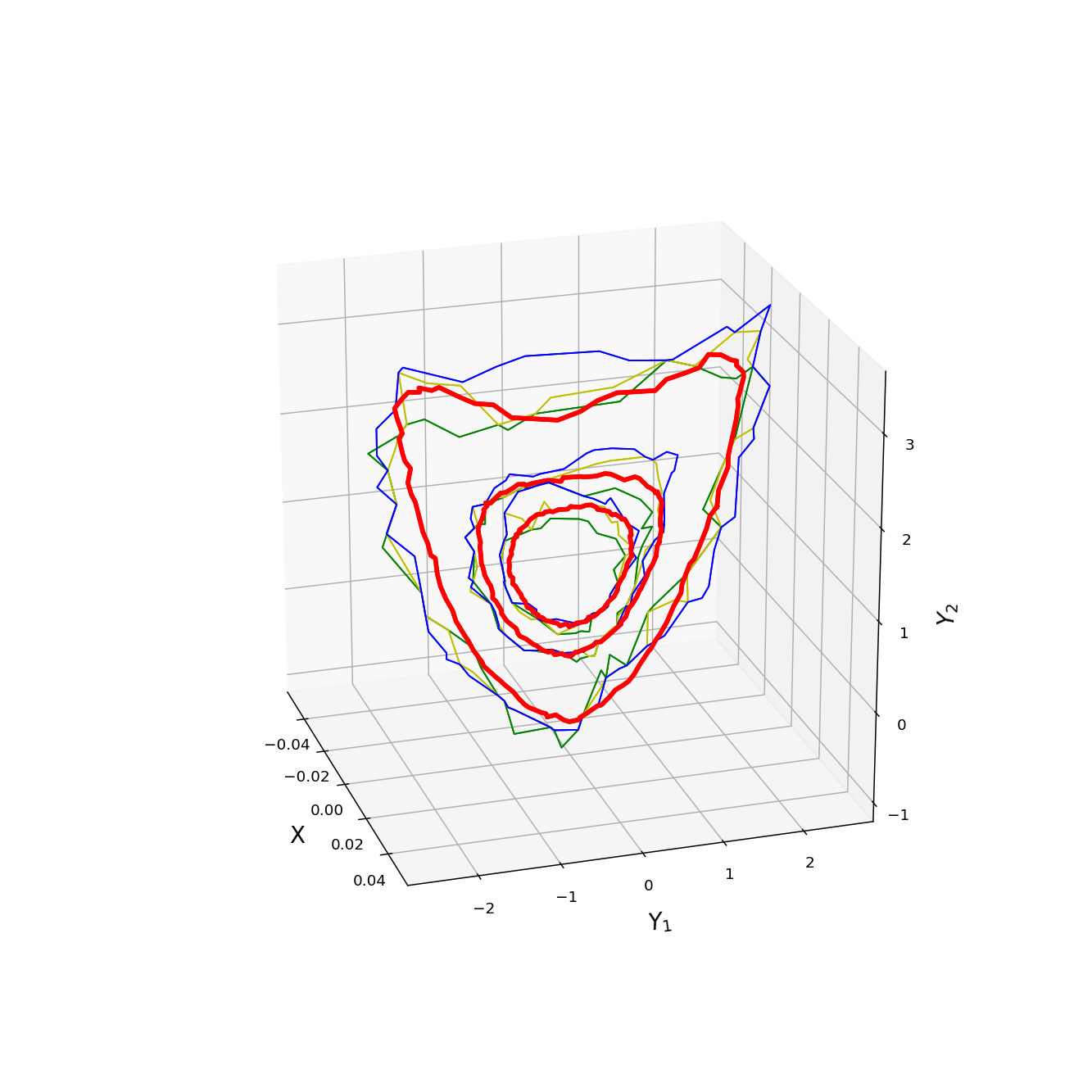} 
    \includegraphics[width=6.5 cm,height=7 cm, trim={6cm 2cm 3cm 5cm},clip]{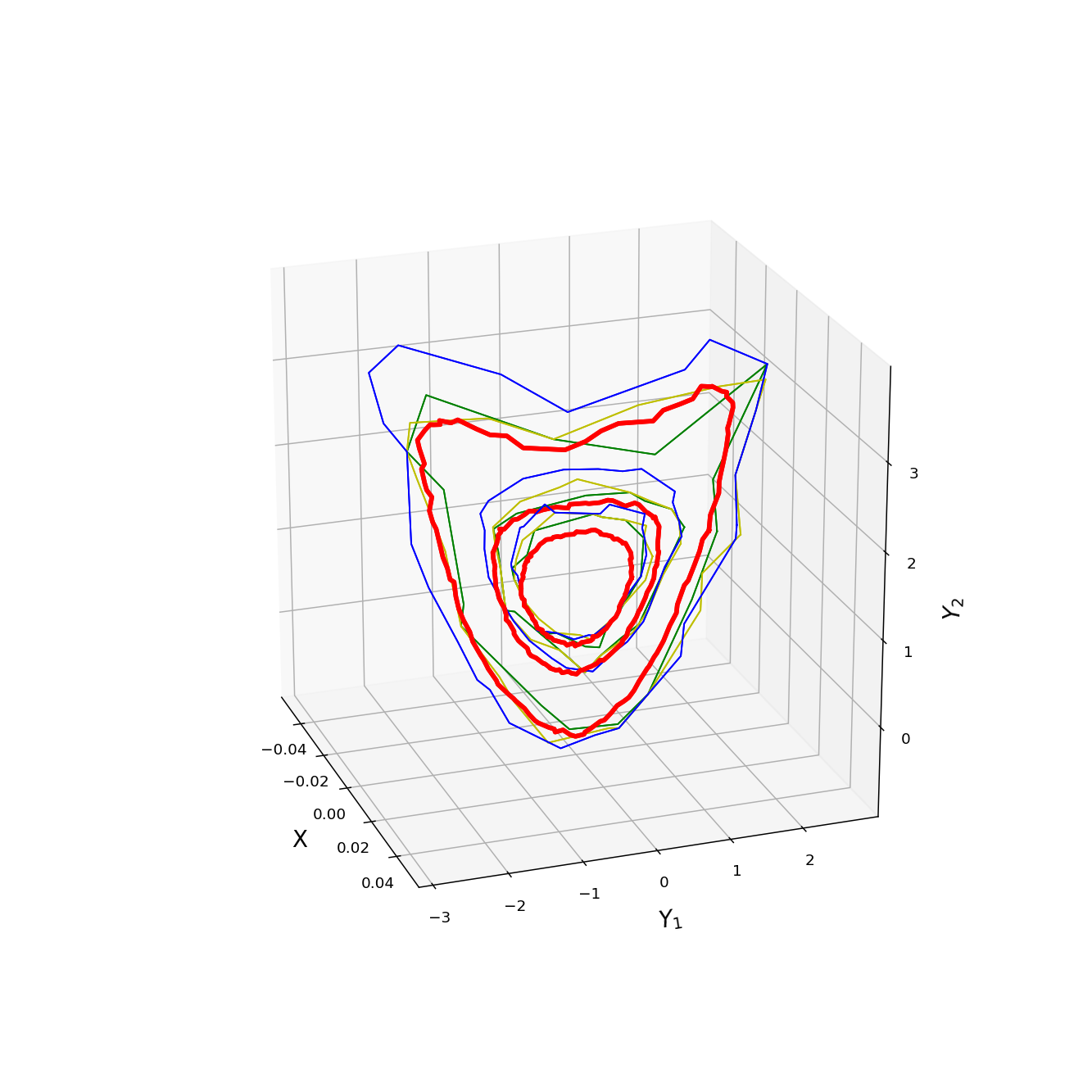}
    \caption{Comparison between the empirical conditional contours~$ \mathcal{C}_{\pm}^{(n)}(\tau \, \big| 0)$ (levels $\tau=0.2,\, 0.4, \, 0.8 $; sample size $n=3,601$) in Model~\eqref{mixtmod}  based on    Gaussian kernel weight functions  (bandwidths $h=0.1$ (green),~$h=~\!0.2$ (yellow), and $h=0.3$ (blue))  (left panel)  and those in Figure~\ref{fig:banana},  based on $k-$nearest neighbors weight functions ($k =226$ (green), $k=485$ (yellow), and $k=901$ (blue); sample size $n=576,040$)    (right panel).   The center-outward quantiles of Figure~\ref{fig:banana} (to be considered as population values) are shown in red. }
    \label{Fig:stabilityMixture}
\end{figure}

Still for Model~\eqref{mixtmod}, Figure \ref{Fig:stabilityMixture} shows, for  sample size $n=3601$ and various choices of the bandwidth $h$ and the neighborhood size~$k$,  the behavior  of the empirical  conditional center-outward contours in $X=0$---and compares them with those of Figure~\ref{fig:banana} (considered as the population contours).
The empirical conditional quantiles are computed for $\tau=0.2,\, 0.4$, and~$0.8$ with Gaussian kernel weights (bandwidths $h=0.1,\, 0.2$, and $0.3$) and the $k-$nearest neighbors weights ($k=226,\, 485$, and~$901$). The influence of the choice of  $h$ and $k$  is clearly seen here:  the bigger $h$ (the bigger $k$), {the smoother  the estimation of the shape of the contours but also,  unfortunately, the worse the estimation of their location. }
\color{black}

\subsection{Some real-data examples}\label{realsec}
\begin{figure}[t!]
    \centering
    \includegraphics[width=6.5 cm,height=7cm, trim={6cm 2cm 3cm 5cm},clip]{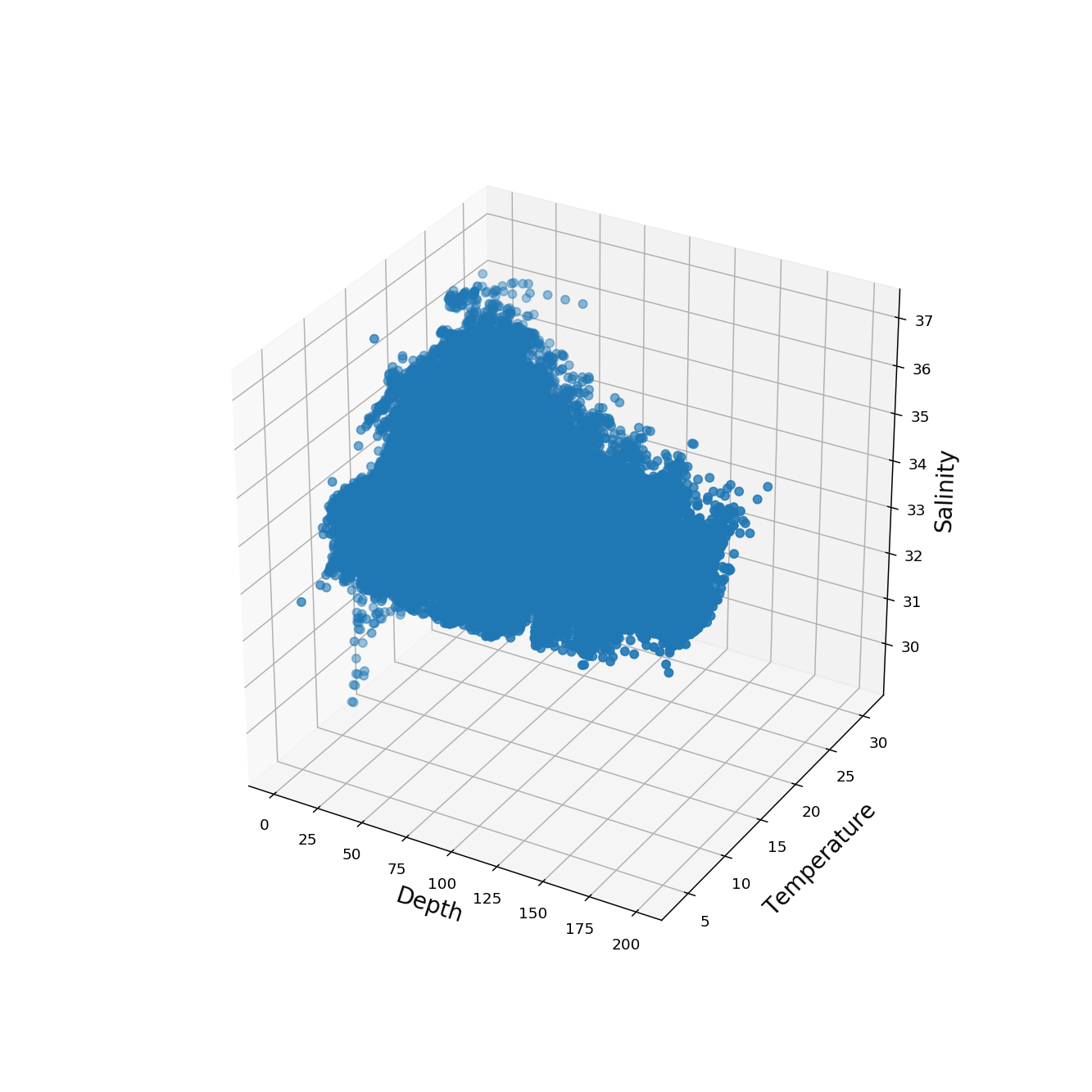} 
    \includegraphics[width=6.5 cm,height=7 cm, trim={6cm 2cm 3cm 5cm},clip]{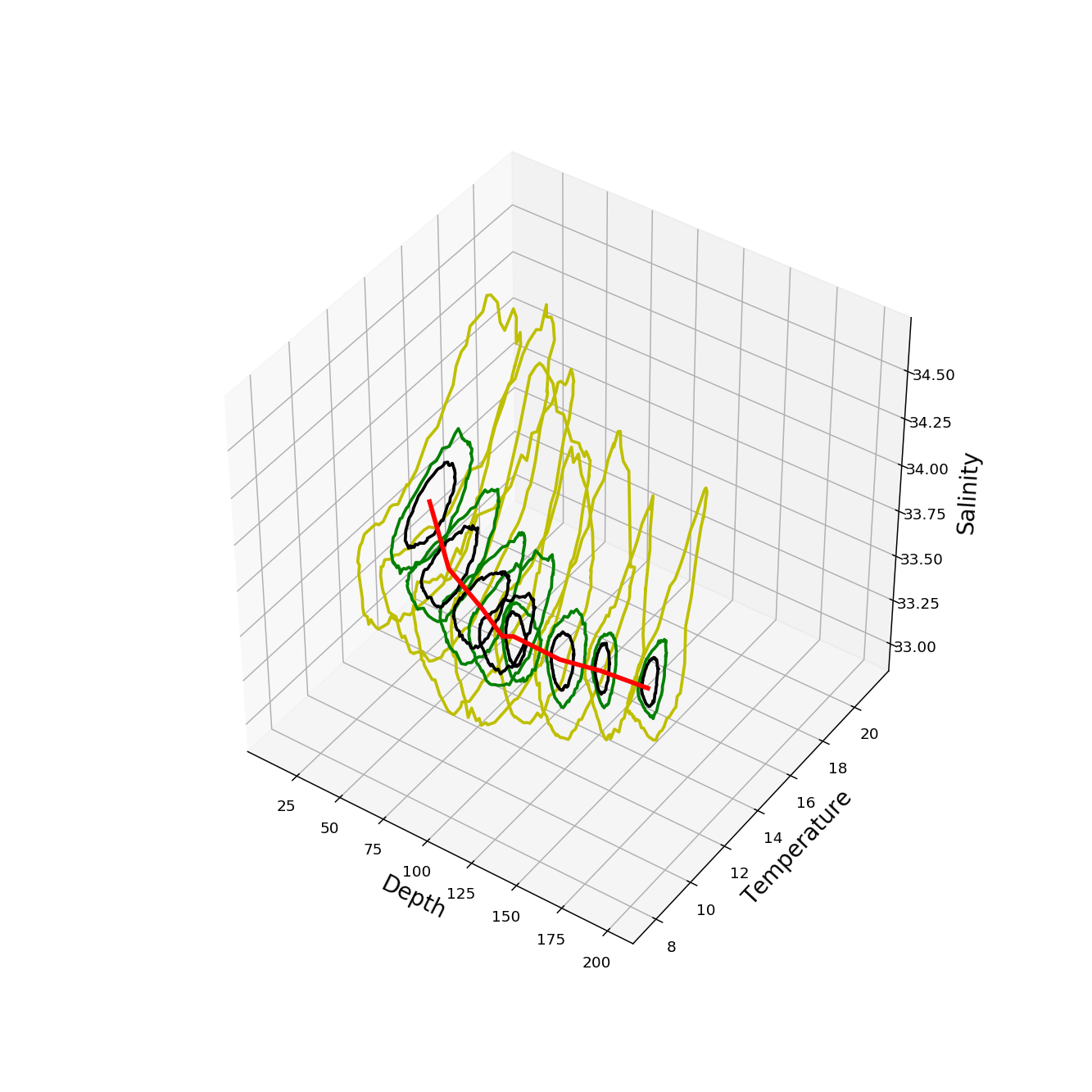}
    \caption{
    CalCOFI dataset. Left panel:  the original dataset (`depth,' `temperature,' `salinity') for $\textit{`depth'}\leq~\!200$ (sample size  for $\textit{`depth'}\leq~\!200$, dropping empty values, is $n=505,829$). Right panel:  the empirical conditional center-outward quantile contours of orders   $\tau=0.2$ (black), 
      $\tau =0.4$  (green), and~$\tau=0.8$  (yellow) and  the empirical conditional center-outward median  
   (red) for the multiple-output regression  of~$(Y_1,Y_2)=~\!(\textit{`temperature,' `salinity')}$ with 
respect to $X=\textit{`depth.'}$ Estimation based on a $k$-nearest neighbors weight function with~$k=6,401$.
    }\vspace{-3mm}
    \label{fig:salinity}
\end{figure}
\begin{figure}[b!]
    \centering
    \includegraphics[width=6.7 cm,height=7cm]{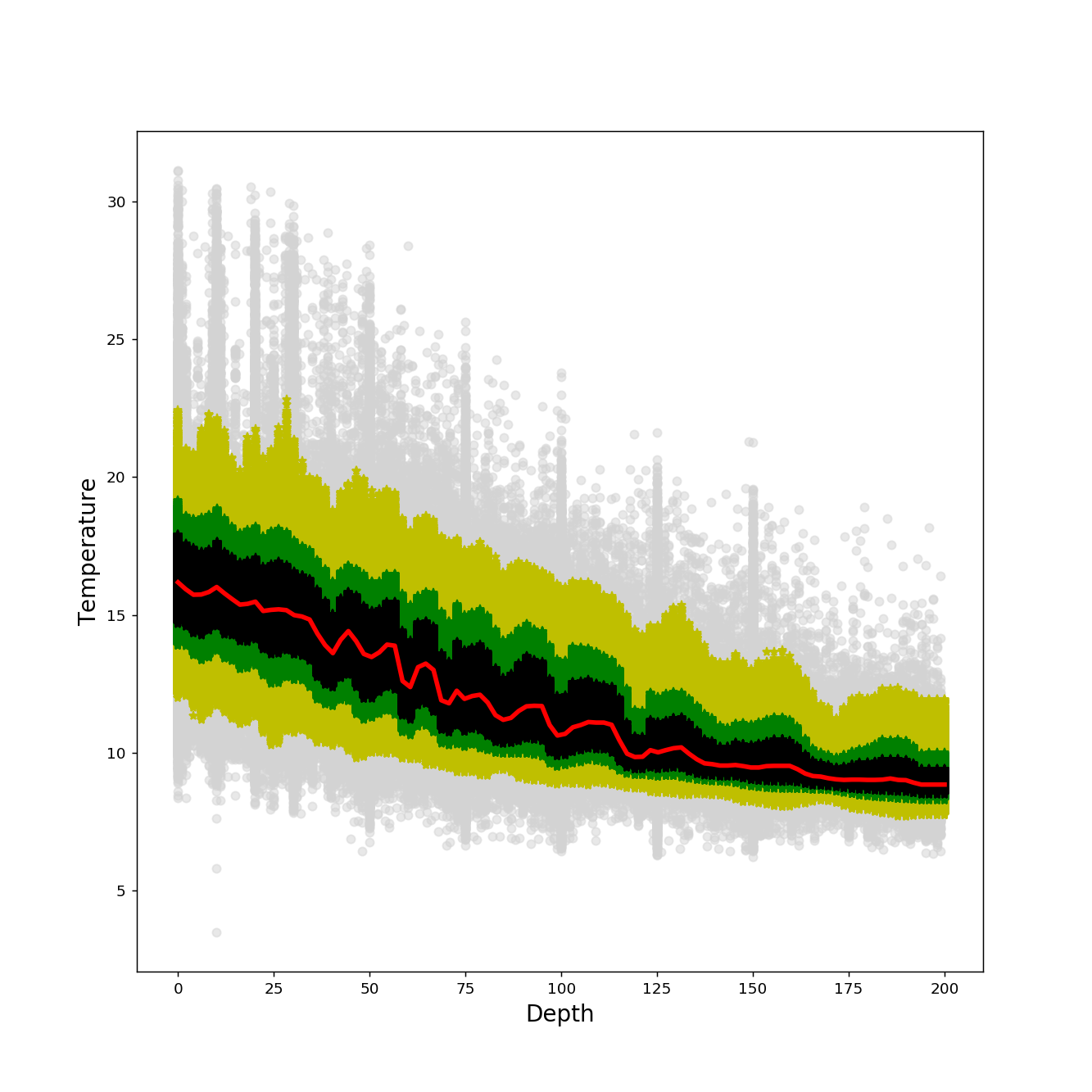}
    \includegraphics[width=6.7 cm,height=7cm ]{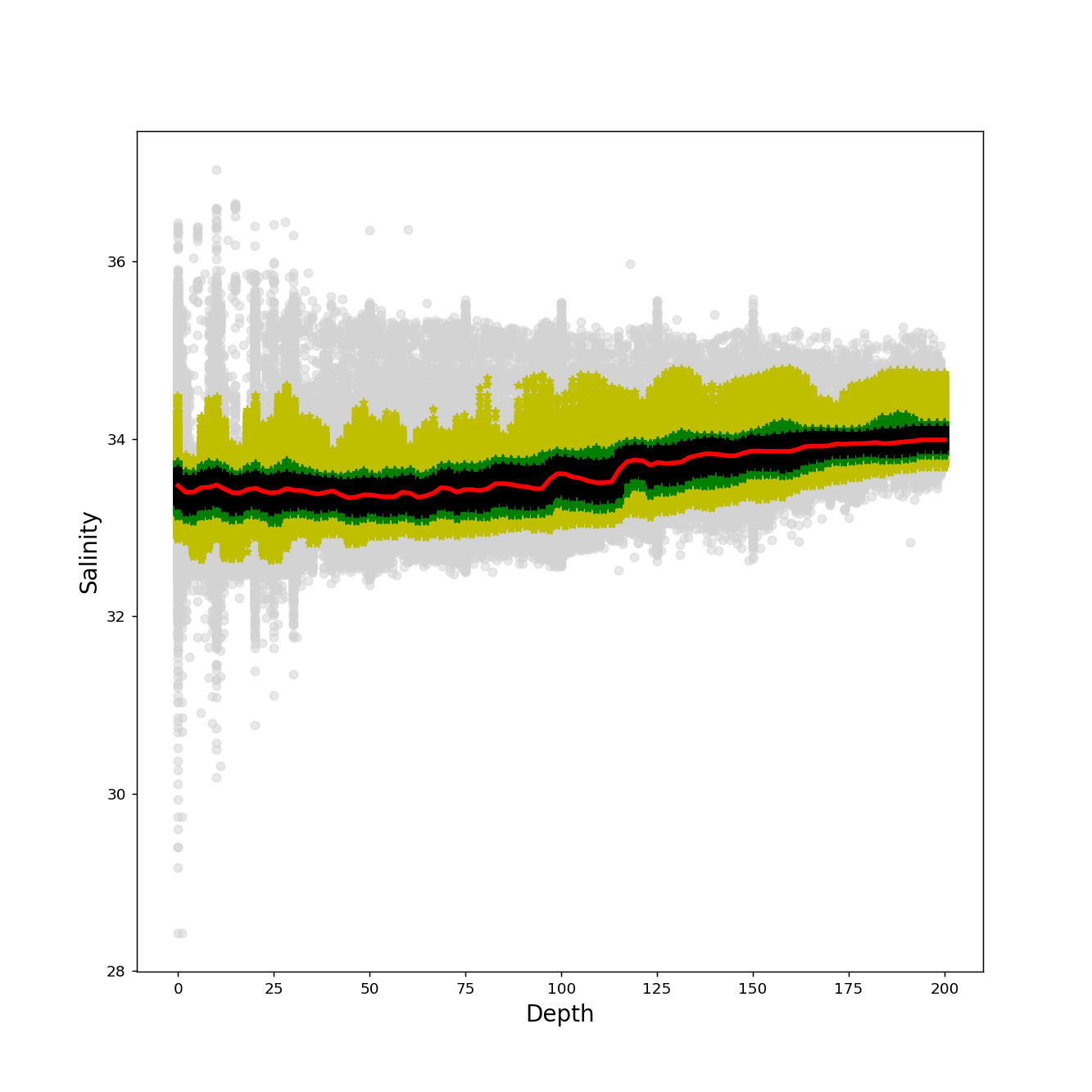} 
    \caption{CalCOFI dataset. Left  panel:  projection of the empirical center-outward quantile regions shown in Figure~\ref{fig:salinity} (orders   $\tau=0.2,\ 0.4,\ 0.8$) and median on the axes (`depth,' `salinity'),  for the multiple-output regression of $(Y_1,Y_2)=(\textit{`temperature,' `salinity'})$ with respect to $X=\textit{`depth.'}$ Right  panel:  projection of the same on the axes (`depth,' `temperature'); see Figure~\ref{fig:salinity} for the color code.}
    \label{fig:salinityproyect}
\end{figure}

\subsubsection{The CalCOFI oceanographic dataset:   depth, temperature, and salinity in the oceans}
 The dataset ``CalCOFI Over 60 Years of Oceanographic Data,'' available at \url{https://www.kaggle.com/sohier/calcofi}, contains the longest (1949-present) and most complete (more than 50,000 sampling stations;  sample size   $n=814,247$) time series of oceanographic and larval fish data  worldwide.  Data collected at depths down to 500 m include temperature, salinity, oxygen, phosphate, silicate, nitrate and nitrite, chlorophyll, transmissometer, PAR, C14 primary productivity, phytoplankton biodiversity, zooplankton biomass, and zooplankton biodiversity. We are focusing here on  the influence of $X=\textit{`depth'}$ (in meters) on the pair $\mathbf{Y}=(\textit{`temperature,' `salinity'})^\prime$ (in degrees and  grams of salt per kilogram of water, respectively).  
 
 Figure~\ref{fig:salinity} shows the corresponding 3D observations and  the estimated conditional center-outward quantile contours obtained from the same method as in Section~\ref{toysec} (nearest neighbors weight function with~$k=6,401$);   Figure~\ref{fig:salinityproyect} shows the projections of the same contours on the $(\textit{`depth,' `salinity'})$ and $(\textit{`depth,' `temperature'})$ axes, respectively. Inspection of these figures reveals a nonlinear center-outward conditional median; heteroskedasticity also appears as the  area of the conditional quantile regions clearly decreases as a function of depth, while a positive dependence between temperatures and salinity, which is present at the surface, gradually disappears as depth increases. The projection plots of Figure~\ref{fig:salinityproyect} also provide clearer views on   marginal dependencies. For example, the decrease of temperature as a function of  depth is monotone and   almost linear, while the dependence on depth of  salinity   is   more complex,  high at shallow depths, lower at medium depths, and higher again at greater depths. However, these marginal analyses, to some degree, are hiding the heteroskedasticity effects (in particular, the dependence on depth of the relation  between salinity and temperature) which are  clearly visible in Figure~\ref{fig:salinity}. Since the dataset is quite large, we used a nearest neighbors   weight function, see the comments about the empirical performance of different weights in Section~\ref{sphcasesec}.

\subsubsection{The Female ANSUR 2 dataset:  stature, foot length and tibial height of female US Army personnel .}
\begin{figure}[t!]
    \centering
    \includegraphics[width=6 cm,height=8cm, trim={6cm 0cm 3cm 0cm},clip ]{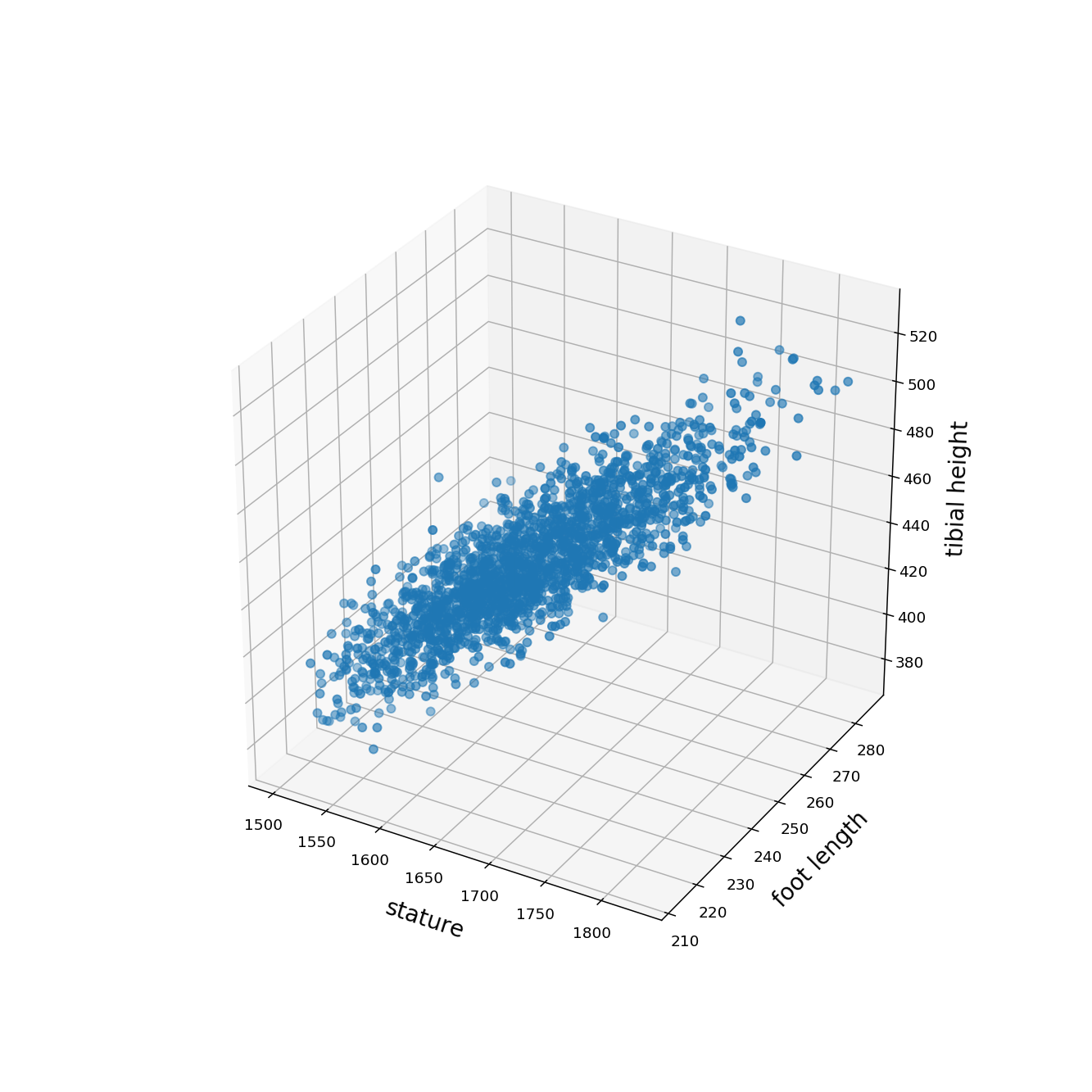} 
    \includegraphics[width=6 cm,height=8cm, trim={6cm 0cm 3cm 0cm},clip ]{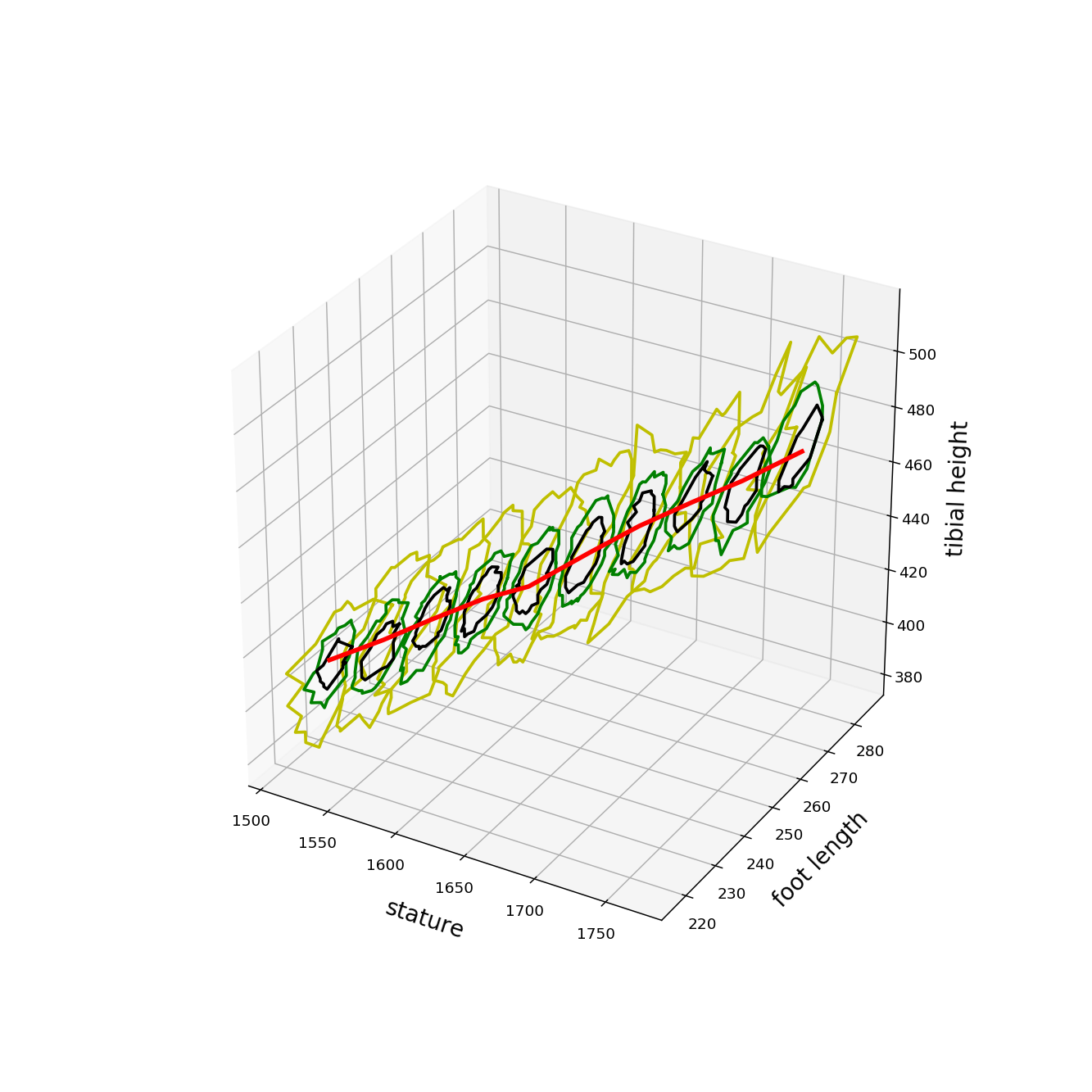}
    \caption{ANSUR 2 dataset (sample size $n=1,986$). Left panel: the original dataset of 
     $X=\textit{`stature'}$ and~$(Y_1,Y_2)=(\textit{`foot length,' `tibial height'})$. Right panel: the empirical conditional center-outward quantile contours of orders $\tau =0.2$ (black),  $\tau =0.4$ (green),   $\tau =0.7$  (yellow) and   the empirical conditional center-outward median   (red)  for the multiple-output regression of $(Y_1,Y_2)$ with respect to $X$; estimation based on  a Gaussian kernel weight function with bandwidth $h=15$.}
    \label{fig:house}
\end{figure}
Our second real-data example involves a smaller sample size $n$. The Female Anthropometric Survey of US Army Personnel (Female ANSUR 2 or Female ANSUR II)  featured in this section    consists  in ninety-three direct measures and 41 derived ones, as well as three-dimensional head, foot, and
whole-body scans of  $n=1,986$  women of the United States'  army. These measurements were collected  between October 4, 2010 and April 5, 2012,  in  May 2014 and in  May 2015; they are
available online at \url{https://www.openlab.psu.edu/ansur2/}.

\begin{figure}[t!]
    \centering
    \includegraphics[width=6 cm,height=7cm, trim={2cm 0cm 2cm 0cm},clip ]{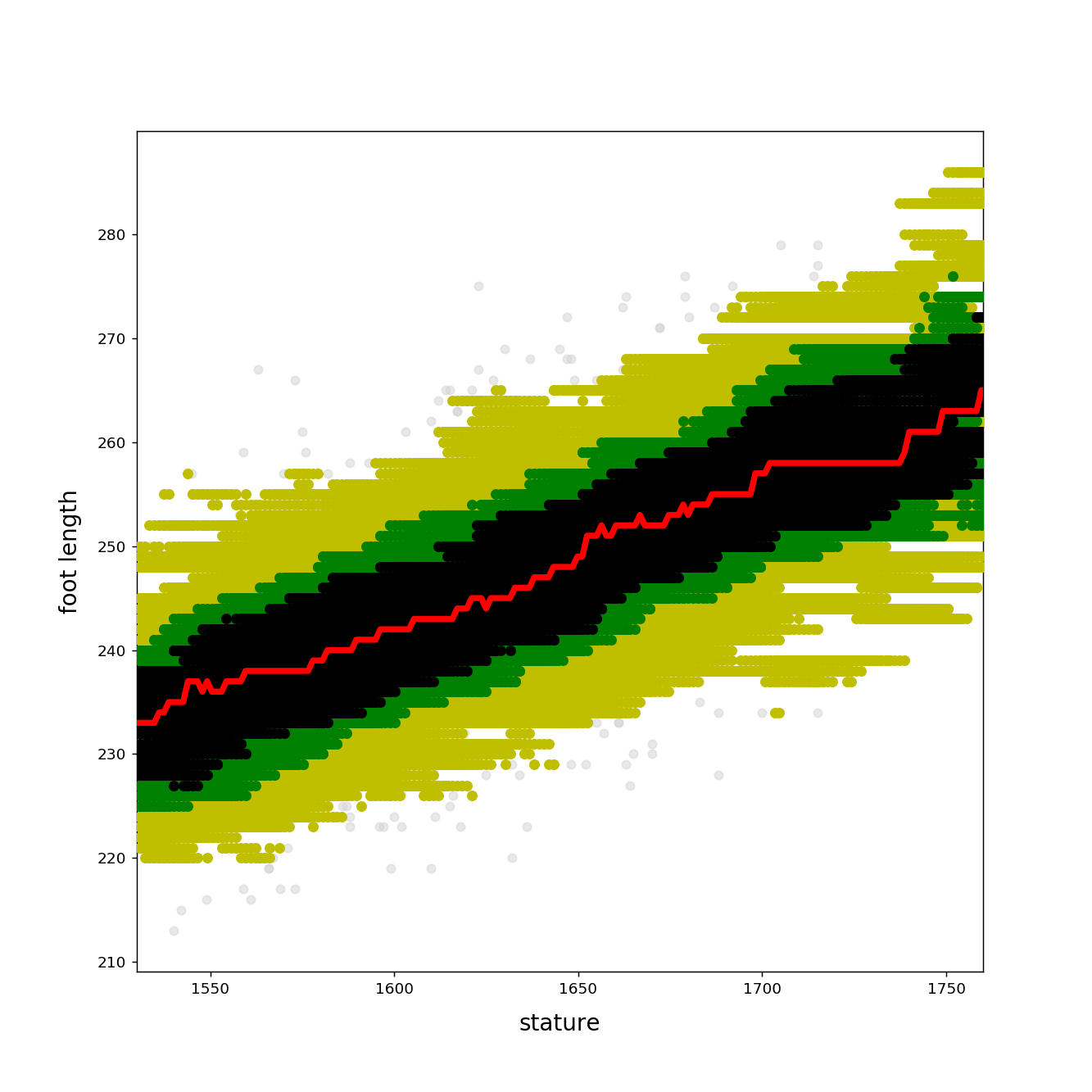} 
    \includegraphics[width=6 cm,height=7cm, trim={2cm 0cm 2cm 0cm},clip ]{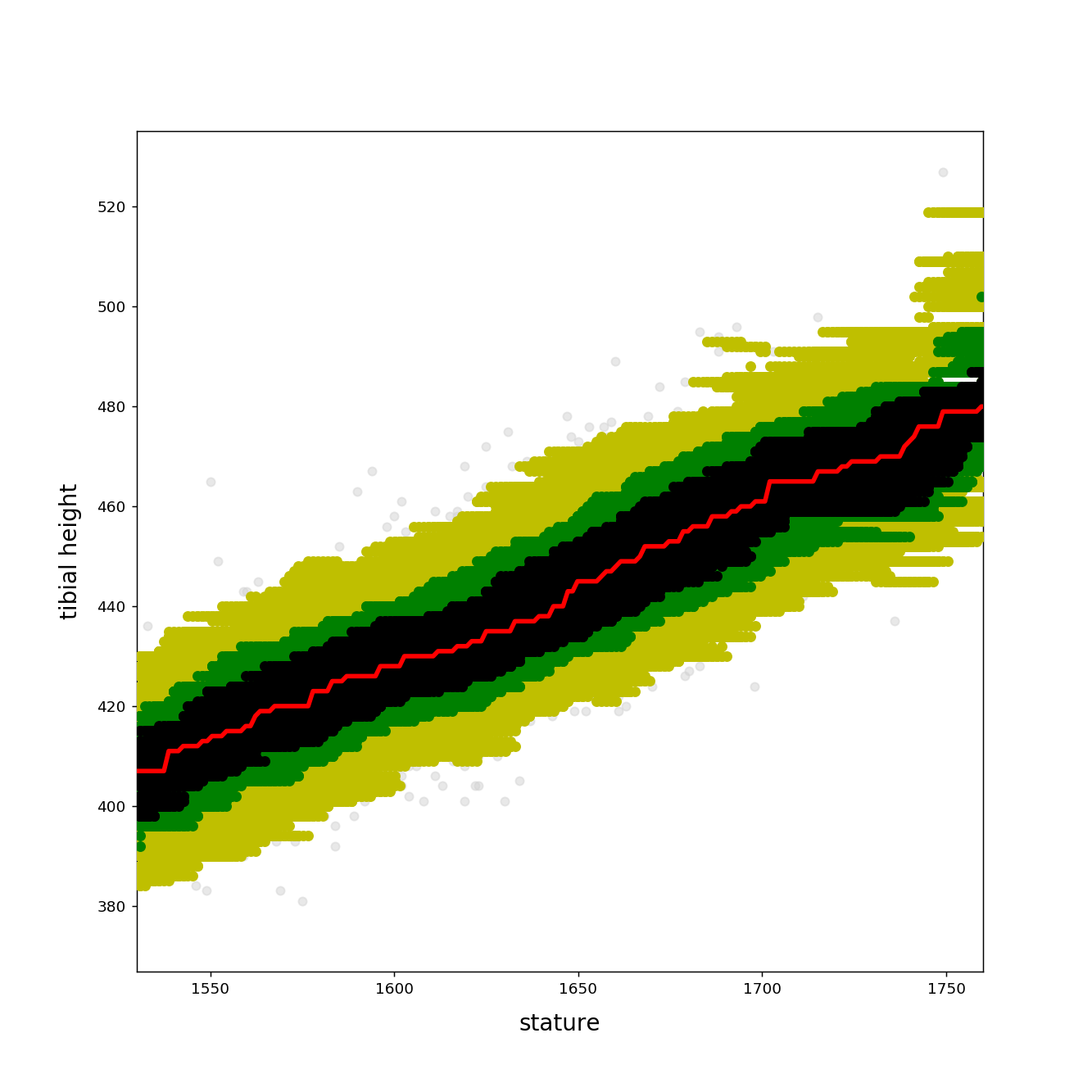}
    \caption{ANSUR 2 dataset (sample size $n=1,986$). Left panel:  projection of the empirical  center-outward quantile regions (orders $\tau=0.2,\ 0.4,\  0.8$) and median on the axes (`stature,' `foot length'), for  the multiple-output regression of $(Y_1,Y_2)=(\textit{`foot length,' `tibial height'})$ with respect to $X=\textit{`stature.'}$ Right  panel:  projection of the same on the axes (`stature,' `tibial height'); see Figure~\ref{fig:house} for the color code.
}
    \label{fig:housepro}
\end{figure}

 We want to analyze the influence of the covariate $X=\textit{`stature'}$ (in centimeters) on the 
  variable of interest $\mathbf{Y}=(\textit{`foot length,' `tibial height'})$ (both in centimeters). Figure~\ref{fig:house} provides a 3D view of the center-outward quantile contours/tubes (levels $\tau= 0.2, 0.4, 0.7$), along with the center-outward regression median (red); Figure~\ref{fig:housepro}  shows the   projections on the axes of the same contours. Inspecting these two figures  reveals the absence of heteroskedasticity, the spherical shape of conditional distributions, and a roughly linear regression. Since the size of the model is not too large, a Gaussian kernel is  convenient. {The bandwidth was chosen as $h=15$, which, up to scale changes, corresponds to the choice  $h=0.2$ in Figure~\ref{fig:toyBananCompareGauss}.}

\section{Some concluding remarks}\label{conclsec}

\subsection{Relation to the recent  literature on numerical optimal transportation}

The estimation of   transport maps beyond the sample points currently is   a  hot topic, and  a fastly developing strand of literature is proposing  such   estimators. However, the objective of most authors is to reach near-optimal convergence rates, for which they typically impose fairly strong assumptions. Some  estimators  (\cite{Rigollet}, \cite{manole2021plugin}) are computationally quite heavy and sometimes numerically almost  infeasible; others  \citep{pooladian2021entropic} use a regularized version (\cite{Cuturi}) of the optimal transport problem to provide   consistent near-optimal estimators, which needs stringent assumptions on the shape of the underlying distributions---such as  being compactly supported, with densities bounded away from $0$ and $\infty$ over their convex supports. This is redhibitory in our case, since the density of $\mathrm{U}_d$ (a choice which plays an essential role   in the interpretation of transports as quantile functions) is unbounded  at $\mathbf{0}$. Other solutions  (\cite{makkuva2020optimal} and \cite{gonzalezsanz2022gan}) are based on deep learning and  neural network methods; they achieve excellent empirical performance, but the lack of theoretical results for the first one, the Lipschitz constraint on   transport maps for the second,  preclude their use in this quantile regression context. 

We  also could estimate the conditional quantiles through the optimal map from the uniform reference measure to ${\rm P}_n^{w(\mathbf{x})}$. From a numerical point of view, however, this would lead to the computation of a semi-discrete optimal transportation plan, which has complexity~$O(n^{d/2})$, hence  is unfeasible even for moderate $d$. While the computational complexity  of our procedure does not depend on the  dimension,   its statistical performance does (see \cite{fournier2015rate}) and, in that sense, we do not escape the curse of dimensionality---up to the case where $\rm P$ is finitely supported, see \cite{delbarrio2021central}. Despite  the fact that the literature on the computation of such maps is   growing quite fastly (see \cite{levy2020fast,Gallou, MEYRON201913, deGoes}), the exisiting methods are restricted to dimension two, sometimes three.  A further  issue is that the solution of the semi-discrete problem  does not produce  quantile {\it contours} but creates  a Voronoi  tessellation of $\mathbb{S}_{d}$,    each piece of which is mapped to a single sample point. 

\subsection{Conclusions and perspectives for further developments} Building on the concepts of center-outward quantiles recently developed in  \cite{Hallin2020DistributionAQ}, we are proposing here a fully nonparametric solution to the problem of nonparametric multiple-output quantile regression. Contrary to earlier attempts, our  solution is enjoying the quintessential property that the (conditional) probability content of its quantile regions is under control irrespective of the underlying distribution. This is only a first step into the multifarious applications of multiple-output quantile regression, though. Due to the minimality of the assumptions  it requires,  a completely agnostic nonparametric approach indeed  is attractive, but also comes at a cost: linear or polynomial quantile regression remain justified whenever  a priori knowledge of the analytical  form of the regression is available and should  be taken advantage of. A center-outward version of the results of  \cite{Carlier2016}, thus, is highly desirable. Single-output quantile regression has been considered in a variety of contexts: survival analysis, longitudinal data, instrumental variable regression, directional, functional, and high-dimensional data,~$\dots$ Quantile regression versions of time-series models such as the quantile autoregressive model  also have been investigated \citep{QAR}. All these applications call for   multiple-output extensions with important real-life consequences; they   should and can be based on the concept of center-outward quantile, regions, and contours and are the subject of our ongoing research.

\bibliographystyle{imsart-nameyear}
\bibliography{Biblioreg}

\begin{thebibliography}{59}

\bibitem[\protect\citeauthoryear{Bagh and Wets}{1996}]{Bagh95convergenceof}
\begin{barticle}[author]
\bauthor{\bsnm{Bagh},~\bfnm{Adib}\binits{A.}} \AND
  \bauthor{\bsnm{Wets},~\bfnm{Roger~J.}\binits{R.~J.}}
(\byear{1996}).
\btitle{Convergence of set-valued mappings: equi-outer semicontinuity}.
\bjournal{Set-Valued Anal.}
\bvolume{4}
\bpages{333--360}.
\end{barticle}
\endbibitem

\bibitem[\protect\citeauthoryear{Breckling and Chambers}{1988}]{Breckling88}
\begin{barticle}[author]
\bauthor{\bsnm{Breckling},~\bfnm{J.}\binits{J.}} \AND
  \bauthor{\bsnm{Chambers},~\bfnm{R.}\binits{R.}}
(\byear{1988}).
\btitle{M-{Q}uantiles}.
\bjournal{Biometrika}
\bvolume{75}
\bpages{761--771}.
\end{barticle}
\endbibitem

\bibitem[\protect\citeauthoryear{Carlier, Chernozhukov and
  Galichon}{2016}]{Carlier2016}
\begin{barticle}[author]
\bauthor{\bsnm{Carlier},~\bfnm{Guillaume}\binits{G.}},
  \bauthor{\bsnm{Chernozhukov},~\bfnm{Victor}\binits{V.}} \AND
  \bauthor{\bsnm{Galichon},~\bfnm{Alfred}\binits{A.}}
(\byear{2016}).
\btitle{{Vector quantile regression: An optimal transport approach}}.
\bjournal{The Annals of Statistics}
\bvolume{44}
\bpages{1165--1192}.
\bdoi{10.1214/15-AOS1401}
\end{barticle}
\endbibitem

\bibitem[\protect\citeauthoryear{Chernozhukov
  et~al.}{2017}]{hallinChernozhukov2017}
\begin{barticle}[author]
\bauthor{\bsnm{Chernozhukov},~\bfnm{Victor}\binits{V.}},
  \bauthor{\bsnm{Galichon},~\bfnm{Alfred}\binits{A.}},
  \bauthor{\bsnm{Hallin},~\bfnm{Marc}\binits{M.}} \AND
  \bauthor{\bsnm{Henry},~\bfnm{Marc}\binits{M.}}
(\byear{2017}).
\btitle{{Monge-Kantorovich depth, quantiles, ranks and signs}}.
\bjournal{The Annals of Statistics}
\bvolume{45}
\bpages{223--256}.
\bdoi{10.1214/16-AOS1450}
\end{barticle}
\endbibitem

\bibitem[\protect\citeauthoryear{Cuturi}{2013}]{Cuturi}
\begin{binproceedings}[author]
\bauthor{\bsnm{Cuturi},~\bfnm{Marco}\binits{M.}}
(\byear{2013}).
\btitle{Sinkhorn distances: lightspeed computation of optimal transport}.
In \bbooktitle{Advances in Neural Information Processing Systems}
(\beditor{\bfnm{C.~J.~C.}\binits{C.~J.~C.}~\bsnm{Burges}},
  \beditor{\bfnm{L.}\binits{L.}~\bsnm{Bottou}},
  \beditor{\bfnm{M.}\binits{M.}~\bsnm{Welling}},
  \beditor{\bfnm{Z.}\binits{Z.}~\bsnm{Ghahramani}} \AND
  \beditor{\bfnm{K.~Q.}\binits{K.~Q.}~\bsnm{Weinberger}}, eds.)
\bvolume{26}.
\bpublisher{Curran Associates, Inc.}
\end{binproceedings}
\endbibitem

\bibitem[\protect\citeauthoryear{de~Goes et~al.}{2012}]{deGoes}
\begin{barticle}[author]
\bauthor{\bparticle{de} \bsnm{Goes},~\bfnm{Fernando}\binits{F.}},
  \bauthor{\bsnm{Breeden},~\bfnm{Katherine}\binits{K.}},
  \bauthor{\bsnm{Ostromoukhov},~\bfnm{Victor}\binits{V.}} \AND
  \bauthor{\bsnm{Desbrun},~\bfnm{Mathieu}\binits{M.}}
(\byear{2012}).
\btitle{Blue noise through optimal transport}.
\bjournal{ACM Trans. pn Graphics}
\bvolume{31}
\bpages{1--11}.
\bdoi{10.1145/2366145.2366190}
\end{barticle}
\endbibitem

\bibitem[\protect\citeauthoryear{Deb and Sen}{2021}]{DebSen21}
\begin{barticle}[author]
\bauthor{\bsnm{Deb},~\bfnm{Nabarun}\binits{N.}} \AND
  \bauthor{\bsnm{Sen},~\bfnm{Bodhisattva}\binits{B.}}
(\byear{2021}).
\btitle{Multivariate rank-based distribution-free nonparametric testing using
  measure transportation}.
\bjournal{Journal of the American Statistical Association}
\bvolume{0}
\bpages{1-16}.
\bdoi{10.1080/01621459.2021.1923508}
\end{barticle}
\endbibitem

\bibitem[\protect\citeauthoryear{del Barrio, Gonz{\'a}lez-Sanz and
  Hallin}{2020}]{BarrioSanzHal}
\begin{barticle}[author]
\bauthor{\bparticle{del} \bsnm{Barrio},~\bfnm{Eustasio}\binits{E.}},
  \bauthor{\bsnm{Gonz{\'a}lez-Sanz},~\bfnm{Alberto}\binits{A.}} \AND
  \bauthor{\bsnm{Hallin},~\bfnm{Marc}\binits{M.}}
(\byear{2020}).
\btitle{A note on the regularity of optimal-transport-based center-outward
  distribution and quantile functions}.
\bjournal{J. Multivar. Anal.}
\bvolume{180}
\bpages{104671}.
\end{barticle}
\endbibitem

\bibitem[\protect\citeauthoryear{del Barrio, González-Sanz and
  Loubes}{2021}]{delbarrio2021central}
\begin{bmisc}[author]
\bauthor{\bparticle{del} \bsnm{Barrio},~\bfnm{Eustasio}\binits{E.}},
  \bauthor{\bsnm{González-Sanz},~\bfnm{Alberto}\binits{A.}} \AND
  \bauthor{\bsnm{Loubes},~\bfnm{Jean-Michel}\binits{J.-M.}}
(\byear{2021}).
\btitle{A central limit theorem for semidiscrete Wasserstein distances}.
\bhowpublished{arXiv:2105.11721}.
\end{bmisc}
\endbibitem

\bibitem[\protect\citeauthoryear{del Barrio and Loubes}{2019}]{BaLo}
\begin{barticle}[author]
\bauthor{\bparticle{del} \bsnm{Barrio},~\bfnm{Eustasio}\binits{E.}} \AND
  \bauthor{\bsnm{Loubes},~\bfnm{Jean-Michel}\binits{J.-M.}}
(\byear{2019}).
\btitle{Central limit theorems for empirical transportation cost in general
  dimension}.
\bjournal{Ann. Probab.}
\bvolume{47}
\bpages{926--951}.
\bdoi{10.1214/18-AOP1275}
\end{barticle}
\endbibitem

\bibitem[\protect\citeauthoryear{Devroye}{1982}]{Devroye1982NecessaryAS}
\begin{barticle}[author]
\bauthor{\bsnm{Devroye},~\bfnm{Luc}\binits{L.}}
(\byear{1982}).
\btitle{Necessary and sufficient conditions for the pointwise convergence of
  nearest neighbor regression function estimates}.
\bjournal{Zeitschrift f{\"u}r Wahrscheinlichkeitstheorie und Verwandte Gebiete}
\bvolume{61}
\bpages{467-481}.
\end{barticle}
\endbibitem

\bibitem[\protect\citeauthoryear{Devroye et~al.}{1994}]{Devroye}
\begin{barticle}[author]
\bauthor{\bsnm{Devroye},~\bfnm{Luc}\binits{L.}},
  \bauthor{\bsnm{Gyorfi},~\bfnm{Laszlo}\binits{L.}},
  \bauthor{\bsnm{Krzyzak},~\bfnm{Adam}\binits{A.}} \AND
  \bauthor{\bsnm{Lugosi},~\bfnm{Gabor}\binits{G.}}
(\byear{1994}).
\btitle{On the strong universal consistency of nearest neighbor regression
  function estimates}.
\bjournal{The Annals of Statistics}
\bvolume{22}
\bpages{1371--1385}.
\end{barticle}
\endbibitem

\bibitem[\protect\citeauthoryear{Feldman, Bates and
  Romano}{2021}]{feldman2021calibrated}
\begin{bmisc}[author]
\bauthor{\bsnm{Feldman},~\bfnm{Shai}\binits{S.}},
  \bauthor{\bsnm{Bates},~\bfnm{Stephen}\binits{S.}} \AND
  \bauthor{\bsnm{Romano},~\bfnm{Yaniv}\binits{Y.}}
(\byear{2021}).
\btitle{Calibrated multiple-output quantile regression with representation
  learning}.
\bhowpublished{arXiv:2110.00816}.
\end{bmisc}
\endbibitem

\bibitem[\protect\citeauthoryear{Figalli}{2018}]{FigalliCenter}
\begin{barticle}[author]
\bauthor{\bsnm{Figalli},~\bfnm{Alessio}\binits{A.}}
(\byear{2018}).
\btitle{On the continuity of center-outward distribution and quantile
  functions}.
\bjournal{Nonlinear Anal.}
\bvolume{177~}
\bpages{413--21}.
\bdoi{10.1016/j.na.2018.05.008}
\end{barticle}
\endbibitem

\bibitem[\protect\citeauthoryear{Fournier and Guillin}{2015}]{fournier2015rate}
\begin{barticle}[author]
\bauthor{\bsnm{Fournier},~\bfnm{Nicolas}\binits{N.}} \AND
  \bauthor{\bsnm{Guillin},~\bfnm{Arnaud}\binits{A.}}
(\byear{2015}).
\btitle{On the rate of convergence in {W}asserstein distance of the empirical
  measure}.
\bjournal{Probability Theory and Related Fields}
\bvolume{162}
\bpages{707--738}.
\end{barticle}
\endbibitem

\bibitem[\protect\citeauthoryear{Gallouët and Mérigot}{2018}]{Gallou}
\begin{barticle}[author]
\bauthor{\bsnm{Gallouët},~\bfnm{Thomas}\binits{T.}} \AND
  \bauthor{\bsnm{Mérigot},~\bfnm{Quentin}\binits{Q.}}
(\byear{2018}).
\btitle{A Lagrangian scheme \`a la Brenier for the incompressible Euler
  equations}.
\bjournal{Found. Comput. Math.}
\bvolume{18}
\bpages{835--865}.
\end{barticle}
\endbibitem

\bibitem[\protect\citeauthoryear{Ghosal and Sen}{2019}]{GhosSen19}
\begin{bmisc}[author]
\bauthor{\bsnm{Ghosal},~\bfnm{Promit}\binits{P.}} \AND
  \bauthor{\bsnm{Sen},~\bfnm{Bodhisattva}\binits{B.}}
(\byear{2019}).
\btitle{Multivariate ranks and quantiles using optimal transportation and
  applications to goodness-of-fit testing}.
\bhowpublished{arXiv:1905.05340}.
\end{bmisc}
\endbibitem

\bibitem[\protect\citeauthoryear{González-Sanz
  et~al.}{2022}]{gonzalezsanz2022gan}
\begin{bmisc}[author]
\bauthor{\bsnm{González-Sanz},~\bfnm{Alberto}\binits{A.}},
  \bauthor{\bparticle{de} \bsnm{Lara},~\bfnm{Lucas}\binits{L.}},
  \bauthor{\bsnm{Béthune},~\bfnm{Louis}\binits{L.}} \AND
  \bauthor{\bsnm{Loubes},~\bfnm{Jean-Michel}\binits{J.-M.}}
(\byear{2022}).
\btitle{GAN estimation of Lipschitz optimal transport maps}.
\bhowpublished{arXiv:2202.07965}.
\end{bmisc}
\endbibitem

\bibitem[\protect\citeauthoryear{Györfi et~al.}{2002}]{gyodfi}
\begin{bbook}[author]
\bauthor{\bsnm{Györfi},~\bfnm{László}\binits{L.}},
  \bauthor{\bsnm{Kohler},~\bfnm{Michael}\binits{M.}},
  \bauthor{\bsnm{Krzyzak},~\bfnm{Adam}\binits{A.}} \AND
  \bauthor{\bsnm{Walk},~\bfnm{Harro}\binits{H.}}
(\byear{2002}).
\btitle{A Distribution-Free Theory of Nonparametric Regression.}
\bseries{Springer Series in Statistics}.
\bpublisher{Springer}.
\end{bbook}
\endbibitem

\bibitem[\protect\citeauthoryear{Hallin}{2017}]{Hallin17}
\begin{bmisc}[author]
\bauthor{\bsnm{Hallin},~\bfnm{Marc}\binits{M.}}
(\byear{2017}).
\btitle{On distribution and quantile functions, ranks and signs in
  $\mathbb{R}^d$: a measure transportation approach}.
\bhowpublished{Available at
  \url{https://ideas.repec.org/p/eca/wpaper/2013-258262.html}}.
\end{bmisc}
\endbibitem

\bibitem[\protect\citeauthoryear{Hallin}{2022}]{HallinReview}
\begin{barticle}[author]
\bauthor{\bsnm{Hallin},~\bfnm{Marc}\binits{M.}}
(\byear{2022}).
\btitle{Measure transportation and statistical decision theory}.
\bjournal{Annual Review of Statistics and its Application}
\bvolume{9}
\bpages{401--424}.
\bdoi{10.1146/annurev-statistics-040220-105948}
\end{barticle}
\endbibitem

\bibitem[\protect\citeauthoryear{Hallin, Hlubinka and
  Hudecová}{2022}]{HallinMANOVA}
\begin{barticle}[author]
\bauthor{\bsnm{Hallin},~\bfnm{Marc}\binits{M.}},
  \bauthor{\bsnm{Hlubinka},~\bfnm{Daniel}\binits{D.}} \AND
  \bauthor{\bsnm{Hudecová},~\bfnm{{\v S}.}\binits{{\v S}.}}
(\byear{2022}).
\btitle{Efficient Fully Distribution-Free Center-Outward Rank Tests for
  Multiple-Output Regression and {MANOVA}}.
\bjournal{Journal of the American Statistical Association}
\bvolume{0}
\bpages{1-43}.
\bdoi{10.1080/01621459.2021.2021921}
\end{barticle}
\endbibitem

\bibitem[\protect\citeauthoryear{Hallin, {La~{V}ecchia} and
  Liu}{2020}]{HallinVARMA}
\begin{barticle}[author]
\bauthor{\bsnm{Hallin},~\bfnm{M.}\binits{M.}},
  \bauthor{\bsnm{{La~{V}ecchia}},~\bfnm{D.}\binits{D.}} \AND
  \bauthor{\bsnm{Liu},~\bfnm{H.}\binits{H.}}
(\byear{2020}).
\btitle{Center-outward R-estimation for semiparametric VARMA models}.
\bjournal{Journal of the American Statistical Association}
\bvolume{0}
\bpages{1-14}.
\bdoi{10.1080/01621459.2020.1832501}
\end{barticle}
\endbibitem

\bibitem[\protect\citeauthoryear{Hallin, {La~{V}ecchia} and
  Liu}{2022}]{HallinBernou}
\begin{barticle}[author]
\bauthor{\bsnm{Hallin},~\bfnm{M.}\binits{M.}},
  \bauthor{\bsnm{{La~{V}ecchia}},~\bfnm{D.}\binits{D.}} \AND
  \bauthor{\bsnm{Liu},~\bfnm{H.}\binits{H.}}
(\byear{2022}).
\btitle{Rank-based testing for semiparametric {VAR} models: a measure
  transportation approach}.
\bjournal{Bernoulli}
\bvolume{0}
\bpages{1-14}.
\end{barticle}
\endbibitem

\bibitem[\protect\citeauthoryear{Hallin, Paindaveine and
  \v{S}iman}{2010}]{Hallin2010}
\begin{barticle}[author]
\bauthor{\bsnm{Hallin},~\bfnm{Marc}\binits{M.}},
  \bauthor{\bsnm{Paindaveine},~\bfnm{Davy}\binits{D.}} \AND
  \bauthor{\bsnm{\v{S}iman},~\bfnm{Miroslav}\binits{M.}}
(\byear{2010}).
\btitle{Multivariate quantiles and multiple-output regression quantiles: from
  L$_1$ optimization to halfspace depth}.
\bjournal{The Annals of Statistics}
\bvolume{38}
\bpages{635--703}.
\bdoi{10.1214/09-aos723}
\end{barticle}
\endbibitem

\bibitem[\protect\citeauthoryear{Hallin and \v{S}iman}{2018}]{Halhandbook}
\begin{binproceedings}[author]
\bauthor{\bsnm{Hallin},~\bfnm{Marc}\binits{M.}} \AND
  \bauthor{\bsnm{\v{S}iman},~\bfnm{Miroslav}\binits{M.}}
(\byear{2018}).
\btitle{Multiple-output quantile regression}.
In \bbooktitle{Handbook of Quantile Regression}
(\beditor{\bfnm{Roger}\binits{R.}~\bsnm{Koenker}},
  \beditor{\bfnm{Victor}\binits{V.}~\bsnm{Chernozhukov}},
  \beditor{\bfnm{Xuming}\binits{X.}~\bsnm{He}} \AND
  \beditor{\bfnm{L.}\binits{L.}~\bsnm{Peng}}, eds.)
\bpages{185--207}.
\bpublisher{CRC Press}.
\end{binproceedings}
\endbibitem

\bibitem[\protect\citeauthoryear{Hallin et~al.}{2015}]{Halin2015}
\begin{barticle}[author]
\bauthor{\bsnm{Hallin},~\bfnm{Marc}\binits{M.}},
  \bauthor{\bsnm{Lu},~\bfnm{Zudi}\binits{Z.}},
  \bauthor{\bsnm{Paindaveine},~\bfnm{Davy}\binits{D.}} \AND
  \bauthor{\bsnm{\v{S}iman},~\bfnm{Miroslav}\binits{M.}}
(\byear{2015}).
\btitle{Local bilinear multiple-output quantile/depth regression}.
\bjournal{Bernoulli}
\bvolume{21}
\bpages{1435--1466}.
\bdoi{10.3150/14-bej610}
\end{barticle}
\endbibitem

\bibitem[\protect\citeauthoryear{Hallin
  et~al.}{2021}]{Hallin2020DistributionAQ}
\begin{barticle}[author]
\bauthor{\bsnm{Hallin},~\bfnm{Marc}\binits{M.}}, \bauthor{\bparticle{del}
  \bsnm{Barrio},~\bfnm{Eustasio}\binits{E.}},
  \bauthor{\bsnm{Cuesta-Albertos},~\bfnm{Juan}\binits{J.}} \AND
  \bauthor{\bsnm{Matrán},~\bfnm{Carlos}\binits{C.}}
(\byear{2021}).
\btitle{{Distribution and quantile functions, ranks and signs in dimension $d$:
  a measure transportation approach}}.
\bjournal{The Annals of Statistics}
\bvolume{49}
\bpages{1139--1165}.
\bdoi{10.1214/20-AOS1996}
\end{barticle}
\endbibitem

\bibitem[\protect\citeauthoryear{Hütter and Rigollet}{2021}]{Rigollet}
\begin{barticle}[author]
\bauthor{\bsnm{Hütter},~\bfnm{Jan-Christian}\binits{J.-C.}} \AND
  \bauthor{\bsnm{Rigollet},~\bfnm{Philippe}\binits{P.}}
(\byear{2021}).
\btitle{{Minimax estimation of smooth optimal transport maps}}.
\bjournal{The Annals of Statistics}
\bvolume{49}
\bpages{1166--1194}.
\bdoi{10.1214/20-AOS1997}
\end{barticle}
\endbibitem

\bibitem[\protect\citeauthoryear{Koenker}{2005}]{koenker_2005}
\begin{bbook}[author]
\bauthor{\bsnm{Koenker},~\bfnm{Roger}\binits{R.}}
(\byear{2005}).
\btitle{Quantile Regression}.
\bseries{Econometric Society Monographs}.
\bpublisher{Cambridge University Press}.
\bdoi{10.1017/CBO9780511754098}
\end{bbook}
\endbibitem

\bibitem[\protect\citeauthoryear{Koenker and Bassett}{1978}]{Koenker1978}
\begin{barticle}[author]
\bauthor{\bsnm{Koenker},~\bfnm{Roger}\binits{R.}} \AND
  \bauthor{\bsnm{Bassett},~\bfnm{Gilbert}\binits{G.}}
(\byear{1978}).
\btitle{Regression quantiles}.
\bjournal{Econometrica}
\bvolume{46}
\bpages{33--50}.
\end{barticle}
\endbibitem

\bibitem[\protect\citeauthoryear{Koenker and Xiao}{2006}]{QAR}
\begin{barticle}[author]
\bauthor{\bsnm{Koenker},~\bfnm{R.}\binits{R.}} \AND
  \bauthor{\bsnm{Xiao},~\bfnm{Z.}\binits{Z.}}
(\byear{2006}).
\btitle{Quantile Autoregression [with Comments, Rejoinder]}.
\bjournal{Journal of the American Statistical Association}
\bvolume{101}
\bpages{980--1006}.
\end{barticle}
\endbibitem

\bibitem[\protect\citeauthoryear{Koenker et~al.}{2018}]{handbook}
\begin{binbook}[author]
\beditor{\bsnm{Koenker},~\bfnm{Roger}\binits{R.}},
  \beditor{\bsnm{Chernozhukov},~\bfnm{Victor}\binits{V.}},
  \beditor{\bsnm{He},~\bfnm{Xuming}\binits{X.}} \AND
  \beditor{\bsnm{Peng},~\bfnm{L.}\binits{L.}}, eds.
(\byear{2018}).
\btitle{Handbook of Quantile Regression}.
\bpublisher{CRC Press}.
\end{binbook}
\endbibitem

\bibitem[\protect\citeauthoryear{Konen and Paindaveine}{2022}]{KonendPd22}
\begin{barticle}[author]
\bauthor{\bsnm{Konen},~\bfnm{D.}\binits{D.}} \AND
  \bauthor{\bsnm{Paindaveine},~\bfnm{D.}\binits{D.}}
(\byear{2022}).
\btitle{{Multivariate $\rho$-quantiles: a spatial approach}}.
\bjournal{Bernoulli}.
\end{barticle}
\endbibitem

\bibitem[\protect\citeauthoryear{Kong and Mizera}{2012}]{Linglong}
\begin{barticle}[author]
\bauthor{\bsnm{Kong},~\bfnm{Linglong}\binits{L.}} \AND
  \bauthor{\bsnm{Mizera},~\bfnm{Ivan}\binits{I.}}
(\byear{2012}).
\btitle{Quantile tomography: using quantiles with multivariate data}.
\bjournal{Statistica Sinica}
\bvolume{22}
\bpages{1589-1610}.
\bdoi{10.5705/ss.2010.224}
\end{barticle}
\endbibitem

\bibitem[\protect\citeauthoryear{Lehmann and Romano}{2005}]{LehRom}
\begin{bbook}[author]
\bauthor{\bsnm{Lehmann},~\bfnm{E.~L.}\binits{E.~L.}} \AND
  \bauthor{\bsnm{Romano},~\bfnm{J.~P.}\binits{J.~P.}}
(\byear{2005}).
\btitle{{Testing Statistical Hypotheses}}.
\bpublisher{Springer}.
\end{bbook}
\endbibitem

\bibitem[\protect\citeauthoryear{Lévy, Mohayaee and von
  Hausegger}{2020}]{levy2020fast}
\begin{bmisc}[author]
\bauthor{\bsnm{Lévy},~\bfnm{Bruno}\binits{B.}},
  \bauthor{\bsnm{Mohayaee},~\bfnm{Roya}\binits{R.}} \AND
  \bauthor{\bparticle{von} \bsnm{Hausegger},~\bfnm{Sebastian}\binits{S.}}
(\byear{2020}).
\btitle{A fast semi-discrete optimal transport algorithm for a unique
  reconstruction of the early Universe}.
\bhowpublished{arXiv:2012.09074}.
\end{bmisc}
\endbibitem

\bibitem[\protect\citeauthoryear{Makkuva et~al.}{2020}]{makkuva2020optimal}
\begin{binproceedings}[author]
\bauthor{\bsnm{Makkuva},~\bfnm{Ashok}\binits{A.}},
  \bauthor{\bsnm{Taghvaei},~\bfnm{Amirhossein}\binits{A.}},
  \bauthor{\bsnm{Oh},~\bfnm{Sewoong}\binits{S.}} \AND
  \bauthor{\bsnm{Lee},~\bfnm{Jason}\binits{J.}}
(\byear{2020}).
\btitle{Optimal transport mapping via input convex neural networks}.
In \bbooktitle{International Conference on Machine Learning}
\bpages{6672--6681}.
\bpublisher{PMLR}.
\end{binproceedings}
\endbibitem

\bibitem[\protect\citeauthoryear{Manole et~al.}{2021}]{manole2021plugin}
\begin{bmisc}[author]
\bauthor{\bsnm{Manole},~\bfnm{Tudor}\binits{T.}},
  \bauthor{\bsnm{Balakrishnan},~\bfnm{Sivaraman}\binits{S.}},
  \bauthor{\bsnm{Niles-Weed},~\bfnm{Jonathan}\binits{J.}} \AND
  \bauthor{\bsnm{Wasserman},~\bfnm{Larry}\binits{L.}}
(\byear{2021}).
\btitle{Plugin estimation of smooth optimal transport maps}.
\bhowpublished{arXiv:2107.12364}.
\end{bmisc}
\endbibitem

\bibitem[\protect\citeauthoryear{McCann}{1995}]{McCann1995ExistenceAU}
\begin{barticle}[author]
\bauthor{\bsnm{McCann},~\bfnm{Robert~J.}\binits{R.~J.}}
(\byear{1995}).
\btitle{Existence and uniqueness of monotone measure-preserving maps}.
\bjournal{Duke Mathematical Journal}
\bvolume{80}
\bpages{309-323}.
\end{barticle}
\endbibitem

\bibitem[\protect\citeauthoryear{Merlo et~al.}{2022}]{Merlo22}
\begin{barticle}[author]
\bauthor{\bsnm{Merlo},~\bfnm{Luca}\binits{L.}},
  \bauthor{\bsnm{Petrella},~\bfnm{Lea}\binits{L.}},
  \bauthor{\bsnm{Salvati},~\bfnm{Nicola}\binits{N.}} \AND
  \bauthor{\bsnm{Tzavidis},~\bfnm{Nikos}\binits{N.}}
(\byear{2022}).
\btitle{Marginal M-quantile regression for multivariate dependent data}.
\bjournal{Computational Statistics \& Data Analysis}
\bpages{107500}.
\bdoi{https://doi.org/10.1016/j.csda.2022.107500}
\end{barticle}
\endbibitem

\bibitem[\protect\citeauthoryear{Meyron}{2019}]{MEYRON201913}
\begin{barticle}[author]
\bauthor{\bsnm{Meyron},~\bfnm{Jocelyn}\binits{J.}}
(\byear{2019}).
\btitle{Initialization procedures for discrete and semi-discrete optimal
  transport}.
\bjournal{Computer-Aided Design}
\bvolume{115}
\bpages{13-22}.
\bdoi{https://doi.org/10.1016/j.cad.2019.05.037}
\end{barticle}
\endbibitem

\bibitem[\protect\citeauthoryear{Paindaveine and
  \v{S}iman}{2011}]{PAINDAVEINE2011193}
\begin{barticle}[author]
\bauthor{\bsnm{Paindaveine},~\bfnm{Davy}\binits{D.}} \AND
  \bauthor{\bsnm{\v{S}iman},~\bfnm{Miroslav}\binits{M.}}
(\byear{2011}).
\btitle{On directional multiple-output quantile regression}.
\bjournal{Journal of Multivariate Analysis}
\bvolume{102}
\bpages{193-212}.
\bdoi{https://doi.org/10.1016/j.jmva.2010.08.004}
\end{barticle}
\endbibitem

\bibitem[\protect\citeauthoryear{Peyré and Cuturi}{2019}]{MAL-073}
\begin{barticle}[author]
\bauthor{\bsnm{Peyré},~\bfnm{Gabriel}\binits{G.}} \AND
  \bauthor{\bsnm{Cuturi},~\bfnm{Marco}\binits{M.}}
(\byear{2019}).
\btitle{Computational Optimal Transport: With Applications to Data Science}.
\bjournal{Foundations and Trends® in Machine Learning}
\bvolume{11}
\bpages{355-607}.
\bdoi{10.1561/2200000073}
\end{barticle}
\endbibitem

\bibitem[\protect\citeauthoryear{Pooladian and
  Niles-Weed}{2021}]{pooladian2021entropic}
\begin{bmisc}[author]
\bauthor{\bsnm{Pooladian},~\bfnm{Aram-Alexandre}\binits{A.-A.}} \AND
  \bauthor{\bsnm{Niles-Weed},~\bfnm{Jonathan}\binits{J.}}
(\byear{2021}).
\btitle{Entropic estimation of optimal transport maps}.
\bhowpublished{arXiv preprint arXiv:2109.12004}.
\end{bmisc}
\endbibitem

\bibitem[\protect\citeauthoryear{Rockafellar}{1970}]{rockafellar1970}
\begin{bbook}[author]
\bauthor{\bsnm{Rockafellar},~\bfnm{R.~Tyrrell}\binits{R.~T.}}
(\byear{1970}).
\btitle{Convex Analysis}.
\bseries{Princeton Mathematical Series}.
\bpublisher{Princeton University Press}, \baddress{Princeton, N. J.}
\end{bbook}
\endbibitem

\bibitem[\protect\citeauthoryear{Rockafellar and Wets}{1998}]{RockWets98}
\begin{bbook}[author]
\bauthor{\bsnm{Rockafellar},~\bfnm{{R. Tyrrell}}\binits{R.}} \AND
  \bauthor{\bsnm{Wets},~\bfnm{Roger J.~B.}\binits{R.~J.~B.}}
(\byear{1998}).
\btitle{Variational Analysis}.
\bpublisher{Springer Verlag}, \baddress{Heidelberg, Berlin, New York}.
\end{bbook}
\endbibitem

\bibitem[\protect\citeauthoryear{Serfling}{2002}]{SerfStatNeerl}
\begin{barticle}[author]
\bauthor{\bsnm{Serfling},~\bfnm{Robert}\binits{R.}}
(\byear{2002}).
\btitle{Quantile functions for multivariate analysis: approaches and
  applications}.
\bjournal{Statistica Neerlandica}
\bvolume{56}
\bpages{214-232}.
\bdoi{https://doi.org/10.1111/1467-9574.00195}
\end{barticle}
\endbibitem

\bibitem[\protect\citeauthoryear{Serfling}{2019a}]{SerflingDepthI}
\begin{bmisc}[author]
\bauthor{\bsnm{Serfling},~\bfnm{Robert}\binits{R.}}
(\byear{2019}a).
\btitle{Depth functions on general data spaces, I. Perspectives, with
  consideration of ``density'' and ``local'' depths''}.
\bhowpublished{Available at \url{https://www.utdallas.edu~serfling}}.
\end{bmisc}
\endbibitem

\bibitem[\protect\citeauthoryear{Serfling}{2019b}]{SerflingDepthII}
\begin{bmisc}[author]
\bauthor{\bsnm{Serfling},~\bfnm{Robert}\binits{R.}}
(\byear{2019}b).
\btitle{Depth functions on general data spaces, II. Formulation and maximality,
  with consideration of the Tukey, projection, spatial, and ``contour''
  depths}.
\bhowpublished{Available at \url{https://www.utdallas.edu~serfling}}.
\end{bmisc}
\endbibitem

\bibitem[\protect\citeauthoryear{Serfling and Zuo}{2000}]{SerfZuo}
\begin{barticle}[author]
\bauthor{\bsnm{Serfling},~\bfnm{Robert}\binits{R.}} \AND
  \bauthor{\bsnm{Zuo},~\bfnm{Yijun}\binits{Y.}}
(\byear{2000}).
\btitle{{General notions of statistical depth function}}.
\bjournal{The Annals of Statistics}
\bvolume{28}
\bpages{461--482}.
\bdoi{10.1214/aos/1016218226}
\end{barticle}
\endbibitem

\bibitem[\protect\citeauthoryear{Shi et~al.}{2021a}]{HallinKendall}
\begin{bmisc}[author]
\bauthor{\bsnm{Shi},~\bfnm{Hongjian}\binits{H.}},
  \bauthor{\bsnm{Drton},~\bfnm{Mathias}\binits{M.}},
  \bauthor{\bsnm{Hallin},~\bfnm{Marc}\binits{M.}} \AND
  \bauthor{\bsnm{Han},~\bfnm{Fang}\binits{F.}}
(\byear{2021}a).
\btitle{Center-outward sign- and rank-based quadrant, Spearman, and Kendall
  tests for multivariate independence}.
\bhowpublished{arXiv:2111.15567}.
\end{bmisc}
\endbibitem

\bibitem[\protect\citeauthoryear{Shi et~al.}{2021b}]{HallinIndAoS}
\begin{barticle}[author]
\bauthor{\bsnm{Shi},~\bfnm{Hongjian}\binits{H.}},
  \bauthor{\bsnm{Drton},~\bfnm{Mathias}\binits{M.}},
  \bauthor{\bsnm{Hallin},~\bfnm{Marc}\binits{M.}} \AND
  \bauthor{\bsnm{Han},~\bfnm{Fang}\binits{F.}}
(\byear{2021}b).
\btitle{On universally consistent and fully distribution-free rank tests of
  vector independence}.
\bjournal{The Annals of Statistics}
\bvolume{0}
\bpages{1-14}.
\end{barticle}
\endbibitem

\bibitem[\protect\citeauthoryear{Sohn, Lee and Yan}{2015}]{NIPS2015}
\begin{binproceedings}[author]
\bauthor{\bsnm{Sohn},~\bfnm{Kihyuk}\binits{K.}},
  \bauthor{\bsnm{Lee},~\bfnm{Honglak}\binits{H.}} \AND
  \bauthor{\bsnm{Yan},~\bfnm{Xinchen}\binits{X.}}
(\byear{2015}).
\btitle{Learning Structured Output Representation using Deep Conditional
  Generative Models}.
In \bbooktitle{Advances in Neural Information Processing Systems}
(\beditor{\bfnm{C.}\binits{C.}~\bsnm{Cortes}},
  \beditor{\bfnm{N.}\binits{N.}~\bsnm{Lawrence}},
  \beditor{\bfnm{D.}\binits{D.}~\bsnm{Lee}},
  \beditor{\bfnm{M.}\binits{M.}~\bsnm{Sugiyama}} \AND
  \beditor{\bfnm{R.}\binits{R.}~\bsnm{Garnett}}, eds.)
\bvolume{28}.
\bpublisher{Curran Associates, Inc.}
\end{binproceedings}
\endbibitem

\bibitem[\protect\citeauthoryear{Stone}{1977}]{stone1977}
\begin{barticle}[author]
\bauthor{\bsnm{Stone},~\bfnm{Charles~J.}\binits{C.~J.}}
(\byear{1977}).
\btitle{Consistent Nonparametric Regression}.
\bjournal{Ann. Statist.}
\bvolume{5}
\bpages{595--620}.
\bdoi{10.1214/aos/1176343886}
\end{barticle}
\endbibitem

\bibitem[\protect\citeauthoryear{Tukey}{1975}]{Tukey75}
\begin{barticle}[author]
\bauthor{\bsnm{Tukey},~\bfnm{J.~W.}\binits{J.~W.}}
(\byear{1975}).
\btitle{Mathematics and the picturing of data}.
\bjournal{Proceedings of the International Congress of Mathematicians, ver,
  1975}
\bvolume{2}
\bpages{523-531}.
\end{barticle}
\endbibitem

\bibitem[\protect\citeauthoryear{van~der Vaart and Wellner}{1996}]{Vart_Well}
\begin{bbook}[author]
\bauthor{\bparticle{van~der} \bsnm{Vaart},~\bfnm{Aad~W.}\binits{A.~W.}} \AND
  \bauthor{\bsnm{Wellner},~\bfnm{Jon~A.}\binits{J.~A.}}
(\byear{1996}).
\btitle{Weak Convergence and Empirical Processes}.
\bpublisher{Springer, New York, NY}.
\end{bbook}
\endbibitem

\bibitem[\protect\citeauthoryear{Villani}{2003}]{Villani2003}
\begin{bbook}[author]
\bauthor{\bsnm{Villani},~\bfnm{C.}\binits{C.}}
(\byear{2003}).
\btitle{Topics in Optimal Transportation}.
\bpublisher{American Mathematical Society}, \baddress{Providence, Rhode
  Island}.
\end{bbook}
\endbibitem

\bibitem[\protect\citeauthoryear{Villani}{2008}]{Villani2008}
\begin{bbook}[author]
\bauthor{\bsnm{Villani},~\bfnm{C.}\binits{C.}}
(\byear{2008}).
\btitle{Optimal Transport: Old and New}.
\bpublisher{Springer-Verlag Berlin Heidelberg}.
\end{bbook}
\endbibitem

\end{thebibliography}
\appendix
\section*{Appendix}

\subsection{Proofs for Section~\ref{Section;3}}
The convergence described of Theorem~\ref{MainTheorem} is based on the topology of set-valued maps, in particular the Graphical convergence (or   Painlev\'e-Kuratowski convergence of the graphs). Recall from \cite{RockWets98} that a sequence of set-valued maps $\{T_n\}_n$ converges \emph{graphically} to another set-valued map $T$ if 
\begin{itemize}
\item[--] the outer limit $\lim\sup_n T_n$---which is the set of $(\mathbf{x},\mathbf{y})\in \R^{d}\times \R^d$ for which there exists a sequence $\{(\mathbf{x}_{n},\mathbf{y}_{n})\}$ with $(\mathbf{x}_{n},\mathbf{y}_{n})\in T_n$ containing a subsequence which converges to $(\mathbf{x},\mathbf{y})$---exists and coincides with $T$ and
\item[--] the inner limit $\lim \inf _n T_n$---which  is the set of $(\mathbf{x},\mathbf{y})\in \R^{d}\times \R^d$  for which there exists a sequence $\{(\mathbf{x}_{n},\mathbf{y}_{n})\}$, with $(\mathbf{x}_{n},\mathbf{y}_{n})\in T_n$, which converges to $(\mathbf{x},\mathbf{y})${---exists and coincides with $T$}. 
\end{itemize}

We start  with 
 some necessary properties of the population and empirical conditional quantiles and 
  some auxiliary results. The following lemma states that the estimated conditional probability converges weakly in probability to its population counterpart. Recall from  Theorem~1.12.4 in \cite{Vart_Well} that weak convergence  can be measured in terms of the bounded Lipschitz norm, which is defined for $\mu,\nu\in\mathcal{P}(\R^d)$ as 
$$ d_{BL}(\mu,\nu)\coloneqq  \sup_{f\in \mathcal{F}_{BL}(\R^{d})}\left|\mathbb{E}_{\mathbf{Z}\sim \mu}(f(\mathbf{Z}))-\mathbb{E}_{\mathbf{W}\sim \nu}(f(\mathbf{W})) \right|,$$
where the class $\mathcal{F}_{BL}$ is the class of functions $f:\R^d\rightarrow\R$ such that $|f(\mathbf{z}_1)-f(\mathbf{z}_2)|\leq |\mathbf{z}_1-\mathbf{z}_2 |$ and $|f(\mathbf{z}_1) |\leq 1$, for all $\mathbf{z}_1,\mathbf{z}_2\in \R^d $.
\begin{Lemma}\label{Lemma_weak_conv}
For any $\epsilon>0$, under the assumptions of Theorem~\ref{MainTheorem}, we have 
\begin{equation*}
\mathbb{P}\left(   d_{BL}(P^{w(\mathbf{X})}_n, {\rm P}_{\mathbf{Y}\, |\mathbf{X}})  >\epsilon\right)\longrightarrow 0.
\end{equation*}
\end{Lemma}

See Section~\ref{ProofLemSec} for the proof.\medskip

Let
$\pi^{(n)}(\mathbf{x})$ be   a solution of \eqref{kanto_reg} 
and 
$$\pi^{*}(\mathbf{x})\coloneqq \left(\mathbf{Id}\times\mathbf{Q}_{\pm}(\ \cdot\ |\mathbf{x})\right)\#  \mathrm{U}_d.  $$
Note that, for each value of  $\mathbf{x}$, 
\begin{enumerate}
    \item[(i)] there exists a sequence of differentiable convex functions $\psi^{(n)}\left(\, \cdot\ \big|  \mathbf{x}\right):\R^d\rightarrow\R\cup\{+\infty\}$, $n\in \N$, such that  
    $$\mathbf{Q}^{(n)}_{w,\pm}\left(\mathbf{u}_j \ \big| \mathbf{x}\right)=\nabla\psi^{(n)}\left(\, \mathbf{u}_j\ \big|  \mathbf{x}\right) \ \text{for}\ \ j=1, \dots, k, \ \text{and}\ n\in \N;$$ 
     \item[(ii)] there exists a  convex function $\psi \left(\, \cdot\ \big|  \mathbf{X}=\mathbf{x})\right):\R^d\rightarrow\R\cup\{+\infty\}$ such that  $$\mathbf{Q}_{\pm}\left(\mathbf{u}\ \big| \mathbf{X}=\mathbf{x}\right)=\nabla\psi\left(\, \mathbf{u}\ \big|  \mathbf{X}=\mathbf{x}\right) \ \text{for $\mathrm{U}_d$-a.e.}\ \mathbf{u}\in \mathbb{S}_{d}.$$ 
\end{enumerate}
Using obvious notation, it holds that $\pi^{*}(\mathbf{x})$ and $\pi^{(n)}(\mathbf{x})$, for $n\in \N$, have cyclically monotone supports. Moreover, as a consequence of Corollary~\ref{Remark}, we have
\begin{align}
    \begin{split}\label{subsetpartial}
        \operatorname{supp}(\pi^{(n)}(\mathbf{x}))\subset \partial\tilde{\psi}^{(n)}\left(\, \cdot\ \big|  \mathbf{x}\right)\ \text{and}\     \operatorname{supp}(\pi^{*}(\mathbf{x}))\subset \partial\psi\left(\, \cdot\ \big| \mathbf{X}= \mathbf{x}\right),
    \end{split}
\end{align}
 possibly for some other sequence  $\tilde{\psi}^{(n)}\left(\, \cdot\ \big|  \mathbf{x}\right):\R^d\rightarrow\R\cup\{+\infty\}$ of  convex functions such that 
\begin{equation}
\label{intersection}
\mathbf{Q}^{(n)}_{w,\pm}\left(\mathbf{u}_j \ \big| \mathbf{x}\right)\in \partial\tilde{\psi}^{(n)}\left(\, \mathbf{u}_j\ \big|  \mathbf{x}\right) \ \text{for}\ \ j=1, \dots, k, \ \text{and}\ n\in \N.
\end{equation}

The following result then follows from Lemma~9 and Corollary~14 in \cite{McCann1995ExistenceAU}.
\begin{Lemma}\label{Lemma:BarrioLoubes}
Let $\mu, \nu \in \mathcal{P}(\R^d)$ be such that $\mu\ll\ell_d$ is supported on a convex set. Let~$\{\mu_n\}_{n\in \N}$ and $\{\nu_n\}_{n\in \N}\subset \mathcal{P}(\R^d) $   converge  weakly as $n\to\infty$ to $\mu$ and $\nu$, respectively. Suppose, moreover, that there exists a sequence of probability measures $\{\pi_n\}_{n\in \N}\subset \mathcal{P}(\R^d\times\R^d)$ with marginals $\mu_n$ and $\nu_n$ such that, for some sequence of convex functions  $\{\phi_n\}_{n\in \N}$, it holds that, for all~$n\in\N$, $\operatorname{supp}(\pi_n)\subset \partial\phi_n$. Then, 
\begin{enumerate}
    \item[(i)]  $\{\pi_n\}_n$  converges weakly as $n\to\infty$ to $
    \pi^*=\left(\mathbf{Id}\times\nabla\phi\right)\# \mu
$, where   $\nabla\phi$ is the gradient of a convex function $ \phi$  pushing  $\mu$ forward to $\nu$, and
    \item[(ii)]  there exists a sequence $\{a_n\}_{n\in \N}\subset\R$ such that, $\mu$-a.e.,  
$\phi_n+ a_n\rightarrow\phi$ and $\partial\phi_n\rightarrow\partial\phi$ graphically as $n\to\infty$.
\end{enumerate}
\end{Lemma}
See Section~\ref{ProofLemSec} for the proof.

\begin{proof}[\textbf{Proof of Theorem~\ref{MainTheorem}}]
Suppose   that there exist  $\mathbf{u}_0\in\mathbb{S}_{d}$ and  $\epsilon_0>0$ such that
\begin{equation*}
  \limsup_{n\to\infty}\mathbb{P}\left(\text{ $\mathbf{Q}^{(n)}_{w,\pm}(\mathbf{u}_0| \mathbf{x})\not\in {\mathbf{Q}}_{\pm}(\mathbf{u}_0|\mathbf{X})+\epsilon_0\mathbb{S}_{d}$ } \right)=\delta>0.\end{equation*}
We can find a subsequence $n_k$ such that
\begin{equation}\label{contradictionalongsubs}
 \lim_{k\to\infty} \mathbb{P}\left(\text{ $\mathbf{Q}^{(n_k)}_{w,\pm}(\mathbf{u}_0| w(\mathbf{X};\mathbf{X}^{[n_k]}))\not\in {\mathbf{Q}}_{\pm}(\mathbf{u}_0|\mathbf{X})+\epsilon_0\mathbb{S}_{d}$ } \right)=\delta>0.\end{equation}
  Note that the space of probability distributions on $\R^{d}$, endowed with the bounded Lipschitz metric, is separable and complete, see Theorem 1.12.4 in \cite{Vart_Well}.  Therefore, the convergence described  in Lemma~\ref{Lemma_weak_conv}
implies that, for the subsequence $n_k$, there exists a further sub sequence $n_{k_i}$ such that the event 
\begin{equation}\label{fromProbatoas}
     \Omega_0=\left(\sup_{f\in \mathcal{F}_{BL}(\R^{d})}\left|\sum_{j=1}^{n_{k_i}} w_j(\mathbf{X};\mathbf{X}^{(n_{k_i})}){f(\mathbf{Y}_j)}- \mathbb{E}(f(\mathbf{Y})|\mathbf{X})\right|\longrightarrow 0\right)
\end{equation}
has probability one. Let     $\mathbf{x}=\mathbf{X}(\omega)$ with $\omega\in \Omega_0$. Then, by Lemma~\ref{Lemma:BarrioLoubes}, we have, as $i\to\infty$, 
\begin{equation*}
   \partial \tilde{\psi}^{(n_{k_i})}\left(\ \cdot\ \big| \mathbf{x}\right)\longrightarrow \mathbf{Q}_{\pm}(\ \cdot\ |\mathbf{X}=\mathbf{x}) \ \ \text{graphically}.
\end{equation*}
This implies, by Theorem 8.3  in \cite{Bagh95convergenceof}, that there exists $i_0\in \N$ such that, for all $i>i_0$,
\begin{equation}\label{a.s.ForSubseq}
     \partial \tilde{\psi}^{(n_{k_i})}\left(\mathbf{u}_0 \big|  \mathbf{x}\right)\subset \mathbf{Q}_{\pm}(\mathbf{u}_0|\mathbf{X}=\mathbf{x})+\epsilon_0\mathbb{S}_{d}.
\end{equation}
From \eqref{intersection}  and the fact that
\eqref{a.s.ForSubseq} holds with probability one, we deduce that 
\begin{equation}\label{a.s.ForSubseq2}
    \mathbb{P}\left(\bigcap_{i_0\in\N}\bigcup_{i\geq i_0}\mathbf{Q}^{(n_{k_i})}_{\pm}(\mathbf{u}_0| \mathbf{X})\not\in \mathbf{Q}_{w,\pm}(\mathbf{u}_0|\mathbf{X})+\epsilon_0\mathbb{S}_{d}\right)=0,
\end{equation}
which implies that
\begin{equation*}
    \limsup_{i\to\infty}\mathbb{P}\left(\mathbf{Q}^{(n_{k_i})}_{\pm}(\mathbf{u}_0| \mathbf{X})\not\in \mathbf{Q}_{w,\pm}(\mathbf{u}_0|\mathbf{X})+\epsilon_0\mathbb{S}_{d}\right)\leq 0.
\end{equation*}
This contradicts \eqref{contradictionalongsubs}.  

The rest of the proof follows from compactness arguments and a refined use of Theorem~8.3. in \cite{Bagh95convergenceof}. Suppose that, for some $q_0\in (0,1)$ and  $\epsilon_0>0$, and along a subsequence $n_k$, we have 
\begin{equation}\label{quqntilesabsurde}
 \lim_{k\to\infty}  \mathbb{P}\left(\text{ $  \mathbb{C}_{\pm}^{(n_k)}(q_0\, \big| \mathbf{X})\not\subset \mathbb{C}_{\pm}(q_0\, \big| \mathbf{X})+\epsilon_0\mathbb{S}_{d}$ } \right)=\delta>0.
\end{equation}
Theorem 8.3. in \cite{Bagh95convergenceof} and    Lemma \ref{Lemma:BarrioLoubes} jointly imply the existence of a further subsequence $n_{k_i}$, such that, for every $\bar{\mathbf{u}}\in\mathbb{S}_{d}$,  there exists $I_{\bar{\mathbf{u}}}\in \N$ and $\lambda_{\bar{\mathbf{u}}}>0$ satisfying 
$$
  \text{ $ \mathbf{Q}^{(n_{k_i})}_{\pm}(\mathbf{u}\, | \mathbf{x})\in {\mathbf{Q}}_{\pm}({\bar{\mathbf{u}}}\, |X=\mathbf{x})+\epsilon_0\mathbb{S}_{d}$ for all $ \mathbf{u}\in \bar{\mathbf{u}}+ \lambda_{\bar{\mathbf{u}}} \mathbb{S}_{d}$ and   all $i>I_{\bar{\mathbf{u}}}$. }
  $$
  
   Note that  $q\, \mathbb{S}_{d}\subset \bigcup_{{\bar{\mathbf{u}}}:\ | {\bar{\mathbf{u}}}\, |\leq q} \bar{\mathbf{u}}+ \lambda_{\bar{\mathbf{u}}} \mathbb{S}_{d}$.  Since $q\, \overline{\mathbb{S}}_{d}$ is compact, there exists a finite co\-vering~$q \,\mathbb{S}_{d}\subset \bigcup_{k=1}^{K_{\epsilon}} \bar{\mathbf{u}}_k+ \delta_{\bar{\mathbf{u}}_k} \mathbb{S}_{d}$. Set  $I_{0}=\max(I_{\bar{\mathbf{u}}_1}, \dots, I_{\bar{\mathbf{u}}_{K_{\epsilon}}})$: then for all $i>I_{\bar{\mathbf{u}}}$, we have
$$\mathbb{C}_{\pm}^{(n_{k_i})}(q_0\, \big| \mathbf{x})\subset \mathbb{C}_{\pm}(q_0\, \big| \mathbf{x})+\epsilon_0\mathbb{S}_{d}, $$
which holds with probability one and, using the same argument as for \eqref{a.s.ForSubseq},     contra\-dicts~\eqref{quqntilesabsurde}.
The same reasoning holds also for the contours.
\end{proof}


\begin{proof}[\textbf{Proof of Theorem~~\ref{MainTheorem_discrete}}] 
Under  Assumption (R),  it follows from \cite{BarrioSanzHal} that, for $\omega\in\Omega_0\subseteq \Omega$ where $\mathbb{P}(\Omega_0)=1$, 
  the center-outward quantile function $\mathbf{Q}_{\pm}(\mathbf{u}\, |\mathbf{X}=\mathbf{x}\coloneqq \mathbf{X}(\omega))$ is a singleton for $\mathrm{U}_d$-almost all values of $\mathbf{u}\in \mathbb{S}_{d}$. Therefore, we adopt here the  slight abuse of notation commented before Theorem~\ref{MainTheorem_discrete} and write~$\mathbf{Q}_{\pm}(\mathbf{u}\, |\mathbf{X}=\mathbf{x})=\{\mathbf{Q}_{\pm}(\mathbf{u}\, |\mathbf{X}$ for $\mathbf{x})\}$. Set $\mathbf{u}\in \mathbb{S}_{d}\setminus \{\mathbf{0}\}$ and note that, since  $\mathbf{Q}_{\pm}(\mathbf{u}\, |\mathbf{X})$ and $\mathbf{Q}^{(n)}_{w,\pm}(\mathbf{u}\, | \mathbf{x})$ are a.s.  singletons, we have, as $n$ and $N$ tend to infinity,
\begin{equation*}
  \mathbb{P}\left( \big|\mathbf{Q}^{(n)}_{w,\pm}(\mathbf{u}\, | \mathbf{x})- {\mathbf{Q}}_{\pm}(\mathbf{u}\, |\mathbf{X})\big|>\epsilon\right)\rightarrow 0.
\end{equation*}
Let $K$ be a compact subset of $\mathbb{S}_{d}\setminus \{\mathbf{0}\} $. In order to establish uniform convergence in $K$,   suppose that the contrary holds. Since the space of continuous functions from $K$ to $\R^d$ (endowed with the topology of uniform convergence) is complete and separable,  there exists  a subsequence $n_k$ such that, for some $\delta>0$, the probability of
\begin{equation}\label{a.s.contradition}
  \Omega'=\left( \sup_{\mathbf{u}\in K}\big|\mathbf{Q}^{(n_k)}_{w,\pm}(\mathbf{u}\, | \mathbf{X})- {\mathbf{Q}}_{\pm}(\mathbf{u}\, |\mathbf{X})\big|\rightarrow \delta\right),
\end{equation}
is one. Set $\omega\in \Omega'$ and consider  $\mathbf{x}=\mathbf{X}(\omega).$  There exists a sequence $\{\mathbf{u}_{n_k}\}\subset K$ such that 
\begin{equation*}
   \big|\mathbf{Q}^{(n_k)}_{w,\pm}(\mathbf{u}_{n_k}\, | \mathbf{x})- {\mathbf{Q}}_{\pm}(\mathbf{u}_{n_k}\, |\mathbf{X}=\mathbf{x})\big|\rightarrow \delta ,
\end{equation*}
for a possibly different $\delta>0$. Since the sequence $\{\mathbf{u}_{n_k}\}_k$ is in  $K$, it admits  at least one point of accumulation $\bar{\mathbf{u}}\in K$.  Hence, 
\begin{align*}
    \liminf_{k\to\infty} (& \big|\mathbf{Q}^{(n_k)}_{w,\pm}(\mathbf{u}_{n_k}\, | \mathbf{x})- {\mathbf{Q}}_{\pm}(\bar{\mathbf{u}}\, |\mathbf{X}=\mathbf{x})\big|\\
    &+\big|{\mathbf{Q}}_{\pm}(\bar{\mathbf{u}}\, |\mathbf{X}=\mathbf{x})- {\mathbf{Q}}_{\pm}(\mathbf{u}_{n_k}\, |\mathbf{X}=\mathbf{x})\big|)\geq \delta,
\end{align*} 
where the second term tends to $0$ by the continuity of $\mathbf{u}\mapsto{\mathbf{Q}}_{\pm}(\mathbf{u}\ |\mathbf{X}=\mathbf{x})$. This implies that 
\begin{equation}
    \label{absrudo_reg}
     \liminf_{k\to\infty}  \big|\mathbf{Q}^{(n_k)}_{w,\pm}(\mathbf{u}_{n_k}\, | \mathbf{x})- {\mathbf{Q}}_{\pm}(\bar{\mathbf{u}}\, |\mathbf{X}=\mathbf{x})\big|\\
    \geq \delta.
\end{equation}
But Lemma \ref{Lemma:BarrioLoubes} entails the Graphical convergence   to ${\mathbf{Q}}_{\pm}(\ \cdot \ |\mathbf{X}=\mathbf{x})$ of $\mathbf{Q}^{(n_k)}_{w,\pm}(\ \cdot\ | \mathbf{x})$, which contradicts \eqref{absrudo_reg}. The desired uniformity over $K$ follows. 

Finally,  the convergence of the contours is a consequence of the previous result on the regions.
\end{proof}

\begin{proof}[{\textbf{Proof of Corollary~\ref{Remarkcontrol}}}]

To prove \eqref{convergenceProbaControl}, let us show that, for any   subsequence $n_k$, there exists a further subsequence converging a.s. To avoid repetitions, assume that  $N=N(n)$ is such that $N\to \infty$ as~$n\to \infty$. 

 Leaving aside the singular points, let us consider the empirical and population {\it quantile~rings}  
$$ \mathbb{C}_{\pm}^{(n)}(\epsilon,\,\tau\, \big|  \mathbf{x})\coloneqq \mathbb{C}_{\pm}^{(n)}(\tau\, \big|  \mathbf{x})\setminus \mathbb{C}_{\pm}^{(n)}(\epsilon\, \big|  \mathbf{x}) $$
and 
$$ \mathbb{C}_{\pm}(\epsilon,\,\tau\, \big|  \mathbf{x})\coloneqq \mathbb{C}_{\pm}(\tau\, \big|  \mathbf{x})\setminus \mathbb{C}_{\pm}(\epsilon\, \big|  \mathbf{x}), $$
respectively. Theorem 3.3 yields,  for all $0<\epsilon<\tau$, as $n \rightarrow \infty$,
\begin{equation}
\label{convergenceProbaControl1}
 \mathbb{P}\left(\mathbf{Y}\in  \mathbb{C}_{\pm}^{(n)}(\epsilon,\,\tau\, \big|  \mathbf{X})\ \text{and} \ \mathbf{Y}\not\in  \mathbb{C}_{\pm}(\epsilon,\,\tau\, \big|  \mathbf{X})\Big|\mathbf{X}\right)\stackrel{{\mathbb P} }{\longrightarrow}  0 .
\end{equation}
Let $\{\epsilon_j\}_{j \in \N}$ be a monotone sequence tending to $0$. For  $j=1$ there exists a further subsequence $n_k^1$, say,  and a subset $\Omega _1$ of $\Omega$ such that $\mathbb{P}(\Omega_1)=1$ such that, for every $\mathbf{x}=\mathbf{X}(\omega)$ with~$\omega\in \Omega_1$,\begin{equation}
\label{convergenceProbaControl2}
 \mathbb{P}\left(\mathbf{Y}\in  \mathbb{C}_{\pm}^{(n_k^1)}(\epsilon_1,\,\tau\, \big|  \mathbf{x})\ \text{and} \ \mathbf{Y}\not\in  \mathbb{C}_{\pm}(\epsilon_1,\,\tau\, \big|  \mathbf{x})\Big|\mathbf{X}=\mathbf{x}\right){\longrightarrow}  0 .
\end{equation}
By definition, $\mathbb{P}\left(\mathbb{C}_{\pm}(\epsilon_1,\,\tau\, \big|  \mathbf{x})\Big|\mathbf{X}=\mathbf{x}\right)=\tau-\epsilon_1;$ therefore, in view of \eqref{convergenceProbaControl2},  
$$\mathbb{P}\left(\mathbb{C}_{\pm}^{(n_{k}^1)}(\epsilon_1,\,\tau\, \big|  \mathbf{x}) \Big|\mathbf{X}=\mathbf{x}\right)\longrightarrow \tau-\epsilon_1.$$
 Repeating the argumentat for $j=2$, there exists a set $\Omega_2\subset \Omega$, with $\mathbb{P}(\Omega_2)=1$, and a subsequence (call it $n_k^2$) of $n_k^1$ such that,  for every $\omega\in \Omega_2$ and $\mathbf{x}=\mathbf{X}(\omega)$, 
$$\mathbb{P}\left(\mathbb{C}_{\pm}^{(n_{k}^2)}(\epsilon_2,\,\tau\, \big|  \mathbf{x}) \Big|\mathbf{X}=\mathbf{x}\right)\longrightarrow \tau-\epsilon_2.$$
This argument can be repeated for each $j \in \N$ and the set $\Omega_0=\bigcap_{j \in \N}\Omega_j$ has probability one. Set $\omega\in \Omega_0\subset\Omega$ and $\mathbf{x}=\mathbf{X}(\omega)$:  then there exists $k_1$ such that 
$$\left|\mathbb{P}\left(\mathbb{C}_{\pm}^{(n_{k_1})}(\epsilon_1,\,\tau\, \big|  \mathbf{x}) \big|\mathbf{X}=\mathbf{x}\right)- \tau\right|<2\,\epsilon_1.$$
Analogously, for each $m$, there exist $k_j$ such that 
$$\left|\mathbb{P}\left(\mathbb{C}_{\pm}^{(n_{k_j})}(\epsilon_j,\,\tau\, \big|  \mathbf{x}) \big|\mathbf{X}=\mathbf{x}\right)- \tau\right|<2\,\epsilon_j.$$
Therefore, we obtain the limit
$$\mathbb{P}\left(\mathbb{C}_{\pm}^{(n_{k_j})}(\epsilon_j,\,\tau\, \big|  \mathbf{x}) \big|\mathbf{X}=\mathbf{x}\right)\longrightarrow \tau$$
and, noticing that $$\mathbb{P}\left(\mathbb{C}_{\pm}^{(n_{k_j})}(\epsilon_j,\,\tau\, \big|  \mathbf{x}) \big|\mathbf{X}=\mathbf{x}\right)\leq \mathbb{P}\left(\mathbb{C}_{\pm}^{(n_{k_j})}(\tau\, \big|  \mathbf{x}) \big|\mathbf{X}=\mathbf{x}\right),$$
also the asymptotic upper bound 
$$\liminf\mathbb{P}\left(\mathbb{C}_{\pm}^{(n_{k_j})}(\tau\, \big|  \mathbf{x}) \big|\mathbf{X}=\mathbf{x}\right)\geq \tau.$$
Now,  Theorem~\ref{MainTheorem} implies, for every subsequence (for which we keep the notation $n$) of $n$ and $j \in \N$, there existence of some $n_j \in \N$ such that
\begin{equation*}
  \mathbb{P}\left(\text{ $ \mathbb{C}_{\pm}^{(n)}(\tau \, \big| \mathbf{X})\not\subset \mathbb{C}_{\pm}(\tau \, \big| \mathbf{X})+\frac{1}{2^j}\mathbb{S}_{d}$ } \right)\leq \frac{1}{2^j}
 \end{equation*}
 for all $n\geq n_j$. The sum 
$$ \sum_{j=1}^{\infty}\mathbb{P}\left(\text{ $ \mathbb{C}_{\pm}^{(n_j)}(\tau \, \big| \mathbf{X})\not\subset \mathbb{C}_{\pm}(\tau \, \big| \mathbf{X})+\frac{1}{2^j}\mathbb{S}_{d}$ } \right)
$$
thus is finite, and the  Borel-Cantelli lemma yields
$$\mathbb{P}\left(\bigcap_{J\in \N}\bigcup_{j\geq J}\text{ $ \mathbb{C}_{\pm}^{(n_j)}(\tau \, \big| \mathbf{X})\not\subset \mathbb{C}_{\pm}(\tau \, \big| \mathbf{X})+\frac{1}{2^j}\mathbb{S}_{d}$ } \right)=0,$$
or, equivalently, ${\mathbb P}(\Omega ^*)=1$, where
$$\Omega^*\coloneqq\left(\bigcup_{J\in \N}\bigcap_{j\geq J}\text{ $ \mathbb{C}_{\pm}^{(n_j)}(\tau \, \big| \mathbf{X})\subset \mathbb{C}_{\pm}(\tau \, \big| \mathbf{X})+\frac{1}{2^j}\mathbb{S}_{d}$ }\right).$$
  Setting $\omega\in \Omega^*\subset\Omega$ and $\mathbf{x}=\mathbf{X}(\omega)$, there exists $J\in \N$ such that
$$ \mathbb{P}\left(\mathbb{C}_{\pm}^{(n_{j})}(\tau\, \big|  \mathbf{x}) \big|\mathbf{X}=\mathbf{x}\right)\leq \mathbb{P}\left(\mathbb{C}_{\pm}(\tau\, \big|  \mathbf{x})+\frac{1}{2^j}\mathbb{S}_{d}  \big|\mathbf{X}=\mathbf{x}\right),$$
for all $j\geq J$. Since $$\mathbb{C}_{\pm}(\tau\, \big|  \mathbf{x})+\frac{1}{2^{j+1}}\mathbb{S}_{d}\subset \mathbb{C}_{\pm}(\tau\, \big|  \mathbf{x})+\frac{1}{2^j}\mathbb{S}_{d}$$
and 
$$ \bigcap_{j \in \N}\mathbb{C}_{\pm}(\tau\, \big|  \mathbf{x})+\frac{1}{2^j}\mathbb{S}_{d} =\mathbb{C}_{\pm}(\tau\, \big|  \mathbf{x}),$$
 we obtain
\begin{equation}
\label{consequence_theorem_weak}
 \limsup\mathbb{P}\left(\mathbb{C}_{\pm}^{(n_{j})}(\tau\, \big|  \mathbf{x}) \big|\mathbf{X}=\mathbf{x}\right)\leq \tau,
\end{equation}
which concludes the proof.
\end{proof}

\noindent\textbf{\textit{Proof of Lemma~\ref{Lemma_k_nearest} }} 
 To prove this lemma, we show that the following   five conditions of Stone's theorem (Theorem~1 in \cite{stone1977}) are satisfied (convergence for $n\to\infty)$:
\begin{enumerate}
\item[(a)] 
 \emph{there exists $C\geq 1$ such that, for any  non-negative measurable function $f$,}    \begin{align*}
        &\mathbb{E}\left( \frac{1}{k}\sum_{j=1}^n \mathbbm{1}_{ \mathbf{X}_{j} \in K_n^k(\mathbf{X})}f(\mathbf{X}_j)\right)\leq C\mathbb{E}\left( f(\mathbf{X})\right) ;
            \end{align*}
 \item[(b)]  $\displaystyle{ \mathbb{P}\left( \frac{1}{k}\sum_{j=1}^n \mathbbm{1}_{ \mathbf{X}_{j} \in K_n^k(\mathbf{X})}\leq 1 \right)=1 \ \ \text{for all}\ \ n\in \N ;}$
  \item[(c)]  $\displaystyle{\frac{1}{k}\sum_{j=1}^n \mathbbm{1}_{ \mathbf{X}_{j} \in K_n^k(\mathbf{X})} \mathbbm{1}_{ |\mathbf{X}_{j}-X|>a}\longrightarrow 0, \ \ \text{in probability, for all $a>0$};}$
  \item[(d)]  $\displaystyle{\frac{1}{k}\sum_{j=1}^n \mathbbm{1}_{ \mathbf{X}_{j} \in K_n^k(\mathbf{X})} \longrightarrow 1\ \ \text{in probability; }}$
  \item[(e)]  $\displaystyle{\max_{i=j,\cdots,n}\frac{1}{k}\mathbbm{1}_{ \mathbf{X}_{j} \in K_n^k(\mathbf{X})} \longrightarrow 0\ \ \text{in probability. }}$
\end{enumerate}
    \begin{proof}[Proof of  (a)]
    Let $\mathbf{X}$ be a random variable   independent of $\mathbf{X}^{(n)}$, with the same distribution as $\mathbf{X}_i$. Set $ \mathbf{u}_0=\mathbf{F}_{\pm}^{(n)}(\mathbf{X})$ and $\mathbf{u}_j=\mathbf{F}_{\pm}^{(n)}(\mathbf{X}_j)$,   $j=1, \dots,n.$ Defining  
    $$K_n^{k,(i)}(\mathbf{X})\coloneqq \{\mathbf{X}_j: \ i\neq j, \ \mathbb{F}_{\pm}^{n} (\mathbf{X}_j)\in N_k(\mathbf{F}_{\pm}^{(n)}(\mathbf{X}_i))\},$$
 note that $\mathbbm{1}_{ \mathbf{X}_{i} \in K_n^{k}(\mathbf{X})}f(\mathbf{X}_i)$  and  $\mathbbm{1}_{ X \in K_n^{k,(i)}(\mathbf{X}_i)}f(\mathbf{X})$ have the same distribution. It follows that 
     \begin{align*}
        \mathbb{E}\left( \frac{1}{k}\sum_{j=1}^n \mathbbm{1}_{ \mathbf{X}_{j} \in K_n^k(\mathbf{X})}f(\mathbf{X}_j)\right)=\frac{1}{k}\sum_{j=1}^n\mathbb{E}\left(  \mathbbm{1}_{ \mathbf{X}_{j} \in K_n^k(\mathbf{X})}f(\mathbf{X}_j)\right)
        =\frac{1}{k}\sum_{j=1}^n\mathbb{E}\left(  \mathbbm{1}_{ X \in K_n^{k,(j)}(\mathbf{X}_j)}f(\mathbf{X})\right),
    \end{align*}
which implies that 
\begin{equation}\label{bound_of_1}
           \mathbb{E}\left( \frac{1}{k}\sum_{j=1}^n \mathbbm{1}_{ \mathbf{X}_{j} \in K_n^k(\mathbf{X})}f(\mathbf{X}_j)\right)\leq \mathbb{E}\left(  f(\mathbf{X})\frac{1}{k}\sum_{j=1}^n \mathbbm{1}_{ X \in K_n^{k,(j)}(\mathbf{X}_j)}\right).
\end{equation}
Since 
 $
        \sum_{j=1}^n \mathbbm{1}_{ X \in K_n^{k}(\mathbf{X}_j)}= \sum_{j=1}^n \mathbbm{1}_{ \mathbf{u}_0 \in N_k({\scriptstyle{\mathfrak G}}_j)}
 \vspace{1mm}$, 
    we can apply Corollary 6.1. in \cite{gyodfi} and conclude that there exists $\lambda_d\in \R$ such that 
 $ \sum_{j=1}^n \mathbbm{1}_{ X \in K_n^{k,(j)}(\mathbf{X}_j)}\leq k \lambda_d.$ 
This and~\eqref{bound_of_1} complete  the proof.
\end{proof}
\begin{proof}[Proof of  (b), (d), and (e)]
Conditions (b) and~(d)  are direct consequences of the properties of the weight function, see Corollary~1 in \cite{stone1977}. As for (e), it follows from  the fact that $k\rightarrow \infty$. 
\end{proof}
\begin{proof}[Proof of (c)]
The following lemma (see Section~\ref{ProofLemSec} for a proof) is a corollary of Lemma 6.1 in \cite{gyodfi}. 
\begin{Lemma}\label{Lemma_converge_u}
Let  ${k}/{n}\rightarrow 0$ as $n\to\infty$. Then,  as $n\to\infty$
\begin{align*}
    \sup_{{\scriptstyle{\mathfrak G}}_i\in N_k({\scriptstyle{\mathfrak G}}_0)}\, |{\scriptstyle{\mathfrak G}}_i-{\scriptstyle{\mathfrak G}}_0 |\longrightarrow 0, \ \ a.s.
\end{align*}
\end{Lemma}
Since $\mathbf{Q}_{\pm}$ is a singleton with probability one, let us assume, without loss of generality, that   $\mathbf{Q}_{\pm}({\scriptstyle{\mathfrak G}}_0)$  and  $\mathbf{Q}_{\pm}^{(n)}({\scriptstyle{\mathfrak G}}_{k})$ are  singletons.  Actually, the set 
$$\bigcap_{k=1}^{\infty}\bigcap_{n=0}^{\infty}\left\{\text{$\mathbf{Q}_{\pm}^{(n)}({\scriptstyle{\mathfrak G}}_{k})$ is a  singleton }\right\}$$
 also has probability one. Within that set,  
\begin{enumerate}
    \item[(i)] $\mathbf{Q}_{\pm}^{(n)}\longrightarrow \mathbf{Q}_{\pm} \ \ \text{graphically as $n\to\infty$}$,
   \item[(ii)] for every $ \mathbf{X}_i\in K_n^{k}(\mathbf{X})$, there exists some ${\scriptstyle{\mathfrak G}}_i\in N_k({\scriptstyle{\mathfrak G}}_0)$ such that $\mathbf{Q}_{\pm}^{(n)}({\scriptstyle{\mathfrak G}}_{i})=\{\mathbf{X}_i \}$, and
    \item[(iii)]  Lemma \ref{Lemma_converge_u} yields  ${\scriptstyle{\mathfrak G}}_{i}\rightarrow{\scriptstyle{\mathfrak G}}_0$ {as $n\to\infty$}.
\end{enumerate}
This, in view of  Proposition 5.33 in \cite{RockWets98}, implies that 
\begin{equation}\label{claim_a.s.}
   \sup_{ \mathbf{X}_i\in K_n^{k}(\mathbf{X})}\, |\mathbf{X}_{i}-\mathbf{X}\, |\longrightarrow 0, \ \ a.s. 
\end{equation}
We are ready now to prove (c). Set $a>0$. Since there are $k$ elements in $ K_n^k(\mathbf{X})$, we have
\begin{align*}
  \mathbb{E}\left( \frac{1}{k}\sum_{j=1}^n \mathbbm{1}_{ \mathbf{X}_{j} \in K_n^k(\mathbf{X})} \mathbbm{1}_{ |\mathbf{X}_{j}-X|>a}\right)&=\mathbb{E}\left( \frac{1}{k}\sum_{\mathbf{X}_{j} \in K_n^k(\mathbf{X})}  \mathbbm{1}_{ |\mathbf{X}_{j}-\mathbf{X}\, |>a}\right)\\
  &=  \mathbb{P}\left(  \sup_{ \mathbf{X}_i\in K_n^{k}(\mathbf{X})}{ |\mathbf{X}_{j}-\mathbf{X}\, |>a}\right)
\end{align*}
which, owing to \eqref{claim_a.s.}, tends to $0$. The desired result follows as  a direct consequence of Markov's inequality.
\end{proof}

\subsection{Proofs of Lemmas~\ref{Lemma_weak_conv},~\ref{Lemma:BarrioLoubes}, and~\ref{Lemma_converge_u}}\label{ProofLemSec}

\begin{proof}[{\textbf{Proof of Lemma~\ref{Lemma_weak_conv}}}]

Let $\delta>0$ and $\epsilon>0$ be arbitrary. Denote by $K\subset \R^d$  a compact set such that $P(\mathbf{Y}\in K)\geq 1-
{\delta \epsilon}/{18}$. Suppose that $ \mathbf{0}\in K$ and define $\mathcal{F}_{K}$ as  the class\linebreak of~$1$-Lipschitz functions $f$ supported on $K$ such that $f(\mathbf{0})=0$.
Such a class, by the Arzelà–Ascoli theorem, is  relatively compact  for $\mathcal{C}(K)$ and the uniform convergence. Then, there exists a sequence $f_1, \dots, f_{N_{\epsilon}}$, such that $\sup_{f\in \mathcal{F}_{K}}\inf_{k=1\cdots N_{\epsilon}}\, || f-f_k||_{\infty}\leq {\epsilon}/{8}$. Therefore, for every~$f\in \mathcal{F}_{K}$, we have
\begin{align*}
    \left|\sum_{j=1}^n w_j(X;\mathbf{X}^{(n)}){f(\mathbf{Y}_j)}-\mathbb{E}(f(\mathbf{Y})|\mathbf{X})\right|\leq \sup_{k=1\cdots N_{\epsilon}}  \left|\sum_{j=1}^n w_j(X;\mathbf{X}^{(n)}){f_k(\mathbf{Y}_j)}-\mathbb{E}(f_k(\mathbf{Y})|\mathbf{X})\right|+\frac{\epsilon}{4}.
\end{align*}
In consequence, there exists $n_0$ such that, for $n\geq n_0$, we have
\begin{align}\label{bound_theorem_1}
\begin{split}
    \mathbb{P}&\left(\sup_{f\in \mathcal{F}_K}\left|\sum_{j=1}^n w_j(X;\mathbf{X}^{(n)}){f(\mathbf{Y}_j)}-\mathbb{E}(f(\mathbf{Y})|\mathbf{X})\right|>\frac{\epsilon}{2}\right)\\
    &\leq \mathbb{P}\left(\sup_{k=1\cdots N_{\epsilon}}  \left|\sum_{j=1}^n w_j(X;\mathbf{X}^{(n)}){f(\mathbf{Y}_j)}-\mathbb{E}(f(\mathbf{Y})|\mathbf{X})\right|>\frac{\epsilon}{4}\right)\\
    &\leq \sum_{k=1}^{N_{\epsilon}} \mathbb{P}\left( \left|\sum_{j=1}^n w_j(X;\mathbf{X}^{(n)}){f(\mathbf{Y}_j)}-\mathbb{E}(f(\mathbf{Y})|\mathbf{X})\right|>\frac{\epsilon}{4}\right)\leq \frac{\delta}{3},
\end{split}
\end{align}
 where the last inequality follows from the fact that the weight function is consistent, Note that  every $f\in \mathcal{F}_{BL}$  can be approximated by $f\mathbbm{1}_{K}$. This yields 
\begin{align}
    \label{Boun_thoerem_3}
    \begin{split}
         &\left|\sum_{j=1}^n w_j(X;\mathbf{X}^{(n)}){f(\mathbf{Y}_j)}-\mathbb{E}(f(\mathbf{Y})|\mathbf{X})\right|\\
    &\leq \sup_{f\in \mathcal{F}_K}\left|\sum_{j=1}^n w_j(X;\mathbf{X}^{(n)}){f(\mathbf{Y}_j)}-\mathbb{E}(f(\mathbf{Y})|\mathbf{X})\right|\\
    &+
    \left|\sum_{j=1}^n w_j(X;\mathbf{X}^{(n)}){f(\mathbf{Y}_j)}\mathbbm{1}_{\R^d\setminus K}(\mathbf{Y}_j)-\mathbb{E}(f(\mathbf{Y})\mathbbm{1}_{\R^d\setminus K}(\mathbf{Y})|\mathbf{X})\right|.
    \end{split}
\end{align}
Inequality    \eqref{bound_theorem_1} provides an upper bound for the first term. The second one, denoted as~$A_n$, is bounded by 
 \begin{align*}
      \left|\sum_{j=1}^n w_j(X;\mathbf{X}^{(n)}){f(\mathbf{Y}_j)}\mathbbm{1}_{\R^d\setminus K}(\mathbf{Y}_j)\right|+ \left| \mathbb{E}(f(\mathbf{Y})\mathbbm{1}_{\R^d\setminus K}(\mathbf{Y})|\mathbf{X})\right|.
 \end{align*}
 Since the weights are positive and $\sup_{\mathbf{x}\in \R^d} |f(\mathbf{x})| \leq 1$, 
  \begin{align*}
     A_n &\leq \sum_{j=1}^n w_j(X;\mathbf{X}^{(n)})\mathbbm{1}_{\R^d\setminus K}(\mathbf{Y}_j)+ \mathbb{E}(\mathbbm{1}_{\R^d\setminus K}(\mathbf{Y})|\mathbf{X})\\
    & \leq  \left|\sum_{j=1}^n w_j(X;\mathbf{X}^{(n)})\mathbbm{1}_{\R^d\setminus K}(\mathbf{Y}_j)-\mathbb{E}(\mathbbm{1}_{\R^d\setminus K}(\mathbf{Y})|\mathbf{X})\right|+ 2\mathbb{E}(\mathbbm{1}_{\R^d\setminus K}(\mathbf{Y})|\mathbf{X}).
 \end{align*}
 Note that the bound does not depend on the function $f$. Consequently,
 \begin{align*}
      \sup_{f\in \mathcal{F}_{BL}}&\left|\sum_{j=1}^n w_j(X;\mathbf{X}^{(n)}){f(\mathbf{Y}_j)}\mathbbm{1}_{\R^d\setminus K}(\mathbf{Y}_j)-\mathbb{E}(f(\mathbf{Y})\mathbbm{1}_{\R^d\setminus K}(\mathbf{Y})|\mathbf{X})\right|\\
      &\leq  \left|\sum_{j=1}^n w_j(X;\mathbf{X}^{(n)})\mathbbm{1}_{\R^d\setminus K}(\mathbf{Y}_j)-\mathbb{E}(\mathbbm{1}_{\R^d\setminus K}(\mathbf{Y})|\mathbf{X})\right|+ 2\mathbb{E}(\mathbbm{1}_{\R^d\setminus K}(\mathbf{Y})|\mathbf{X}).
 \end{align*}
 Taking expectations on both sides we obtain 
  \begin{align*}
     &\mathbb{E}\left(\sup_{f\in \mathcal{F}_{BL}}\left|\sum_{j=1}^n w_j(X;\mathbf{X}^{(n)}){f(\mathbf{Y}_j)}\mathbbm{1}_{\R^d\setminus K}(\mathbf{Y}_j)-\mathbb{E}(f(\mathbf{Y})\mathbbm{1}_{\R^d\setminus K}(\mathbf{Y})|\mathbf{X})\right|\right)\\
      &\leq   \mathbb{E}\left|\sum_{j=1}^n w_j(X;\mathbf{X}^{(n)})\mathbbm{1}_{\R^d\setminus K}(\mathbf{Y}_j)-\mathbb{E}(\mathbbm{1}_{\R^d\setminus K}(\mathbf{Y})|\mathbf{X})\right|+ 2\mathbb{P}\left(\mathbf{Y}\notin K\right).
 \end{align*}
Since the weights are consistent, there exists $n_1$ such that   \begin{align*}
       \mathbb{E}\left|\sum_{j=1}^n w_j(X;\mathbf{X}^{(n)})\mathbbm{1}_{\R^d\setminus K}(\mathbf{Y}_j)-\mathbb{E}(\mathbbm{1}_{\R^d\setminus K}(\mathbf{Y})|\mathbf{X})\right|\leq \frac{\delta\epsilon}{12}
 \end{align*}
 for all $n>n_1$. Since $\mathbb{P}\left(\mathbf{Y}\notin K\right)\leq {\delta\epsilon}/{12}$, if $n>n_1$, then 
 \begin{align}
    \label{bound_theorem_2}
    \mathbb{E}\left(\sup_{f\in \mathcal{F}_{BL}}\left|\sum_{j=1}^n w_j(X;\mathbf{X}^{(n)}){f(\mathbf{Y}_j)}\mathbbm{1}_{\R^d\setminus K}(\mathbf{Y}_j)-\mathbb{E}(f(\mathbf{Y})\mathbbm{1}_{\R^d\setminus K}(\mathbf{Y})|\mathbf{X})\right|\right)\leq  \frac{\delta \epsilon}{3}.
\end{align}
 Using Markov's inequality in \eqref{bound_theorem_2}, we obtain
 \begin{align}
    \label{bound_theorem_4}
    \mathbb{P}\left(\sup_{f\in \mathcal{F}_{BL}}\left|\sum_{j=1}^n w_j(X;\mathbf{X}^{(n)}){f(\mathbf{Y}_j)}\mathbbm{1}_{\R^d\setminus K}(\mathbf{Y}_j)-\mathbb{E}(f(\mathbf{Y})\mathbbm{1}_{\R^d\setminus K}(\mathbf{Y})|\mathbf{X})\right|>\frac{\epsilon}{2}\right)\leq  \frac{2\delta}{3}
\end{align}
 for all $n>n_1$. Finally using \eqref{bound_theorem_1}, \eqref{bound_theorem_4} and \eqref{Boun_thoerem_3}, we conclude that 
  \begin{align}
    \label{bound_theorem_5}
    \mathbb{P}\left(\sup_{f\in \mathcal{F}_{BL}}\left|\sum_{j=1}^n w_j(X;\mathbf{X}^{(n)}){f(\mathbf{Y}_j)}-\mathbb{E}(f(\mathbf{Y})|\mathbf{X})\right|>\epsilon\right)\leq  \delta
\end{align}
for all $n>N=\max(n_0,n_1)$.\medskip
\end{proof}

\begin{proof}[{\textbf{ Proof of Lemma~\ref{Lemma:BarrioLoubes}}}]
Due to the fact that finite second-order moments are not required in our setting, Theorem~2.8 in \cite{BaLo} does not directly apply. Their proof, however, relies on the weak convergence of the couplings (the joint measures solving the Kantorovich problem). In our case, we can prove a similar result using Lemma~9 in \cite{McCann1995ExistenceAU}. Indeed, since the sequences $\{ \mu_n\}_n$ and $\{ \nu_n\}_n$ are tight with respect to  weak convergence, the same result holds also for the class $\Gamma(\mu_n,\nu_n)$ of probabilities on $\R^d\times \R^d$ with  marginals $\{ \mu_n \}_n$ and $\{ \nu_n\}_n$, see Lemma~4.4 in \cite{Villani2008}. Note that all the measures 
$
    \pi_n
$,  $n\in \N$,
belong to $\Gamma(\mu_n,\nu_n)$. Denote by  $\pi\in \mathcal{P}(\R^d\times \R^d)$ the  weak limit of $
  \{  \pi_n\}_{n}
$ along a subsequence; for simplicity, we keep the index $n$ for the subsequence. Lemma~9  (ii) in \cite{McCann1995ExistenceAU} implies that the marginals of $\pi$  are $\mu$ and $\nu$. Moreover,   the support of~$\pi_n$ is  cyclically monotone: indeed,   it is contained in the subdifferential of $\phi_n$. Therefore, using Lemma~9  (i) in \cite{McCann1995ExistenceAU},  we know that also $\pi$ is supported on a cyclically monotone set.  Corollary~14 in \cite{McCann1995ExistenceAU} yields---since $\mu$  is uniformly continuous with respect to the Lebesgue measure---the existence of a unique measure with cyclically monotone support  and  marginals  $\mu$ and $\mathrm{U}_d$. As a consequence, $\pi=\left(\mathbf{Id}\times\phi\right)\# \mu
$. Since this holds along  all possible subsequence, we have the weak convergence of $\left(\mathbf{Id}\times\phi_n\right)\# \mu_n
$ to $\left(\mathbf{Id}\times\phi\right)$. At this point, to conclude the proof of Lemma~\ref{Lemma:BarrioLoubes}, we can repeat verbatim the rest of the proof of Theorem~2.8 in \cite{BaLo}. 
\end{proof}
\begin{proof}[{\textbf{ Proof of Lemma~\ref{Lemma_converge_u}}}]
Note that, for all $\epsilon>0$,
$
    \sup_{{\scriptstyle{\mathfrak G}}_i\in N_k({\scriptstyle{\mathfrak G}}_0)}\, |{\scriptstyle{\mathfrak G}}_i-{\scriptstyle{\mathfrak G}}_0 |>\epsilon$ if and only if~$ \sum_{i=1}^n\mathbbm{1}_{|{\scriptstyle{\mathfrak G}}_i- {\scriptstyle{\mathfrak G}}_0|<\epsilon}<  {k}/{n}$. Since 
    $$\inf_{\mathbf{u}\in \mathbb{S}_m}\sum_{i=1}^n\mathbbm{1}_{|{\scriptstyle{\mathfrak G}}_i- {\scriptstyle{\mathfrak G}}_0|<\epsilon}\longrightarrow \inf_{\mathbf{u}\in {\mathbb S}_m} {\rm U}_m(\mathbf{u}+\epsilon\, {\mathbb S}_m)>0 $$
    and ${k}/{n}\to 0$ as $n\to\infty$,  
$$
   \sup_{{\scriptstyle{\mathfrak G}}_0\in {\mathbb S}_m} \sup_{{\scriptstyle{\mathfrak G}}_i\in N_k({\scriptstyle{\mathfrak G}}_0)}\, |{\scriptstyle{\mathfrak G}}_i-{\scriptstyle{\mathfrak G}}_0 |\longrightarrow 0\quad\text{as $n\to\infty$}.$$
\end{proof}
\end{document}